\documentclass[a4paper,fleqn,usenatbib]{mnras}

\usepackage{newtxtext,newtxmath}

\usepackage[T1]{fontenc}
\usepackage{ae,aecompl}

\usepackage{graphicx}	
\usepackage{amsmath}	
\usepackage{amssymb}	

\usepackage{epsfig}
\usepackage{multirow}
\usepackage{float}
\def\mnras{MNRAS}
\def\apj{ApJ}
\def\aj{AJ}
\def\apjl{ApJL}
\def\apjs{ApJS}
\def\aap{A\& A}
\def\pasp{PASP}
\def\pasj{PASJ}
\def\nat{Nature}

\def\xmm{{\it XMM-Newton}}

\def\rxj{RX J0439.6-5311}

\title[RX J0439.6-5311. I. Soft Excess and Inner Accretion Flow]{Super-Eddington QSO RX J0439.6-5311. I. Origin of the Soft X-ray Excess and Structure of the Inner Accretion Flow}
\author[C. Jin, et al.]{
Chichuan Jin$^{1}$\thanks{E-mail: chichuan@mpe.mpg.de},
Chris Done$^{2}$,
Martin Ward$^{2}$
\\
$^{1}$Max-Planck-Institut f\"{u}r Extraterrestrische Physik, Giessenbachstrasse, D-85748 Garching, Germany\\
$^{2}$Centre for Extragalactic Astronomy, Department of Physics, University of Durham, South Road, Durham DH1 3LE, UK\\
}

\date{prepared for MNRAS}

\pubyear{2017}

\begin{document}
\label{firstpage}
\pagerange{\pageref{firstpage}--\pageref{lastpage}}
\maketitle

\begin{abstract}
We report the results from a recent 133 ks \xmm\ observation of a highly
super-Eddington narrow-line Type-1 QSO \rxj. This source has one
of the steepest AGN hard X-ray slopes, in
addition to a prominent and smooth soft X-ray excess. Strong
variations are found throughout the 0.3 to 10 keV energy range on all
time-scales covered by the observation, with the soft excess mainly showing low frequency
variations below 0.1 mHz while the hard X-rays show stronger
variability at higher frequencies. We perform a full set of spectral-timing
analysis on the X-ray data, including a simultaneous modelling of the time-average
spectra, frequency-dependent RMS and covariance spectra, lag-frequency and
lag-energy spectra. Especially, we find a significant time-lag signal in the low frequency band,
which indicates that the soft X-rays lead the hard by $\sim$4 ks, with a broad continuum-like profile
in the lag spectrum. Our analysis strongly supports the model where the
soft X-ray excess is dominated by a separate low temperature, optically thick
Comptonisation component rather than relativistic reflection or a
jet. This soft X-ray emitting region is several tens or hundreds of $R_{\rm g}$ away from
the hot corona emitting hard X-rays, and is probably associated with a
geometrically thick (`puffed-up') inner disc region.
\end{abstract}




\section{Introduction}
\label{sec-intro}
Narrow-Line Seyfert 1 (NLS1) galaxies are an intriguing subclass of
Active Galactic Nuclei (AGN) (Osterbrock \& Pogge 1985; Boroson \&
Green 1992), consistent on average with having lower black hole masses
and higher mass accretion rates compared with typical Seyfert 1s.
They typically have steep 2-10~keV spectra (Brandt, Mathur \& Elvis 1997), and
even steeper spectra at lower energies forming a prominent soft
X-ray excess below $\sim$ 2 keV (e.g. Boller, Brandt \& Fink 1996; Leighly
1999; Boroson 2002). Gallo (2006) proposed
that NLS1s could be split into two types. There are `complex' NLS1s
which show deep dips in their X-ray light curves, during which their
hard X-ray spectra become harder and contain strong features around
the Fe K$\alpha$ line, and `simple' NLS1s which do not show these
features.  So far the most robust AGN Quasi Periodic Oscillation (QPO)
detection is also in the `simple' NLS1 RE J1034+396 (Gierli\'{n}ski et
al. 2008; Alston et al. 2014).

While there is general consensus that the high energy 2-10~keV
emission in AGN is from the Compton up-scattering by high temperature,
optically thin electrons in a corona, the origin of the soft X-ray
excess is less clear. There are two main models proposed for this,
namely the highly relativistically smeared, partially ionised reflection model
(e.g. Miniutti \& Fabian 2004; Ross \& Fabian 2005; Fabian \& Miniutti
2005; Crummy et al 2006), and the low temperature, optically thick
Comptonisation model (Laor et al. 1997; Magdziarz et al. 1998;
Gierli\'{n}ski \& Done 2004; Mehdipour et al 2011; Done et al. 2012).
Both models can fit the spectra equally well over the classic
0.3-10~keV X-ray bandpass, and both require some fine-tuning of
parameters: Comptonisation models all give very similar temperatures
(Czerny et al. 2003; Gierli\'{n}ski \& Done 2004; Porquet et al. 2004) while
reflection models give similar ionisation states (Done \& Nayakshin 2007).

Nonetheless, the models can be separated with variability or higher
energy data. Noda et al. (2011) use the fast variability in the broad
line Seyfert 1 (BLS1) Mkn 509 to show that on the shortest timescales
there is a constant component in the spectrum which has a shape
compatible with thermal Comptonisation and not with
reflection. Similarly, including high energy data shows that
a thermal Comptonisation component is better than reflection in terms of fitting
the soft X-ray excess in BLS1s (Boissay et al. 2014; Matt et al. 2014;
Mehdipour et al. 2015). Lohfink et al. (2016) similarly find the soft
X-ray excess in the BLS1 Fairall 9 is dominated by an additional
component (though they model it with a very steep power law rather
than thermal Comptonisation). They also require some additional
relativistically smeared reflection, but this disappears when more
physical models are used to describe the narrow torus reflection
component in these data (Yaqoob et al. 2016). Thus in BLS1s there is
mounting evidence that the soft X-ray excess is an additional thermal
Comptonisation component.

However, in NLS1s the situation is more open. The key observation is
that the new spectral-timing techniques reveal a soft lag in the data,
consistent with the reflection geometry as reflected photons have
longer light paths than those from the hot corona. This predicts a
reverberation time-lag for the reflected emission which depends on the
height of the corona and the black hole mass (Fabian et al. 2009;
Zoghbi et al. 2010; Zoghbi \& Fabian 2011; Zoghbi, Uttley \& Fabian 2011;
Kara et al. 2013a, b, c; Fabian et al. 2013; Uttley et
al. 2014; Fabian et al. 2014; Chiang et al. 2014). While the reflected
spectrum is dominated by the iron line in the 6-7~keV bandpass, this
reverberation lag is as strong or stronger at soft energies for
partially ionised reflection, and current instruments have so much
more effective area below 2~keV that this is much easier to detect as
a soft lag than as an iron line lag (though the latter h$R_{\rm sx}$as also been
seen: De Marco et al. 2013; Kara et al. 2016).  

However, the reflection-dominated model often requires extreme
parameters, with the X-ray source being very close to the event horizon of a
high spin black hole. This is inconsistent with the lag-frequency
spectra seen from these objects, as these are actually dominated by a
soft lead at low frequencies, which only switches to the reverberation
soft lag for the fastest variability timescales
(e.g. Alston, Done \& Vaughan 2014; Jin, Done \& Ward 2016).
A soft lead on timescales much longer than the soft lags requires that the hard
X-rays respond to changes in the soft X-ray flux on timescales which
are longer than the light travel time between the two regions.  These
longer lags are generally interpreted as propagation, where
fluctuations in the accretion flow propagate down to progressively
smaller radii which emit harder spectra (e.g. Lyubarskii 1997; Kotov,
Churazov \& Gilfanov 2001). This is most naturally produced in a
radially stratified flow, which requires that a large fraction of the
soft X-ray excess is intrinsically produced in the flow rather than
being reflected (but see Chainakun \& Young 2016 for an alternative
dual lamppost model).

Thus it seems most likely that in NLS1s there is a mix of both
intrinsic (soft leads) and reprocessed (soft lags) emission making up
the soft X-ray excess. Gardner \& Done (2014) do a full
spectral-timing model and find that they can model the lag-frequency
spectrum in the `simple' NLS1 PG 1244+026 by a model where fluctuations
in the thermal Comptonisation inner disc region propagate down to the
high energy corona. The high energy X-rays reflect from the same inner
disc, producing a small, moderately (not extremely) smeared reflection
component. This is not sufficient to produce the observed soft lag,
but many of the illuminating photons are not reflected, so instead are
absorbed in the disc. These should thermalise, producing a (quasi)
blackbody soft component which reverberates with the harder X-rays,
producing the observed lag. This model can also be generalised to fit
the much shorter observed soft lag of a few tens of seconds seen in
the `complex' NLS1s if much of the extreme variability is associated
with occultation events (Gardner \& Done 2015).  

Both `simple' and `complex' NLS1s may well be accreting at
super-Eddington rates (Done \& Jin 2016), in which case the
occultations can easily be explained as arising from the clumpy disc
wind which should be produced at such mass accretion rates (Ohsuga \&
Mineshige 2011; Jiang et al. 2014; Takeuchi et al. 2014; Hagino et al.
2016; Done \& Jin 2016). This also provides a way to unify the `simple'
and `complex' NLS1s if the differences in their X-ray spectra and
variability are caused by differences in the viewing angle relative to
the clumpy wind. We note that complex absorption has long been
suggested as the origin of the extreme iron features associated with
the `complex' NLS1s (Turner et al. 2007; Miller et al. 2007; Sim et
al. 2010; Tatum et al. 2012).

\begin{figure}
\includegraphics[bb=25 144 518 648, scale=0.45]{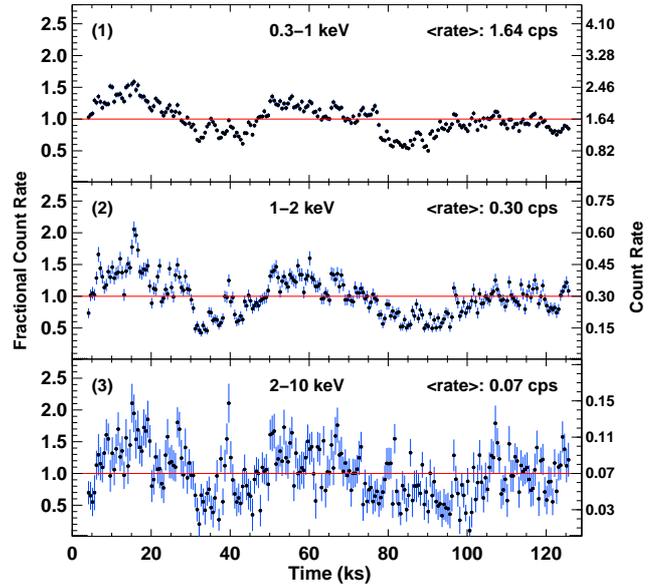}
\caption{Light curves of \rxj\ observed by the \xmm\ EPIC-pn camera in the 0.3-1, 1-2 and 2-10 keV bands (background subtracted, binned with 500 s). The left y-axis is in the unit of fractional count rate relative to the mean count rates (<rate>), while the right y-axis is the absolute count rate. High background periods at the beginning and end of the observation have been masked out.}
\label{fig-lc}
\end{figure}

The number of NLS1s which are bright enough and/or have long
enough observations to allow a detailed spectral-timing analysis is
very limited. Here we present the analysis of a recent 133 ks
\xmm\ observation of another `simple' NLS1, namely \rxj.
This source is a relatively nearby, Type-1 narrow-line Quasi-Stellar Object (QSO)
($z = 0.243$, Thomas et al. 1998), with a Galactic gas column density of
$N_{\rm H}=7.45\times10^{19}$ cm$^{-2}$ (Kalberla et al. 2005) and no
significant intrinsic extinction (Grupe et al. 2010), indicating a very
clear line of sight. The H$\beta$ FWHM of \rxj\ was found to be
only $700\pm140$~km~s$^{-1}$, which is the narrowest among all the 110
soft X-ray selected AGN in Grupe et al. (2004) (also see Bian \& Zhao
2004). Based on the {\it Swift} observation, Grupe et al. (2010) report
a soft X-ray slope of $\sim$2.2, a single-epoch black hole mass of
$3.9\times10^{6}$ M$_{\odot}$ and an extreme Eddington ratio of 12.9 which is
derived from the broadband spectral energy distribution (SED).
Therefore, \rxj\ has a very similar X-ray spectral shape to the
`simple' NLS1s such as PG 1244+026 and RE J1034+396,
but with an even more extreme mass
accretion rate, potential more comparable to 1H 0707-495 (Done \& Jin
2016). The study of this source allows us to further understand the
soft excess mechanism in these unobscured, highly accreting AGN, and to
identify additional ubiquitous properties among these sources.

\begin{figure*}
\includegraphics[scale=0.6,angle=90]{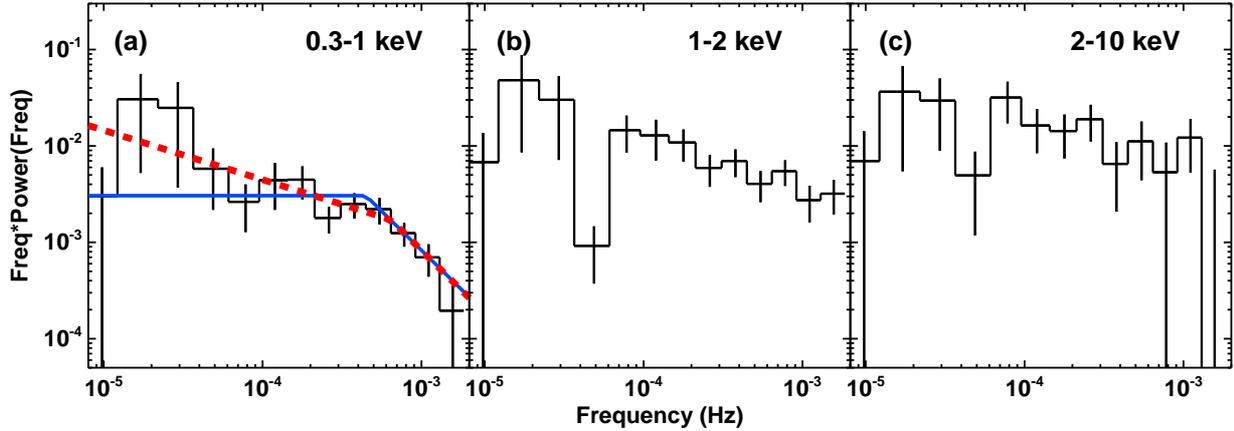}
\caption{The PSD of \rxj\ in various X-ray bands. These are produced by the {\sc powspec} tool (FTOOLS v6.19), with the Poisson noise subtracted, rebinned with a geometrical step of 1.4, and plotted in $f{\cdot}P(f)~vs~f$. The normalisation is chosen such that the integration of the PSD over a specific frequency band gives the fractional excess variance. In Panel-a the blue solid line and red dash line are two best-fit broken power law models. The blue line's slope is fixed at $\alpha=-1$ below the break frequency (see Section~\ref{sec-powspec}).}
\label{fig-powspec}
\end{figure*}

This paper is organised as follows. Firstly we describe the latest
\xmm\ observation of \rxj\ and our data reduction procedures. Then we
present the source's variability properties in Section 3. A detailed
spectral timing modelling is present in Section 4, in order to
separate the various components and to understand their origins. In
Section 5 we report results from our X-ray inter-band coherence and
covariance analysis. The inter-band time-lag analysis is present in Section 5.
Discussion of the soft X-ray excess and its potential
connection with the region of inner disc which we claim is puffed up,
is given in Section 7. Finally, Section 8 summaries the main results
of this work. We adopt a flat universe model for the luminosity
distance with the Hubble constant H$_{0} = 72$ km s$^{-1}$ Mpc$^{-1}$,
$\Omega_{\rm \Lambda} = 0.73$ and $\Omega_{\rm M} = 0.27$.

\section{\xmm\ Observation and Data Reduction}
\label{sec-obs}
\xmm\ (Jansen et al. 2001) observed \rxj\ on 2016-02-12 for a continuous duration of 133 ks (PI: C. Jin). All three European Photon Imaging Cameras (EPIC) (pn, MOS1, MOS2) were operated in the {\it Imaging} data-mode, with the EPIC-pn camera in the {\it PrimeLargeWindow} mode with the {\it thick} filter, and the two MOS cameras in the {\it PrimePartialW3} mode with the {\it thin} filter. The two Reflection Grating Spectrometer (RGS) cameras were both in the {\it Spectroscopy} data-mode. The Optical Monitor (OM) was in the {\it Imaging+Fast} data-mode with exposures in six optical/UV filters (V, B, U, UVW1, UVM2, UVW2).

We followed the standard procedures of using the {\sc SAS} software (v15.0.0) and the latest calibration files to reduce the data. The {\tt epproc} and {\tt emproc} tasks were used to reprocess the EPIC data and create the event files. The {\tt rgsproc} and {\tt omfchain} tasks were used to reprocess the RGS and OM data. For the EPIC data, we defined a source extraction region of 80 arcsec centred at the position of \rxj, and extracted background from a nearby region of the same size without any point source contamination. By checking the background light curve above 12 keV, we found that this observation is almost free from high background flares except for some short periods at the beginning and end of the exposure. As a result we excluded all contaminated data from the first 4 ks and last 8 ks using the {\tt tabgtigen} task. The filtered observation has a continuous 121 ks duration with 113 ks exposure in EPIC-pn (94.9\% live time) and 120 ks exposure in MOS1 and MOS2 (99.5\% live time). The background subtracted mean source count rates are 2.01 counts per second (cps), 0.64 cps and 0.61 cps in EPIC-pn, MOS1 and MOS2, respectively, which are all well below the pile-up threshold for the chosen observing mode and filter. We further checked that there is no photon pile-up in any of the three EPIC cameras by running the {\tt epatplot} task\footnote{In the {\it PrimeLargeWindow} mode, EPIC-pn has a 3 cps threshold for 2.5\% flux-loss. We noticed that the source count rate may exceed 3 cps slightly during short flaring peaks of the first 30 ks. As a double check for the pile-up effect, for the first 30 ks, we re-ran the SAS {\tt epatplot} task and found no significant pileup effect. We also compared the spectra before and after excluding the central 10 arcsec Point Spread Function (PSF) area and found no significant differences either, so we can conclude that pile-up should not affect our analysis.}. In the subsequent analysis, we only adopted good events (FLAG=0) with PATTERN $\le$ 4 for EPIC-pn (i.e. single and double patterns) and PATTERN $\le$ 12 for MOS1 and MOS2 (i.e. single, double, triple, quadruple patterns)\footnote{https://xmm-tools.cosmos.esa.int/external/xmm\_user\_support/\\documentation/uhb/epic\_evgrades.html}.

The EPIC light curves and spectra were extracted from the source and background regions separately using the {\tt evselect} task. These regions were chosen to avoid the CCD area where the copper instrumental background is high (Freyberg et al. 2004). The {\tt epiclccorr} task was used to perform the background subtraction and {\it Absolute} corrections on the source light curves for various instrumental factors. The {\tt rmgfen}, {\tt arfgen} and {\tt backscale} tasks were used to create response and auxiliary files and to calculate the scaling factor. The RGS spectra were extracted using the {\tt rgsproc} task. All the EPIC and RGS spectra were grouped with at least 20 counts per bin using the {\tt grppha} tool ({\sc FTOOLS} v6.19). Spectral fittings were performed using the {\sc Xspec} (v12.9.0u) package (Arnaud 1996).

\section{X-ray Variability Analysis}
\label{sec-var}
Variability studies can provide crucial information concerning the mechanism of the soft X-ray emission and to help break the degeneracy of spectral components. This clean and uninterrupted 120 ks \xmm\ observation of \rxj\ reveals the characteristics of its energy-dependent variability in great detail.

\subsection{EPIC Light Curves}
\label{sec-lc}
We first compared the light curves extracted from the 0.3-1, 1-2 and 2-10 keV bands. After applying all the filtering, correction and background subtraction (see Section~\ref{sec-obs}), the intrinsic source variability can be visualised in these light curves. Fig.\ref{fig-lc} shows that \rxj\ exhibits strong variability in all three X-ray bands. The count rate varies by $\pm$50\% over tens of ks. For timescales of ks and shorter, the 2-10 keV band shows a factor of 2 variability, while the 0.3-1 keV band is more stable on these short timescales. This is confirmed by the intrinsic fractional root-mean-square (RMS) variability (Edelson et al. 2002; Markowitz, Edelson \& Vaughan 2003; Vaughan et al. 2003), which is $22.4\pm0.2$\%, $28.7\pm0.4$\%, $47.5\pm1.1$\% for the 0.3-1, 1-2 and 2-10 keV bands, respectively\footnote{RMS errors were calculated using Equation B2 in Vaughan et al. (2003).}.

\subsection{Power Spectra Density (PSD)}
\label{sec-powspec}
A PSD quantifies the variability power in different frequency bands. We first combined light curves from all three EPIC cameras to maximise the signal-to-noise (S/N) using the {\tt lcmath} tool ({\sc FTOOLS} v6.19), and then calculated the PSD from the combined light curve using the {\tt powspec} tool. The normalisation of {\tt powspec} was chosen to be -2, which allowed it to produce a white noise subtracted PSD, the integration of which gives the excess variance. The resultant PSDs are binned with a geometrical step of 1.4 and plotted in Fig.\ref{fig-powspec}.

These PSDs show significant variability across a wide frequency band
($10^{-5}-10^{-3}$ Hz) covered by the \xmm\ observation, extending
from the soft up to the hard X-rays. There is no detection of periodic
signals, which is not surprising as QPOs in AGN are very rare, and
not always easy to detect even when present (Middleton et al. 2009; Alston
et al. 2014)

For the high frequency
band of $f\ge10^{-4}$ Hz, the hard X-rays show stronger RMS
variability than seen in the soft X-rays. This is also observed in
several other NLS1s of high mass accretion rates (e.g. PG 1244+026, Jin et
al. 2013; RE J1034+396, Middleton et al. 2009; Ark 564, M$^{c}$Hardy
et al. 2007). There is a high frequency rollover in the 0.3-1 keV band
PSD (Fig.\ref{fig-powspec}a). A broken power law fit to the PSD
indicates a best-fit break frequency of
$\nu_{\rm b}=6.4^{+4.7}_{-2.7}~\times10^{-4}$ Hz (1 $\sigma$ confidence
level, red dashed line in Fig.\ref{fig-powspec}a). Assuming a power
law form of $P(f)\propto f^{\alpha}$, we find
$\alpha_{1}=-1.51^{+0.19}_{-0.19}$ (the 90\% confidence range is
[-1.81, -1.19]) and $\alpha_{2}=-2.64^{+0.87}_{-1.36}$ below and above
the break frequency. This indicates a similar PSD shape to some other
NLS1s such as Ark 564 around the high frequency break (M$^{c}$Hardy et
al. 2007). If the slope below the break frequency is fixed at -1, we find
$\nu_{\rm b}=4.4^{+1.1}_{-4.4}~\times10^{-4}$ Hz (blue
solid line in Fig.\ref{fig-powspec}a) which is slightly lower
than in the previous fit, with $\alpha_{1}=-2.62^{+1.31}_{-0.70}$ which
is consistent with the previous value. The S/N of PSDs of the higher
energy bands is not sufficient to provide strong constraints on the
broken power law model.

\begin{figure}
\includegraphics[bb=54 210 558 648, scale=0.45]{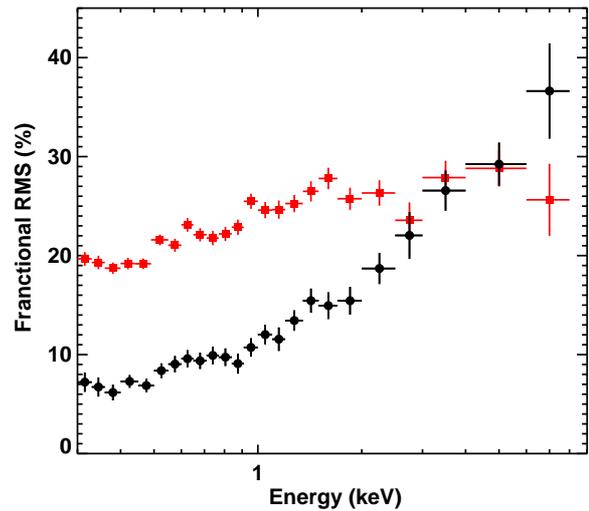}
\caption{The energy-dependent RMS fractional variability in the low frequency band ($8.3\times10^{-6}-10^{-4}$ Hz, red square points) and high frequency band ($10^{-4}-1.25\times10^{-3}$ Hz, black circular points). }
\label{fig-rms}
\end{figure}

\subsection{Frequency-dependent RMS Spectra}
\label{sec-rms}
The X-ray variability not only has an energy dependence, but also has a clear frequency dependence as already seen in the PSD. This can be revealed directly by the frequency-dependent RMS spectra (Ar\'{e}valo et al. 2008; Middleton et al. 2009; Jin et al. 2013, here after: J13). We adopted the prescription in Ar\'{e}valo et al. (2008) to calculate the frequency-dependent RMS from the PSD in every energy bin, and used the equations given in Poutanen, Zdziarski \& Ibragimov (2008) to calculate the error. To reduce the number of zero-count bins in the light curve, which would otherwise bias the RMS, we combined light curves from all three EPIC cameras and chose a binning time of 400 s. With careful division of energy bins, all the light curves below 4 keV have $\le 5$ zero-count bins out of 305 bins (i.e. $<$ 2\%), while the 4-6 keV light curve has 9 zero-count bins (3\%) and 6-8 keV light curve has 70 zero-count bins (23\%). The light curve above 8 keV has more than 50\% zero/negative-count bins after background subtraction, so we did not use it in the timing analysis. Each light curve has bins of 400 s and 120 ks long, which corresponds to a frequency range of $8.3\times10^{-6}-1.25\times10^{-3}$ Hz. Since the PSDs in Fig.\ref{fig-powspec} show the soft X-rays have smaller RMS than the hard X-rays above $10^{-4}$ Hz, we chose to divide the frequency range into two sub-ranges at $10^{-4}$ Hz, and refer to $f\ge1\times10^{-4}$ Hz as the high frequency (HF) band and $f<1\times10^{-4}$ Hz as the low frequency (LF) band. 

Similar to other `simple' NLS1s such as RX J0136.9-3510 (Jin et al. 2009), RE J1034+396 (Middleton et al. 2009), PG 1244+026 (J13) and RX J1140.1-0307 (Jin, Done \& Ward 2016), the HF RMS of \rxj\ increases steeply from soft to hard X-rays, suggesting a strong dilution to the HF variability in the soft excess (Fig.\ref{fig-rms}). The LF RMS spectrum shows strong variability across the entire X-ray band with no obvious trend, which is different from PG 1244+026 where the LF RMS was strongly suppressed in the hard X-ray band (J13), and also different from RE J1034+396 where the LF RMS is small across the entire X-ray band (Middleton et al. 2009). These RMS spectra imply that there must be at least two major X-ray components with different variability behaviours dominating the soft and hard X-ray bands separately, which we will model in the next section.

\begin{figure*}
\centering
\begin{tabular}{cc}
\includegraphics[bb=30 124 540 612,scale=0.41]{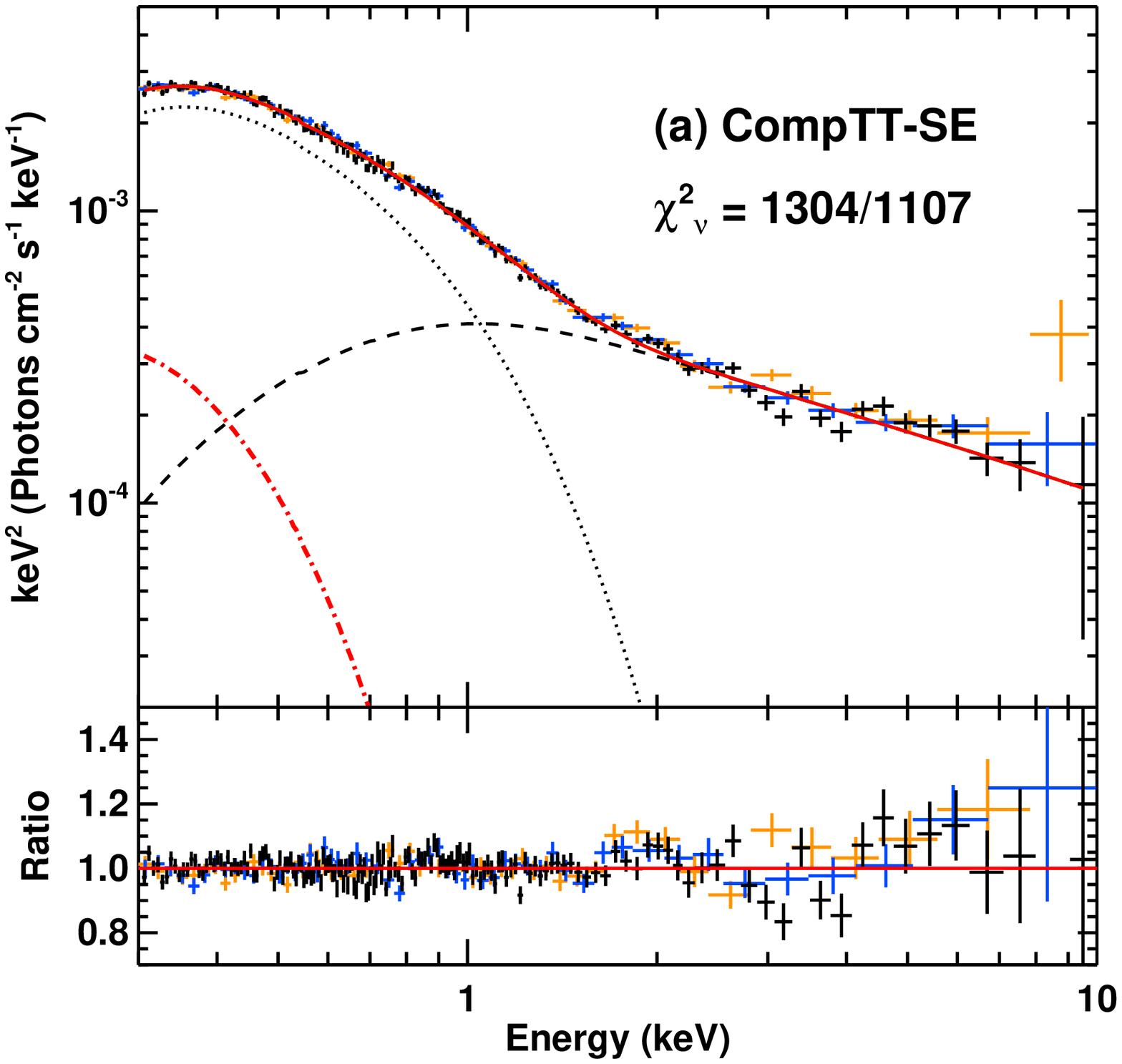} &
\includegraphics[bb=30 124 540 612,scale=0.41]{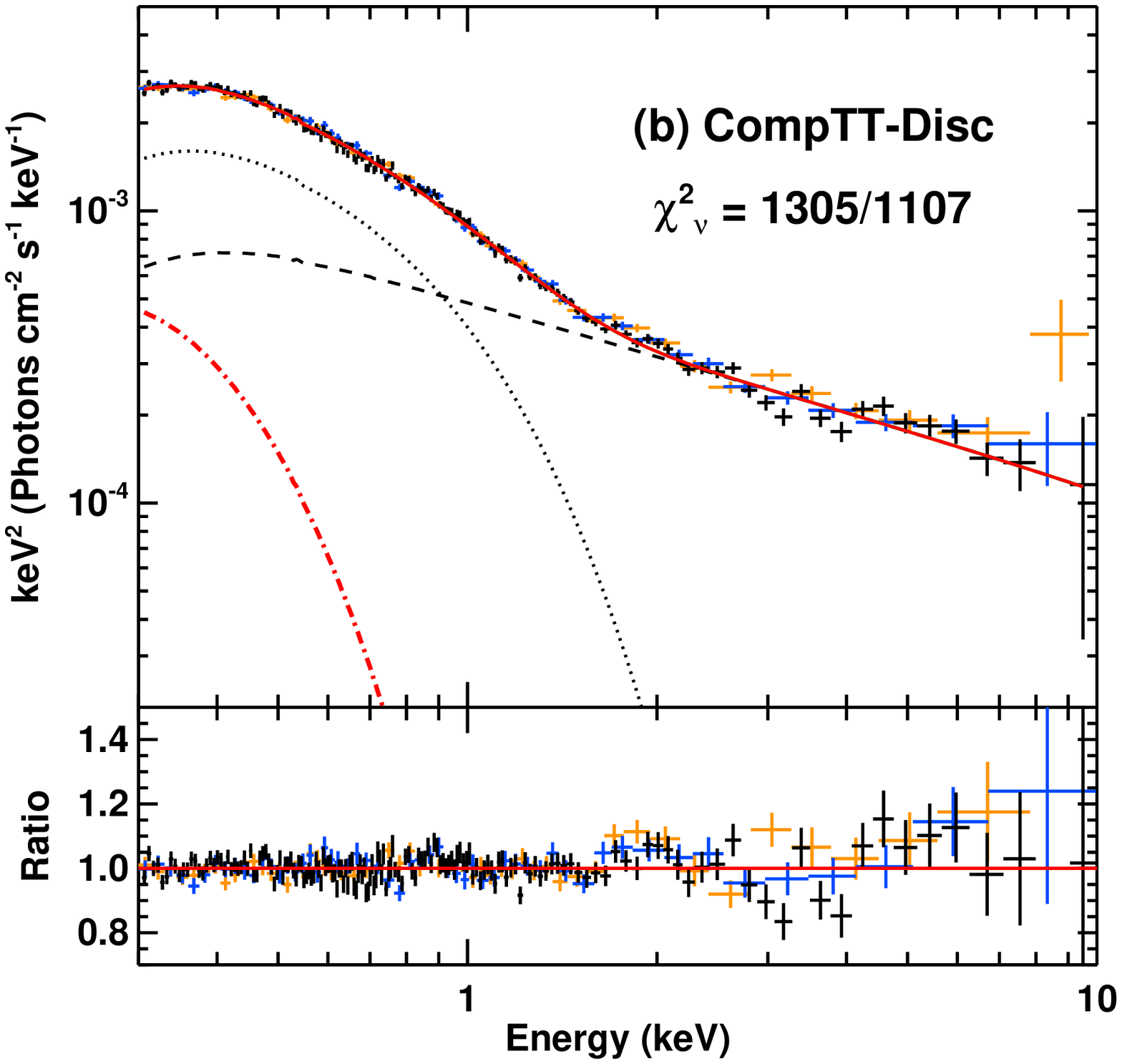} \\
\includegraphics[bb=30 124 540 612,scale=0.41]{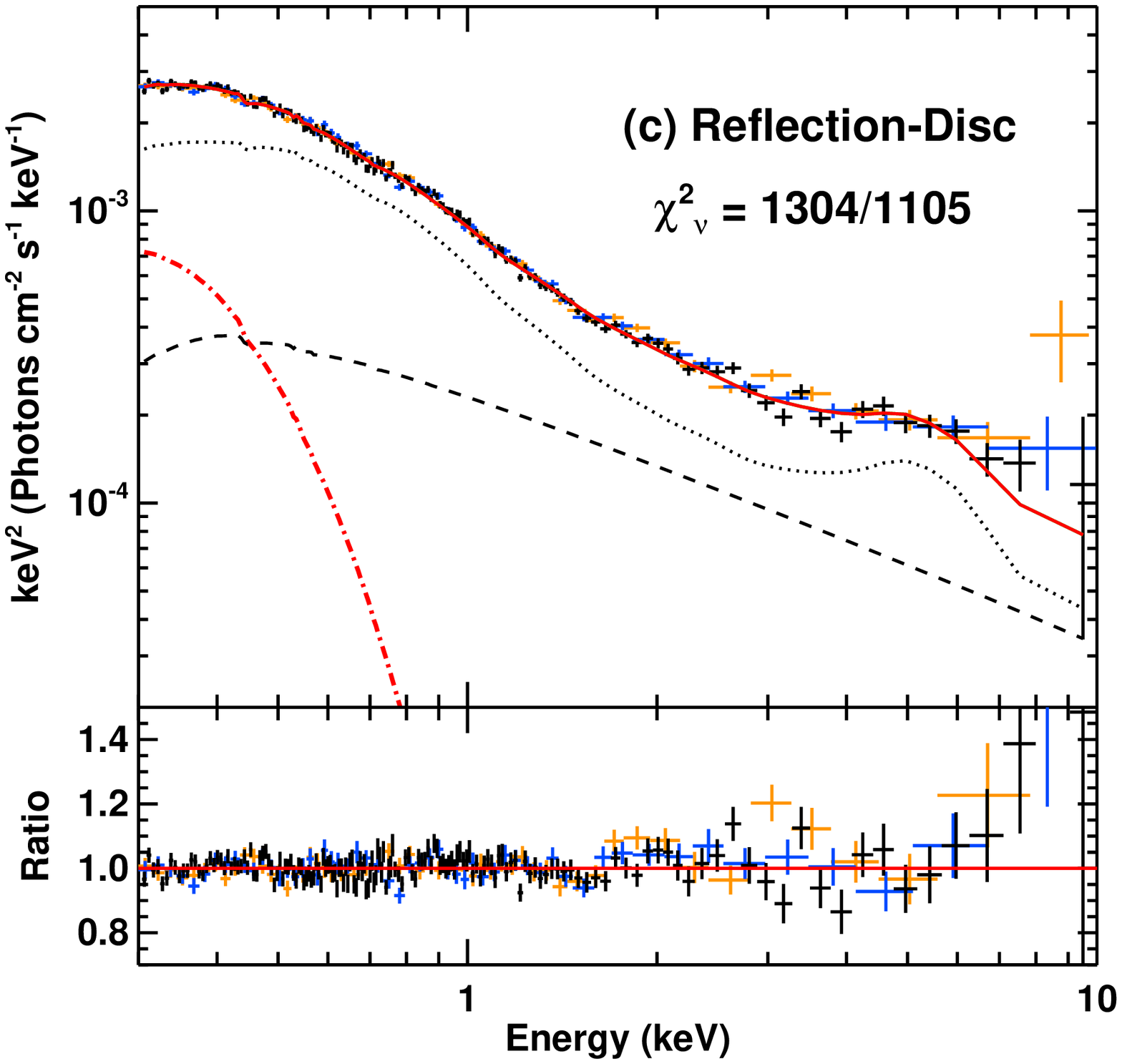} &
\includegraphics[bb=30 124 540 612,scale=0.41]{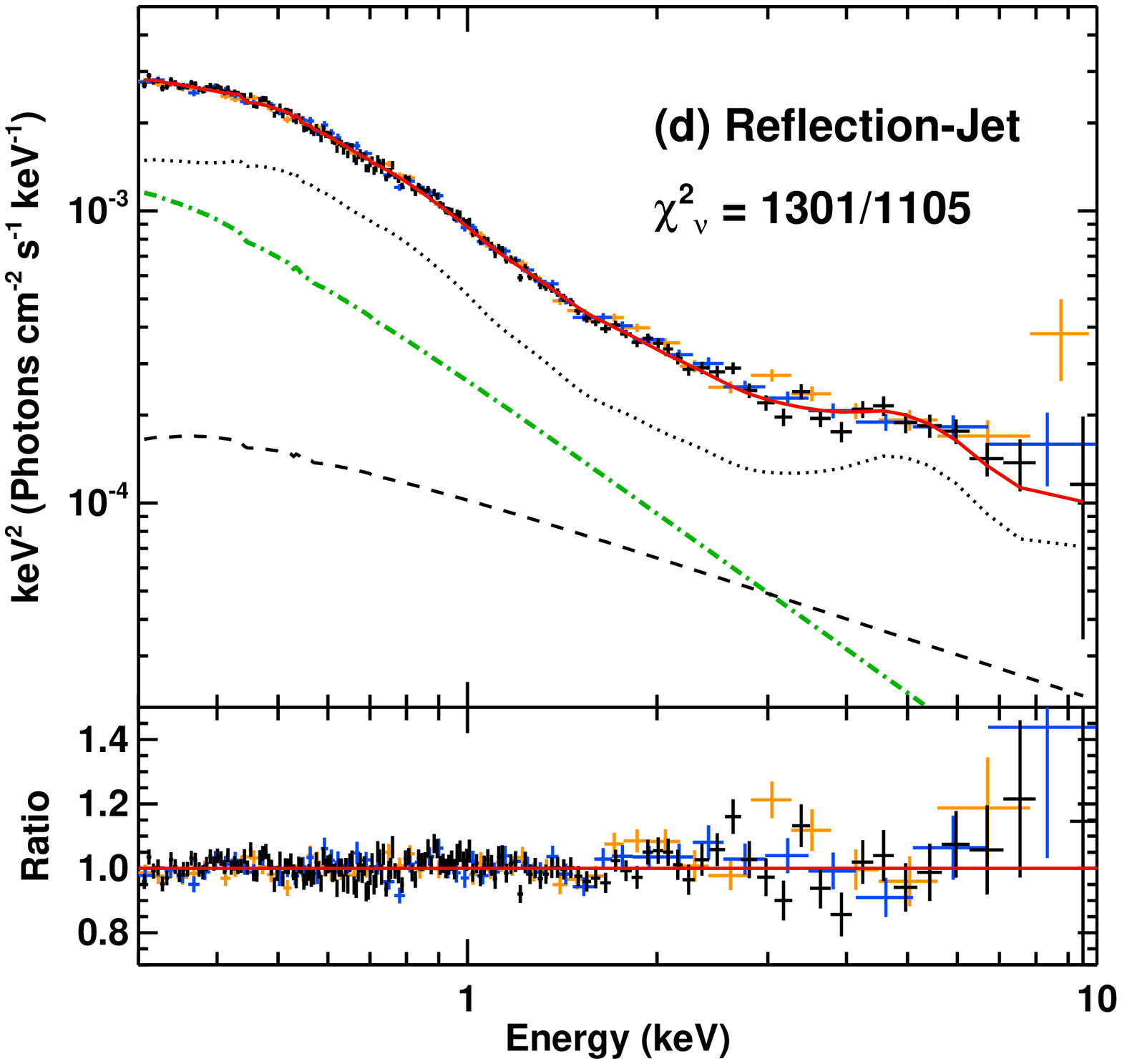} \\
\end{tabular}
\caption{Modelling the time-averaged EPIC spectra, including the EPIC-pn spectrum in black, MOS1 in blue and MOS2 in orange. Panel-a: a disc component ({\sc diskbb}, red dash-dot curve), plus a soft X-ray Comptonisation component ({\sc comptt}, dotted curve), plus a hard X-ray Comptonisation component ({\sc nthcomp} dash curve) whose seed photons are provided by the soft X-ray component. Panel-b: the same model components as in panel-a, but the seed photons of {\sc nthcomp} come from the inner disc. Panel-c: {\sc diskbb} (red dash-dot curve), plus {\sc nthcomp} (dash curve), plus a relativistic reflection component ({\sc kdblur}$\cdot${\sc rfxconv}, dotted curve). Panel-d: similar to panel-c, but replacing {\sc diskbb} with a power law jet component (green dash-dot curve) and replacing {\sc nthcomp} with a simple power law model. The best-fit parameters can be found in Table~\ref{tab-specfit}.}
\label{fig-specfit1}
\end{figure*}

\begin{table*}
 \centering
   \caption{Best-fit parameters in
     Fig.\ref{fig-specfit1}. The upper and lower limits are for the 90\%
     confidence range. In the Reflection-Jet model, {\sc powerlaw1} is for the jet emission, and {\sc powerlaw2} is for the intrinsic corona emission. {\it low} ({\it up}) indicates the parameter's lower (upper) limit. The disc inclination angle is fixed at 30$\degr$.}
     
     \begin{tabular*}{\textwidth}{@{}cc@{}}
     
\begin{tabular}{@{}clll@{}}
\hline
  Model & Component & Parameter & Value\\
\hline
{\sc CompTT}   & (Fig.\ref{fig-specfit1}a) & \multicolumn{2}{l}{$\chi^{2}_{\rm \nu}~=~1304/1107~=~1.18$} \\
{\sc  -SE}       & {\sc tbnew}     & $N_{\rm H}$ (10$^{22}$ cm$^{-2}$) & 0~$^{+0.011}_{-0}$\\
             & {\sc diskbb}     & $T_{\rm in}$ (keV)                   & 0.075~$^{+0.011}_{-0.006}$\\
             & {\sc diskbb}     & norm                       & 2.53~$^{+4.19}_{-2.53}\times 10^{3}$\\
             & {\sc nthcomp}   & $\Gamma$                   & 2.62~$^{+0.05}_{-0.06}$\\
             & {\sc nthcomp}   & $kT_{\rm seed}$ (keV)          & tied to kT$_{\rm e}$\\
             & {\sc nthcomp}   & norm                       & 4.16~$^{+0.45}_{-0.46}\times 10^{-4}$ \\
             & {\sc comptt}    & $kT_{\rm e}$ (keV)              & 0.22~$^{+0.02}_{-0.02}$\\
             & {\sc comptt}    & $\tau$                     & 16.6~$^{+8.0}_{-1.8}$\\
             & {\sc comptt}    & norm                       &0.12~$^{+0.03}_{-0.07}$\\
             & {\sc const}     & (MOS1) & 1.005 $^{+0.007}_{-0.007}$ \\
             & {\sc const}     & (MOS2) & 1.009 $^{+0.008}_{-0.008}$ \\
\hline
\end{tabular} 

&
\begin{tabular}{@{}clll@{}}
\hline
  Model & Component & Parameter & Value\\
\hline
{\sc CompTT}   & (Fig.\ref{fig-specfit1}b) & \multicolumn{2}{l}{$\chi^{2}_{\rm \nu}~=~1305/1107~=~1.18$} \\
{\sc -Disc }            & {\sc tbnew}     & $N_{\rm H}$ (10$^{22}$ cm$^{-2}$) & 0~$^{+0.013}_{-0}$\\
             & {\sc diskbb}     & $T_{\rm in}$ (keV)                   & 0.075~$^{+0.004}_{-0.008}$\\
             & {\sc diskbb}     & norm                       & 3.54~$^{+3.12}_{-3.54}\times 10^{3}$\\
             & {\sc nthcomp}   & $\Gamma$                   & 2.62~$^{+0.06}_{-0.07}$\\
             & {\sc nthcomp}   & $kT_{\rm seed}$ (keV)          & tied to $kT_{\rm in}$\\
             & {\sc nthcomp}   & norm                       & 4.90~$^{+0.32}_{-0.40}\times 10^{-4}$ \\
             & {\sc comptt}    & $kT_{\rm e}$ (keV)              & 0.22~$^{+0.03}_{-0.03}$\\
             & {\sc comptt}    & $\tau$                     & 17.0~$^{+19.8}_{-3.0}$\\
             & {\sc comptt}    & norm                       &0.082~$^{+0.03}_{-0.06}$\\
             & {\sc const}     & (MOS1) & 1.005 $^{+0.007}_{-0.007}$ \\
             & {\sc const}     & (MOS2) & 1.009 $^{+0.008}_{-0.008}$ \\
\hline
\end{tabular} \\

\begin{tabular}{@{}clll@{}}
\hline
  Model & Component & Parameter & Value\\
  \hline
{\sc Reflection}   & (Fig.\ref{fig-specfit1}c) & \multicolumn{2}{l}{$\chi^{2}_{\rm \nu}~=~1304/1105~=~1.18$} \\
{\sc -Disc  }   & {\sc tbnew}     & $N_{\rm H}$ (10$^{22}$ cm$^{-2}$) & 0.017~$^{+0.010}_{-0.009}$\\
             & {\sc diskbb}     & $T_{\rm in}$ (keV)                   & 0.073~$^{+0.007}_{-0.008}$\\
             & {\sc diskbb}     & norm                       & 9.19~$^{+6.66}_{-4.04}\times 10^{3}$\\
             & {\sc nthcomp}   & $\Gamma$                   & 2.83~$^{+0.03}_{-0.03}$\\
             & {\sc nthcomp}   & $kT_{\rm seed}$ (keV)          & tied to $kT_{\rm in}$\\
             & {\sc nthcomp}   & norm                       & 2.41~$^{+0.50}_{-0.65}\times 10^{-4}$\\
             & {\sc kdblur}    & Index                      & 5.36~$^{+0.58}_{-0.24}$\\
             & {\sc kdblur}    & $R_{\rm in}$ ($R_{\rm g}$)           & 2.56~$^{+0.21}_{-0.19}$\\
             & {\sc rfxconv}   & $^\dagger$rel\_refl ($\Omega/{2\pi}$)           & -3.70~$^{+2.22}_{-0.38}$\\
             & {\sc rfxconv}   & $Fe_{\rm abund}$ (Solar)                & 1.38~$^{+0.08}_{-0.40}$\\
             & {\sc rfxconv}   & log $\xi$                   & 3.41~$^{+0.06}_{-0.05}$\\
             & {\sc const}     & (MOS1) & 0.962~$^{+0.008}_{-0.008}$ \\
             & {\sc const}     & (MOS2) & 0.966~$^{+0.008}_{-0.008}$ \\
\hline
\end{tabular}

&
\begin{tabular}{@{}clll@{}}
\hline
  Model & Component & Parameter & Value\\
  \hline
{\sc Reflection}   & (Fig.\ref{fig-specfit1}d) & \multicolumn{2}{l}{$\chi^{2}_{\rm \nu}~=~1301/1105~=~1.18$} \\
{\sc  -Jet  }    & {\sc tbnew}     & $N_{\rm H}$ (10$^{22}$ cm$^{-2}$) & 0.012~$^{+0.010}_{-0.006}$\\
             & {\sc powerlaw1}     & $\Gamma$                   & 3.55~$^{+0.31}_{-0.11}$\\
             & {\sc powerlaw1}     & norm                       & 2.71~$^{+0.48}_{-0.73}\times 10^{-4}$\\
             & {\sc powerlaw2}   & $\Gamma$                   & 2.70~$^{+0.03}_{-0.03}$\\
             & {\sc powerlaw2}   & norm                       & 1.06~$^{+0.66}_{-0.33}\times 10^{-4}$\\
             & {\sc kdblur}    & Index                      & 5.76~$^{+0.50}_{-0.36}$\\
             & {\sc kdblur}    & $R_{\rm in}$ ($R_{\rm g}$)           & 2.44~$^{+0.14}_{-0.07}$\\
             & {\sc rfxconv}   & rel\_refl ($\Omega/{2\pi}$)                     & -8.52~$^{+3.25}_{-1.48~low}$\\
             & {\sc rfxconv}   & $Fe_{\rm abund}$ (Solar)                & 1.00~$^{+0.11}_{-0.10}$\\
             & {\sc rfxconv}   & log $\xi$                   & 3.12~$^{+0.08}_{-0.18}$\\
             & {\sc const}     & (MOS1) & 0.970~$^{+0.008}_{-0.007}$ \\
             & {\sc const}     & (MOS2) & 0.974~$^{+0.008}_{-0.008}$\\
\hline
\end{tabular}  \\

   \end{tabular*}

\begin{flushleft}
Notes. $kT_{\rm seed}$ and $kT_{\rm e}$ are the temperature of the seed photon and electron, separately. $\tau$ is the scattering optical depth. `ref\_refl' is the relative reflection normalisation in units of $\Omega/{2\pi}$. $\dagger$A negative `ref\_refl' allows the {\sc rfxconv} model to only return the reflection spectrum (see footnote~\ref{ft-rfxconv}). $\xi$ is the ionisation parameter defined as the ratio of number density between the ionising photons and free electrons. {\sc diskbb} normalisation is defined as $(R_{\rm in}/D_{10})^2~cos~\theta$, where $R_{\rm in}$ is the apparent inner disc radius in km, $D_{10}$ is the source distance in units of 10 kpc, $\theta$ is the disc inclination angle. {\sc nthcomp} and {\sc powerlaw} normalisations are given in units of $\rm photons~keV^{-1}cm^{-2}s^{-1}$ at 1 keV.
\end{flushleft}
 \label{tab-specfit}
\end{table*}

\section{Time-averaged X-ray Spectra}
\label{sec-specfit-mean}

\subsection{EPIC Spectra}
Based on previous studies of similar NLS1s such as RE J1034+396 (Middleton et al. 2009)
and PG 1244+026 (J13), we perform spectral fittings on the time-averaged spectra
by employing two distinct physical scenarios,
namely the Comptonisation-dominated scenario (Laor et al. 1997) and
the reflection-dominated scenario (Fabian \& Miniutti 2005).
Some other models were also proposed in the literatures,
such as the smeared absorption model (Gierli\'{n}ski \&
Done 2004, 2006; but see Schurch \& Done 2007) and partial covering
absorption model (Miller et al. 2007; Reeves et al. 2008; Sim et
al. 2010; Tatum et al. 2012), but these models were less favoured by
other similar NLS1s in terms of both spectral shape and variability
grounds (Miniutti et al. 2009; Ai et al. 2011; J13).

To increase the spectral constraints, we fit all three EPIC spectra
simultaneously, and adopt a free constant to account for their small
normalisation differences. In the EPIC-pn spectrum we noticed some absorption features within 8-9 keV which was not found in the two MOS spectra, and was not found before the background subtraction, so these feature are likely caused by the subtraction Cu K$\alpha$ instrumental background when the source count rate is too low above 8 keV, so we excluded the EPIC-pn data within 8-9 keV to avoid this background contamination. 
Galactic extinction along the line of sight
to \rxj\ is $N_{\rm H}=7.45\times10^{-19}$ cm$^{-2}$ (Kalberla et al. 2005),
which is modelled with the {\sc tbnew} model in {\sc Xspec} using
cross-sections of Verner et al. (1996) and the interstellar medium (ISM)
abundances of Wilms, Allen \& McCray (2000). Host galaxy extinction is
also modelled with {\sc tbnew} by setting the redshift to 0.243. In
both the Comptonisation and reflection scenarios, hard X-rays come
from the Compton up-scattering of seed photons by electrons in the hot
corona. We model this X-ray component with the {\sc nthcomp} model
(Zdziarski, Johnson \& Magdziarz 1996) and fix the electron
temperature at 100 keV (J13). Since \rxj\ has a high mass accretion
rate and a relatively low black hole mass, its accretion disc emission
is probably hot enough to extend into the soft excess (Done et
al. 2012, 2013; DJ16). Thus we add a {\sc diskbb} model to the soft
X-ray band. However, the disc and coronal power law are not sufficient to
provide a good fit to the soft excess (Done et al. 2012),
and so it requires another model component. Below we present the
spectral fitting of Comptonisation- and reflection-dominated models
for the soft excess. Note that the physical mechanisms of these two
models do not conflict with each other, and so both processes can
contribute to a single spectrum at the same time (Gardner \& Done 2015).
Our objective is to understand what is the dominant emission mechanism
responsible for the soft excess in \rxj.

\subsubsection{Comptonisation-dominated Models}
\label{sec-ave-compt}
In the Comptonisation model, the soft excess mainly arise from the
Compton up-scattering by electrons with lower temperature and higher
opacity than those in the hot corona, which can be modelled with the
{\sc comptt} model (Titarchuk 1994). Then the total model in {\sc
  Xspec} is {\sc diskbb}+{\sc comptt}+{\sc nthcomp} multiplied by two
{\sc tbnew} components and a free constant.

Another degree of freedom in this model is the origin of seed photons
for the hard X-ray corona, which can be either from the thermal disc emission which can be hot enough to reach the soft X-ray band at small radii (Done et al. 2012), or from the
dominant soft excess component, depending on the geometry of the inner disc region which
is still not clear. Firstly, we assume the soft X-ray Comptonisation
occurs in a region between the disc and hot corona, and so seed
photons for the hot corona are from the soft excess (hereafter: the
CompTT-SE model). This model can fit all three EPIC spectra reasonably
well with $\chi^2_v=1304/1107$ (Fig.\ref{fig-specfit1}a and
Table~\ref{tab-specfit}). No intrinsic extinction is required by the
spectral fitting. The hard X-ray Comptonisation shows a very steep
spectral shape with the photon index $\Gamma=2.62^{+0.05}_{-0.06}$. The
soft X-ray Comptonisation requires an electron temperature of
$0.22^{+0.02}_{-0.02}$ keV and an optical depth of $\tau=16.6^{+8.0}_{-1.8}$,
similar to that observed in the soft X-ray excess in all AGN
(Gierli\'{n}ski \& Done 2004). The temperature of the inner
thermal disc is found to be $75^{+11}_{-6}$ eV, but its normalisation
is poorly constrained. Removal of this {\sc diskbb} component causes little change to the $\chi^2$.
Then we assume the hard X-ray
Comptonisation receives the required seed photons from the inner thermal disc emission
(hereafter: the CompTT-Disc model). This model also produces a good
fit to the EPIC spectra ($\chi^2_v=1305/1107$,
Fig.\ref{fig-specfit1}b). The best-fit parameters of the other two
components are similar to those in the CompTT-SE model. Therefore,
the time-averaged spectra cannot distinguish between these models for
the origin of seed photons for the hard X-ray Comptonisation
component.

\subsubsection{Reflection-dominated Models}
\label{sec-ave-refl}
In the reflection scenario the soft excess is dominated by an ionised
reflection component, and the seed photons for the hot corona come
from the inner disc (hereafter: the Reflection-Disc model). We use the
{\sc rfxconv} model
(Ross \& Fabian 2005, recoded by Kolehmainen, Done \& D\'{i}az Trigo 2011)
to calculate the reflection spectrum and then
smear it using the {\sc kdblur} model (Laor et al. 1991).
The {\sc rfxconv} model combines the ionised disc table model from Ross \& Fabian (2005) with the Compton reflection model from Magdziarz \& Zdziarski (1995)\footnote{\label{ft-rfxconv}see https://heasarc.gsfc.nasa.gov/xanadu/xspec/manual/node278.html}. Its parameters include the redshift, the relative reflection normalisation (ref\_refl $=\Omega/2\pi$) which determines the relative strength between the input spectrum and the reflection spectrum, the Iron abundances in the unit of Solar abundances, inclination angle and the ionisation parameter ($log~\xi$). A strong constraint of the disc inclination should come from the modelling of Iron K$\alpha$ emission line profile (especially the blue-wing above 6.4 keV), but the S/N of our data is clearly too low to resolve this line profile. Since \rxj\ is one of the most highly super-Eddington NLS1s with clean line-of-sight, a low inclination angle is preferred by the disc wind geometry (Gardner \& Done 2015; DJ16; Hagino et al. 2016). Besides, Nandra et al. (1997) studied the Iron K$\alpha$ line profile for a sample of Seyfert 1 galaxies observed by Advanced Satellite for Cosmology and Astrophysics (ASCA), and found that their inclination angles were clustered at $\sim30\degr$ (also see Fabian et al. 2000). Thus we fixed the inclination angle of \rxj\ at 30$\degr$.

This Reflection-Disc model produces as good a fit with a similar
$\chi^2$ to the Comptonisation models (Fig.\ref{fig-specfit1}c and
Table~\ref{tab-specfit}). The fitting requires the intrinsic corona
emission to have an extremely steep slope of $\Gamma=2.83^{+0.03}_{-0.03}$ with
a seed photon temperature of $kT\_bb=73^{+7}_{-8}$ eV. The smooth soft
excess requires highly relativistic smearing with
$R_{\rm in}=2.56^{+0.21}_{-0.19}~R_{\rm g}$ and an emissivity index of
$q=5.36^{+0.58}_{-0.24}$ in order to suppress all sharp line features
in the reflection spectrum. Then the small $R_{\rm in}$ implies a black
hole spin of $a\gtrsim0.86$. The reflecting material is highly ionised
($log~\xi =3.41^{+0.06}_{-0.05}$) and has a large covering factor of
$\Omega/{2\pi}=3.70^{+0.38}_{-2.22}$, with $1.38^{+0.08}_{-0.40}~\times$
Solar iron abundance. Therefore, the physical scenario can be summarised
as that the X-ray corona is very close to the black hole which is fast spinning.
The resulting very strong gravitational field would cause a major light
bending effect, so that most of the corona emission is illuminating
and reflected by the highly ionised inner disc.

However, a simple reflection-dominated model was found to have
difficulties in explaining the lack of soft X-ray time-lag in PG
1244+026 whose X-ray properties are similar to \rxj\ (Alston, Done, \&
Vaughan 2014). So Kara et al. (2014) proposed an additional spectral
component associated with the synchrotron emission from a jet to account
for part of the soft excess, thereby diluting the time-lag signal in
this energy band. However, Gardner \& Done (2014) show this is
incompatible with the observed soft lead at low frequencies if the
fluctuations propagate from the hard X-ray corona close to the black
hole, up to the jet, since this gives a soft lag. 
Nonetheless, a more complex, dual point lamppost model, Chainakun \& Young (2016)
is able to produce the lag, so we 
also apply this jet model to \rxj\ (hereafter: the Reflection-Jet
model) due to its X-ray spectral similarity to PG 1244+026 (though unlike
PG 1244+026 it has no detected radio emission, see Section~\ref{sec-discussion})
In Kara et al. (2014), this
jet component is modelled with a steep power law to approximate the
synchrotron tail in the soft X-ray band. Therefore, we replace the
disc component with a steep power law, and also replace the {\sc
  nthcomp} model with a simple power law in order to be fully
consistent with the model adopted in Kara et al. (2014). This model
gives $\chi^2_v=1301/1105$, which is equally good as the other
models (Fig.\ref{fig-specfit1}d). The reflection component is still
highly smeared with $R_{\rm in}=2.44^{+0.14}_{-0.07}~R_{\rm g}$, but the corona
emission has a slightly flatter slope of
$\Gamma=2.70\pm0.03$, because the jet component accounts for
part of the soft excess.

For comparison, we also ran the spectral fitting for an inclination angle of 60$\degr$ and put the results in Fig.\ref{app-fig-specfit1} and Table~\ref{app-tab-specfit}. A higher inclination produces a more bumpy reflection spectrum, which is clearly not favoured by the smoothness of the observed spectra, thus in order to improve the fitting {\sc Xspec} will reduce the contribution of reflection component in the smooth soft excess while maintaining the intensity above 4 keV. This explains the maximal black hole spin, smaller hard X-ray power law photon index, higher iron abundance and smaller reflection fraction as shown by the best-fit parameters.

\begin{figure}
\centering
\includegraphics[bb=30 124 540 612,scale=0.43]{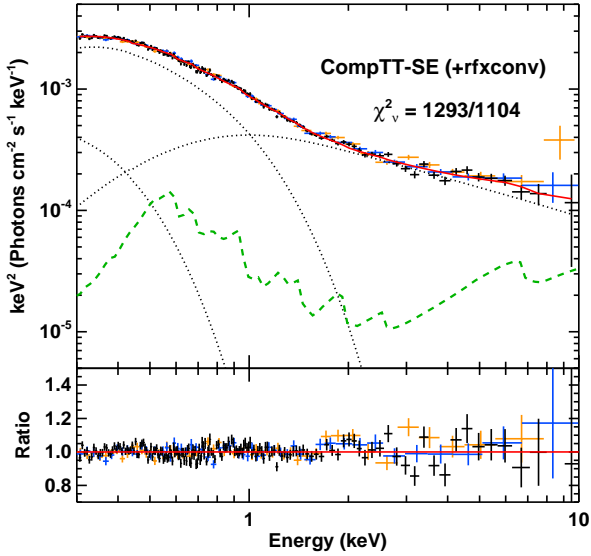}
\caption{Adding a reflection component to the CompTT-SE model in Fig.\ref{fig-specfit1}a. The dash green curve shows the ionised reflection component. The best-fit parameters are in Table~\ref{tab-comptrefl}.}
\label{fig-comptrefl}
\end{figure}

\begin{table}
 \centering
   \caption{Best-fit parameters in 
     Fig.\ref{fig-comptrefl}. The inclination angle is fixed at 30$\degr$. `low' indicates that the parameter reaches its lower limit during the fitting. Units of the normalisations are the same as given in Table~\ref{tab-specfit}.}
\begin{tabular}{@{}clll@{}}
\hline
  Model & Component & Parameter & Value\\
\hline
{\sc CompTT}   & (Fig.\ref{fig-comptrefl}) & \multicolumn{2}{l}{$\chi^{2}_{\rm \nu}~=~1293/1104~=~1.17$} \\
{\sc -SE }            & {\sc tbnew}     & $N_{\rm H}$ (10$^{22}$ cm$^{-2}$) & 0.007 $^{+0.024}_{-0.007}$\\
{\sc (+rfxconv)}             & {\sc diskbb}     & $T_{\rm in}$ (keV)                   & 0.075 $^{+0.007}_{-0.005}$\\
             & {\sc diskbb}     & norm                       & 3.00 $^{+8.85}_{-0~low}$ $\times10^{3}$\\
             & {\sc nthcomp}   & $\Gamma$                   & 2.72 $^{+0.05}_{-0.07}$\\
             & {\sc nthcomp}   & $kT_{\rm seed}$ (keV)          & tied to $kT_{\rm e}$\\
             & {\sc nthcomp}   & norm                       & 4.21 $^{+0.32}_{-0.38}$ $\times10^{-4}$\\
             & {\sc comptt}    & $kT_{\rm e}$ (keV)              & 0.22 $^{+0.01}_{-0.01}$\\
             & {\sc comptt}    & $\tau$                     & 16.5 $^{+7.2}_{-1.6}$\\
             & {\sc comptt}    & norm                       &0.12 $^{+0.04}_{-0.07}$\\
	     & {\sc kdblur}    & Index                      & 3.0 fixed\\
             & {\sc kdblur}    & $R_{\rm in}$ ($R_{\rm g}$)           & 2.80 $^{+2.74}_{-1.80~low}$\\
             & {\sc rfxconv}   & rel\_refl ($\Omega/{2\pi}$)        & -1.17 $^{+0.54}_{-0.54}$\\
             & {\sc rfxconv}   & $Fe_{\rm abund}$ (Solar)       & 1.0 fixed\\
             & {\sc rfxconv}   & log $\xi$                   & 1.34 $^{+0.25}_{-0.17}$\\
             & {\sc const}     & (MOS1) & 1.002 $^{+0.008}_{-0.007}$\\
             & {\sc const}     & (MOS2) & 1.006 $^{+0.008}_{-0.008}$\\
\hline
   \end{tabular}
 \label{tab-comptrefl}
\end{table}

\subsubsection{Reflection Component in the CompTT-SE Model}
\label{sec-comptrefl}
Fig.\ref{fig-specfit1}a shows some positive residuals above 4 keV,
implying an extra reflection component in the CompTT-SE model. This
reflection can be due to ionised material associated with the inner
disc and/or the soft X-ray Comptonisation region (Gardner \& Done
2014). We attempt to model this by adding an ionised reflection component
({\sc rfxconv}) smoothed by the relativistic effects using {\sc
  kdblur}. A lower limit of 3000 is set for the normalisation of the
disc component in order to retain its contribution to the soft
X-rays. Since the reflection component is weak, not all the parameters
can be well constrained, and so we freeze the power law emissivity
index and outer radius of {\sc kdblur} at their default values of 3 and
100 $R_{\rm g}$. We also fix the Fe abundance at the solar abundance. The
inclusion of this reflection component improves the fitting by
$\Delta\chi^2=11$ for including three additional free parameters
(2.5$\sigma$ significance, Table~\ref{tab-comptrefl}), with the
main improvement being the fitting above 4 keV
(Fig.\ref{fig-comptrefl}). The material has low ionisation with
$log~\xi=1.34^{+0.25}_{-0.17}$ and a covering factor of $\Omega/2\pi=1.17^{+0.54}_{-0.54}$.
The other parameters are similar to those found in the previous CompTT-SE
fitting, including a large photon index of $\Gamma=2.72^{+0.05}_{-0.07}$.
We also find the best-fit $R_{\rm in}=2.80^{+2.74}_{-1.80}~R_{\rm g}$,
corresponding to a spin parameter of $a=0.82^{+0.18}_{-0.68}$ which is poorly constrained.
However, if we allow the Fe abundance to be a free
parameter, the best-fit parameters would be $Fe_{\rm abund}=3.7$, $R_{\rm in}=4.35~R_{\rm g}$ and
$\Gamma=2.74$ with an improved $\Delta\chi^2=6$ for one additional
free parameter (2.4$\sigma$ significance), the rest parameters are
little changed. Furthermore, if we assume that the reflecting material
is mainly associated with the soft X-ray component, which is several
tens to hundreds of $R_{\rm g}$ away from the hot corona as suggested by the time-lag
analysis (see Section~\ref{sec-timelag}), then we can place a lower
limit of 10 $R_{\rm g}$ for $R_{\rm in}$. Then we find $R_{\rm in}$ reaches its lower
limit of 10 $R_{\rm g}$ and a worse fit by $\Delta\chi^2=2$ compared to the original value, and the other
parameters are all similar to those found previously, except that the
covering factor decreases slightly to $\Omega/2\pi=0.80$.

The above results from the spectral fitting can be understood as
follows. Firstly, the excess flux above 4 keV requires a reflection
component, while the soft excess is so smooth that it does not favour
any sharp line features\footnote{The smoothness of the soft excess is also confirmed by the two RGS spectra, where we found no significant sharp emission/absorption lines, which also rules out any significant warm absorbers along the line-of-sight.},
and so a small $R_{\rm in}$ is required to smear all the
line features in the reflection spectrum. A low ionisation state is
also favoured in order to further reduce the intensity of these
lines. When we allow the Fe abundance to be a free parameter, the
fitting increases the abundance to account for the excess flux above 4 keV, and
also reduces the strength of the reflection component in the soft
excess, and so a small $R_{\rm in}$ is no longer required. When a lower
limit of 10 $R_{\rm g}$ is set for $R_{\rm in}$, the line features in the
reflection spectrum are more sharp, and so the fitting
reduces the flux of the reflection component by reducing its covering
factor, accompanied by a worse $\chi^2$. Therefore, we conclude that
the small $R_{\rm in}$ is mainly due to the contradiction between the sharp line
features in the reflection spectrum and the observed smooth soft excess, rather
than due to a broad Iron K$\alpha$ feature.

\begin{figure*}
\centering
\begin{tabular}{cc}
\includegraphics[bb=30 124 540 612,scale=0.43]{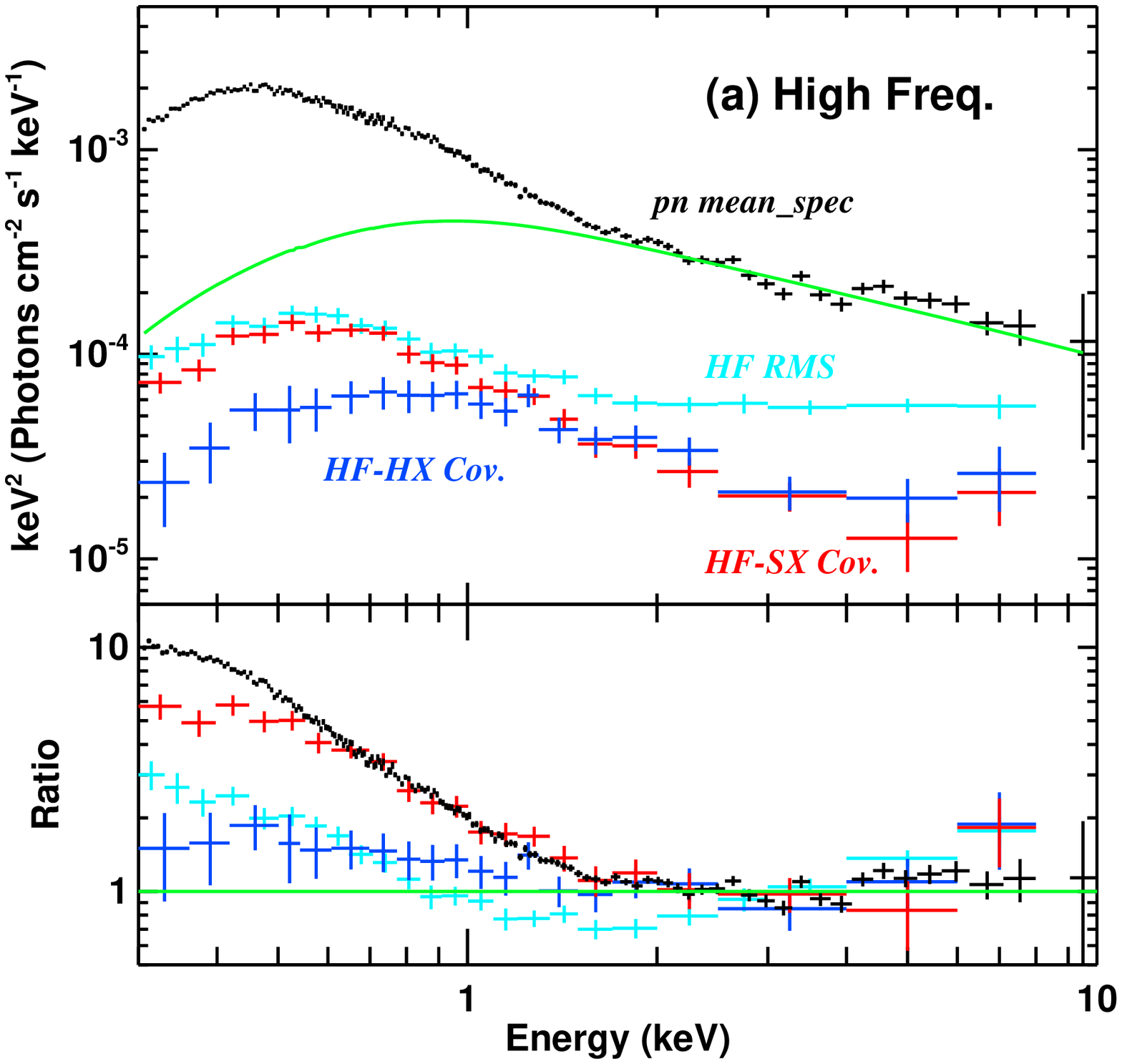} &
\includegraphics[bb=30 124 540 612,scale=0.43]{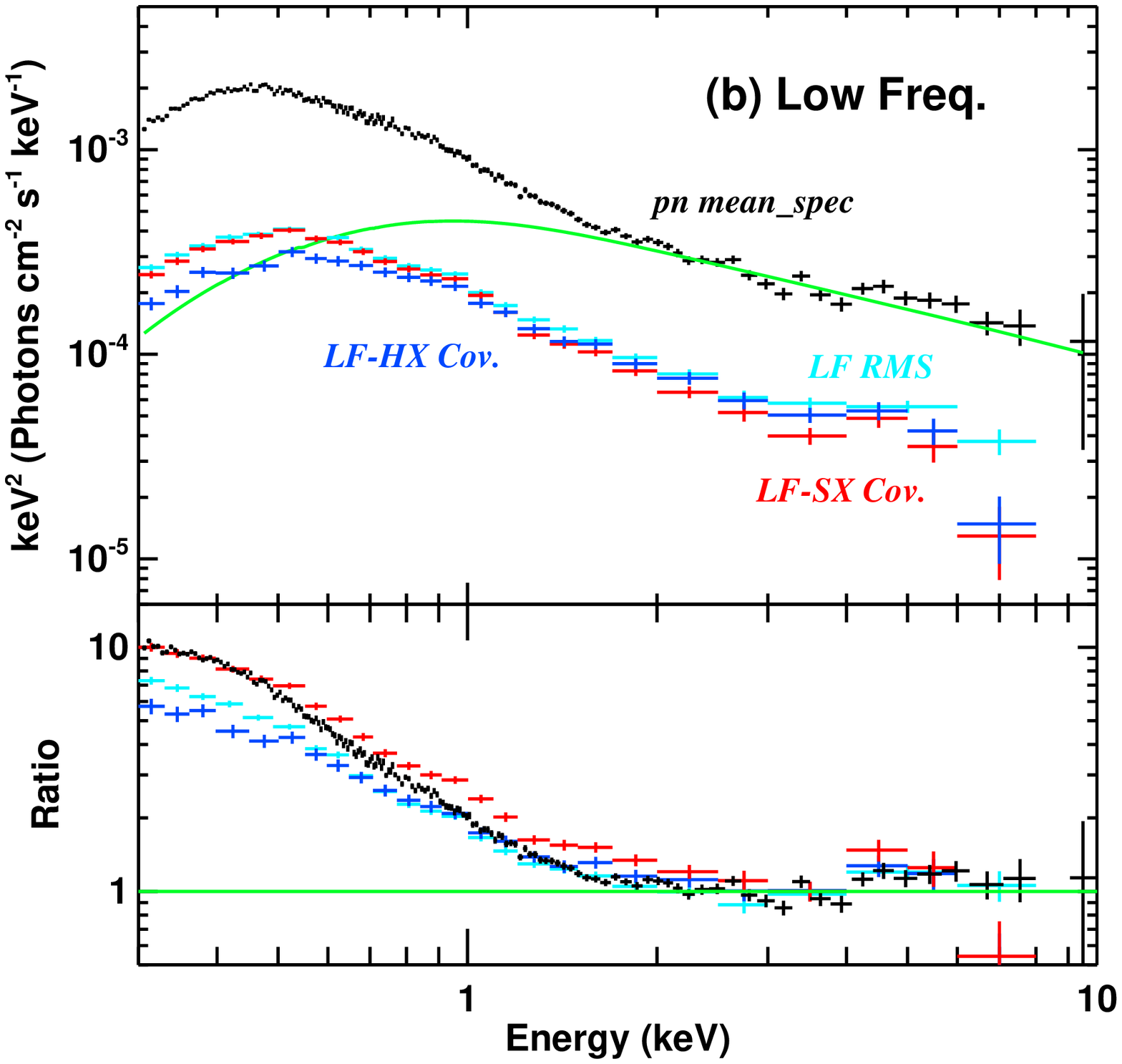}
\end{tabular}
\caption{A direct comparison between the time-averaged spectrum, RMS and covariance spectra from EPIC-pn for the high-frequency band (Panel-a) and the low-frequency band (Panel-b). In each panel, the green solid line shows the hard X-ray Comptonisation component in the CompTT-SE model in Fig.\ref{fig-specfit1}a. `Ratio' panels show the strength of the soft excess in every spectrum relative to the extrapolation of the best-fit Comptonisation model within the 2-10 keV band, assuming the same photon index as in the time-averaged spectrum.}
\label{fig-rms-cov}
\end{figure*}

\section{Combined spectral and variability analysis}
\subsection{Frequency-dependent RMS and Covariance Spectra}
\label{sec-covspec}
Previous spectral analysis has suggested that the
time-averaged spectra are highly degenerate to different spectral
models, and so we need to use information on variability in order to
break the spectral degeneracy and so distinguish between these
models. Compared to the RMS spectra which show the total intrinsic
variability in every energy bin, a covariance spectrum shows all
spectral components that vary coherently with a chosen energy band (i.e. the
reference band). Therefore, it would have an identical shape (but with
smaller error bars) as the RMS spectrum if there is only one
variability pattern over the entire energy band. This technique was
first introduced by Wilkinson \& Uttley (2009) to increase the S/N of
the RMS spectrum (also see the review by Uttley et al. 2014), and then
was adopted by Middleton, Uttley \& Done(2011) to disentangle the soft excess
from the hard X-ray power law in RE J1034+396. J13 further explored
the capability of this technique by producing the
frequency/reference-band dependent covariance spectra, which showed
that the soft excess in PG 1244+026 strongly favoured the
Comptonisation origin rather than reflection. \rxj\ is similar
to PG 1244+026 in terms of their X-ray spectra. It is important to understand
if their X-ray variabilities are also similar.
We apply various spectral timing techniques to explore the variability
properties of \rxj\ in both energy and frequency domains.

Firstly, we multiply the fractional RMS to the time-averaged spectrum of
EPIC-pn to derive the absolute RMS spectra. For the covariance
spectra, we choose 0.3-1 keV as the soft X-ray reference band
(hereafter: SX) and 2-10 keV as the hard X-ray reference band
(hereafter: HX). The frequency band is divided into the LF band and HF
band, the same as for the RMS spectra. So in total there are
four covariance spectra, which can be identified as HF-HX, HF-SX, LF-HX
and LF-SX. To produce covariance spectra, a band-limited light curve
is derived by applying a frequency filter in the Fourier domain. Then
the prescription in Uttley et al. (2014) is used to calculate their
fractional covariances in every energy bin relative to the reference
band. Note that energy bins inside the reference band are excluded
from the reference band before calculating the covariance. Finally, we
multiply the fractional covariances to the time-averaged spectra of
the EPIC-pn to derive the covariance spectra (see J13 for further
details). This procedure is repeated for different frequency and
reference bands to produce all four versions of covariance spectra.

Fig.\ref{fig-rms-cov} shows the comparison between the time-averaged
spectrum and the RMS and covariance spectra in both LF and HF
bands. In order to compare the strength of the soft excess, we use the
{\sc nthcomp} model component in the best-fit CompTT-SE model in
Fig.\ref{fig-specfit1}a to fit the variability spectra within the 2-10 keV
band with a fixed photon index, and then extrapolate the model down to
lower energies to show the soft excess (`Ratio' panels in
Fig.\ref{fig-rms-cov}). In the HF band (Fig.\ref{fig-rms-cov}a), we
can see that the normalisations of the HF-SX and HF-HX covariance
spectra are both lower than the HF RMS spectra, which should be due to
the Poisson noise contamination to the HF covariance. However, we also
find different shapes among these spectra. For example, the HF RMS
spectrum is clearly flatter than the time-averaged spectra and
covariance spectra above 2 keV, which implies some uncorrelated
variability in the hard X-rays. The HF-SX covariance spectrum has a
similar shape to the time-averaged spectra except below 0.6 keV,
indicating the presence of an extra component with no HF variability,
e.g. an inner disc component. The HF-HX covariance spectrum has the
weakest soft excess compared to the other spectra in
Fig.\ref{fig-rms-cov}a, indicating that only a small fraction of the
HF variability in the hard X-ray component is correlated with the soft
X-ray components. Similar results can be found in the LF band
(Fig.\ref{fig-rms-cov}b) where the LF covariance is less affected by
the Poisson fluctuation. But the LF RMS spectrum shows a similar steep
spectral slope as the other spectra above 2 keV, indicating no extra
uncorrelated LF variability in the hard X-rays. Since the differences
among these variability spectra are quite distinct, we can use them to
test different models and so break the spectral degeneracy.

\begin{table}
 \centering
   \caption{The best-fit parameters of the CompTT-SE model to the time-averaged spectrum and all four covariance spectra from EPIC-pn in Fig.\ref{fig-cov-compttse}. The upper and lower limits give the 90\% confidence range. Units of the normalisations are the same as given in Table~\ref{tab-specfit}.}
     \begin{tabular}{@{}clll@{}}
\hline
  Model & Component & Parameter & Value\\
\hline
  {\sc CompTT-SE} & (Fig.\ref{fig-cov-compttse}) &\multicolumn{2}{l}{$\chi^{2}_{\rm \nu}~=~842/721~=~1.17$} \\
   mean-spec          & {\sc tbnew}     & $N_{\rm H}$ (10$^{22}$ cm$^{-2}$) & 0~$^{+0.034}_{-0}$\\
             & {\sc diskbb}     & $T_{\rm in}$ (keV)                   & 0.080~$^{+0.004}_{-0.004}$\\
             & {\sc diskbb}     & norm                       & 5.41~$^{+1.79}_{-1.26}\times 10^{3}$\\
             & {\sc nthcomp}   & $\Gamma$                   & 2.67~$^{+0.06}_{-0.06}$\\
             & {\sc nthcomp}   & $kT_{\rm seed}$ (keV)          & tied to $kT_{\rm e}$\\
             & {\sc nthcomp}   & norm                       & 4.28~$^{+0.36}_{-0.37}\times 10^{-4}$ \\
             & {\sc comptt}    & $kT_{\rm e}$ (keV)              & 0.21~$^{+0.01}_{-0.01}$\\
             & {\sc comptt}    & $\tau$                     & 18.6~$^{+1.43}_{-1.29}$\\
             & {\sc comptt}    & norm                       &9.12~$^{+1.3}_{-1.1}\times 10^{-2}$\\
   cov-LF-HX          & {\sc nthcomp}   & norm     & 1.16~$^{+0.10}_{-0.10}\times 10^{-4}$ \\
             & {\sc comptt}    & norm                       &1.65~$^{+0.20}_{-0.17}\times 10^{-2}$\\
   cov-HF-HX         & {\sc nthcomp}   & norm      & 4.01~$^{+0.40}_{-0.39}\times 10^{-5}$ \\
             & {\sc comptt}    & norm                       &1.78~$^{+0.41}_{-0.41}\times 10^{-3}$\\
   cov-LF-SX          & {\sc nthcomp}   & norm     & 9.84~$^{+1.05}_{-1.04}\times 10^{-5}$ \\
             & {\sc comptt}    & norm                       &2.23~$^{+0.23}_{-0.20}\times 10^{-2}$\\
   cov-HF-SX         & {\sc nthcomp}   & norm     & 3.90~$^{+0.46}_{-0.45}\times 10^{-5}$ \\
             & {\sc comptt}    & norm                       &6.16~$^{+0.77}_{-0.67}\times 10^{-3}$\\
   \hline
   \end{tabular}
 \label{tab-cov-compttse}
\end{table}

\begin{figure*}
\centering
\begin{tabular}{cc}
\includegraphics[bb=30 124 540 612,scale=0.43]{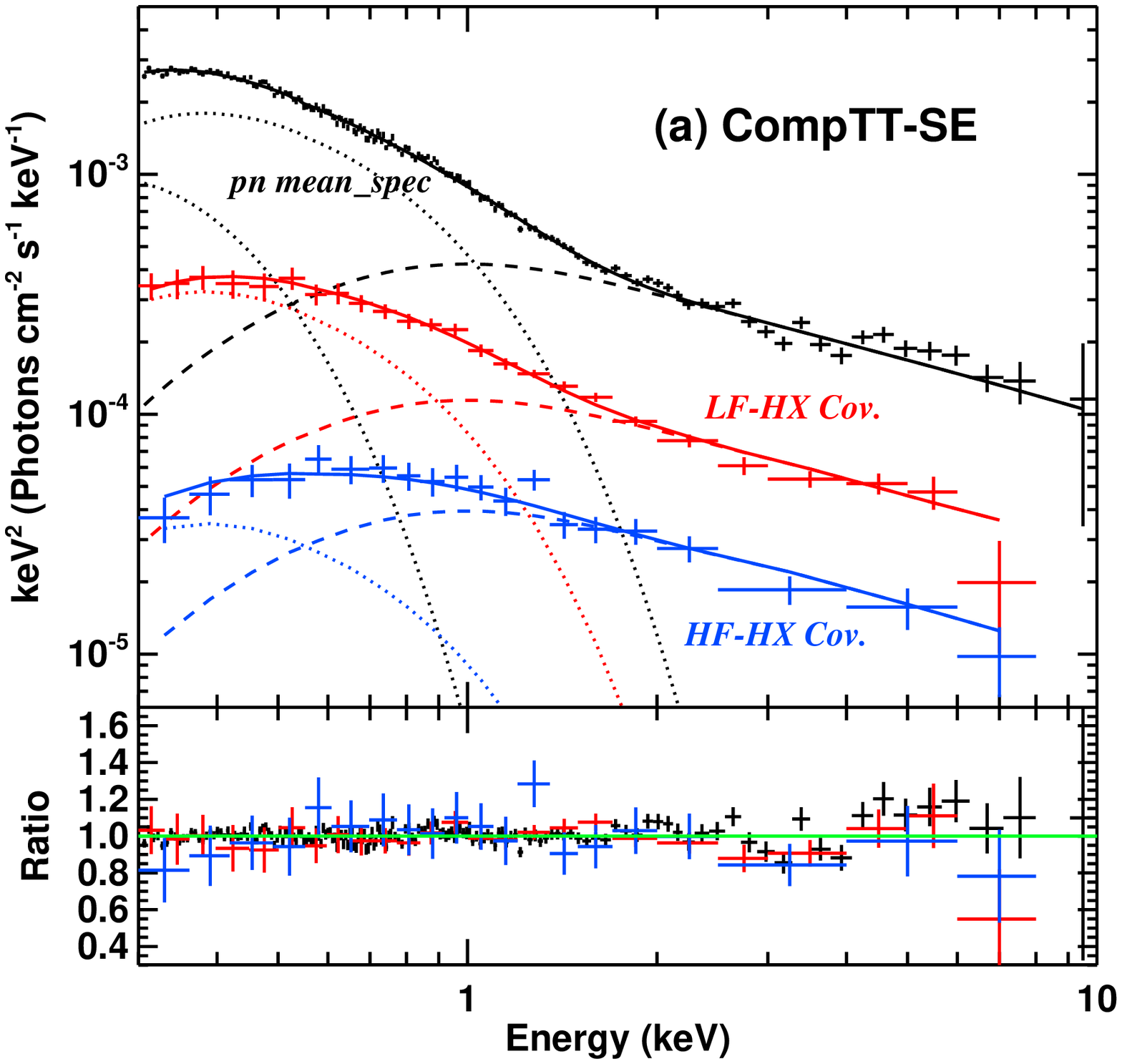} &
\includegraphics[bb=30 124 540 612,scale=0.43]{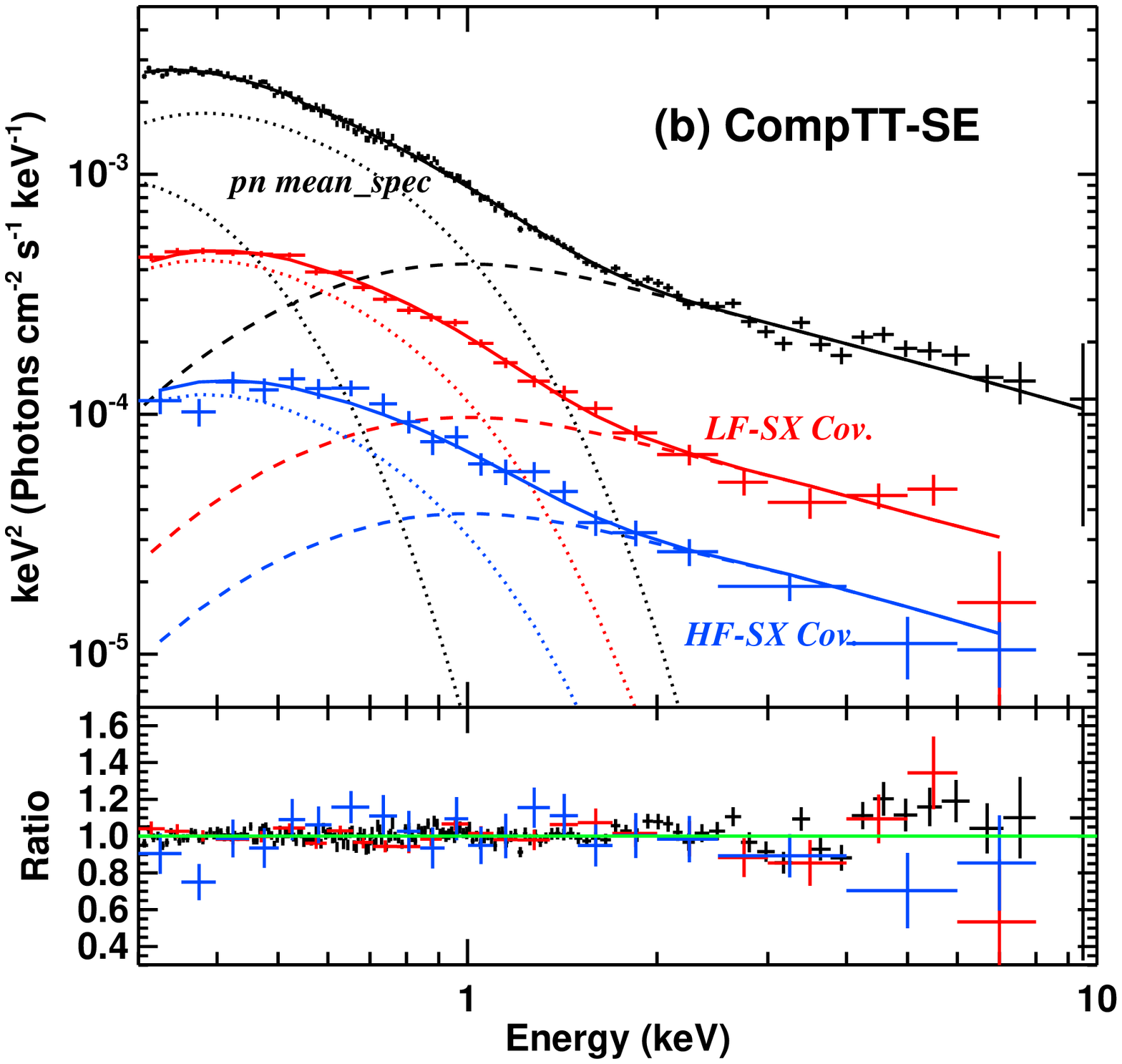} \\
\end{tabular}
\caption{Applying the CompTT-SE model to the EPIC-pn time-averaged spectrum (black) and all four types of covariance spectra. For clarity we plot the spectra in two panels. Panel-a shows the HF-HX (blue) and LF-HX (red) covariance spectra. Panel-b shows the HF-SX (blue) and LF-SX (red) covariance spectra. It is clear that the CompTT-SE model can produce good fits to all the spectra (see Section~\ref{sec-covspec}).}
\label{fig-cov-compttse}
\end{figure*}

\begin{figure}
\includegraphics[bb=30 124 540 612, scale=0.43]{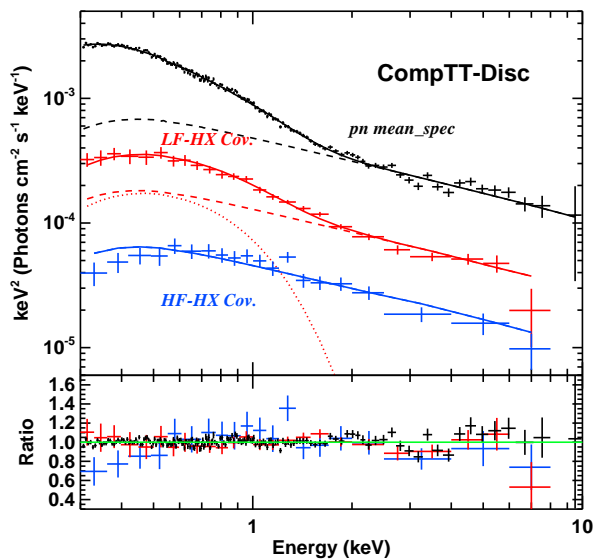}
\caption{Similar to Fig.\ref{fig-cov-compttse}, but applying the CompTT-Disc model. The HF-HX covariance spectrum is not as well fitted below 1 keV.}
\label{fig-cov-compttdisc}
\end{figure}

\begin{figure*}
\centering
\begin{tabular}{cc}
\includegraphics[bb=30 124 540 612,scale=0.43]{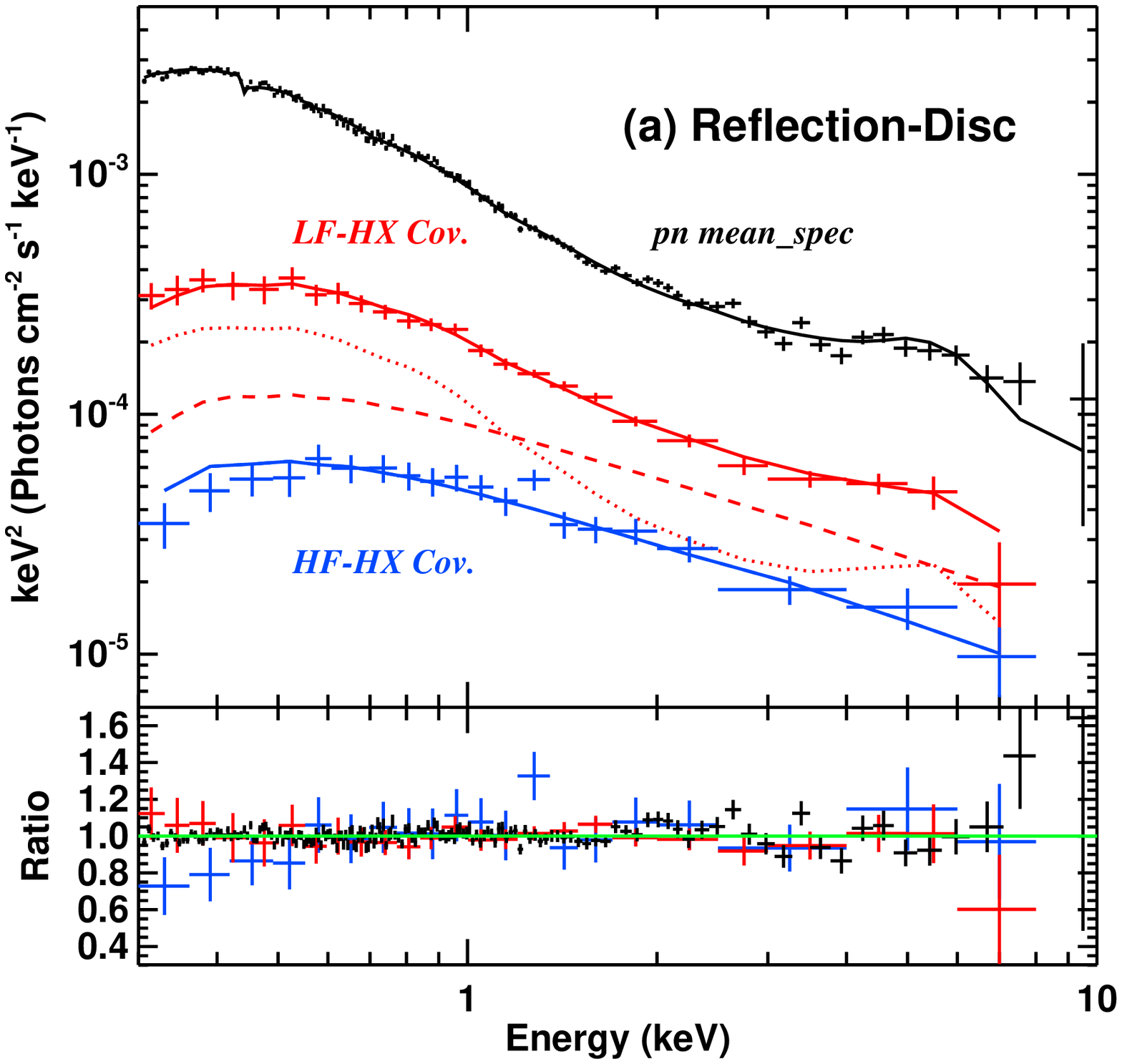} &
\includegraphics[bb=30 124 540 612,scale=0.43]{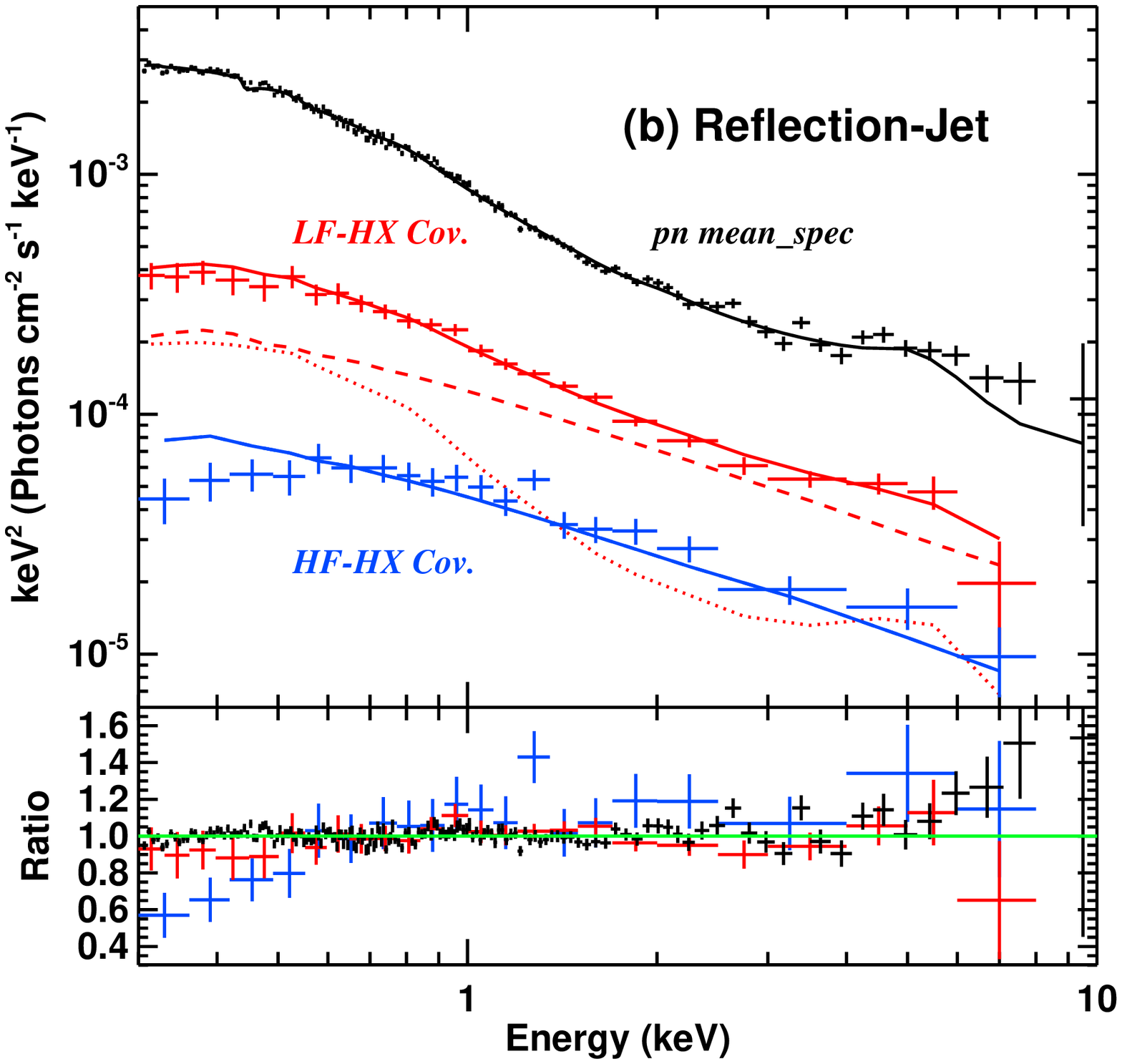} \\
\end{tabular}
\caption{Applying the reflection-dominated models to the EPIC-pn time-averaged spectrum (black), HF-HX covariance spectrum (blue) and LF-HX covariance spectrum (red). Panel-a shows the results for the Reflection-Disc model. Panel-b shows the results for the Reflection-Jet model. Both models have some difficulties in fitting the HF-HX covariance spectrum below $\sim$1 keV.}
\label{fig-cov-refl}
\end{figure*}

\begin{figure*}
\centering
\begin{tabular}{cc}
\includegraphics[bb=30 124 540 612,scale=0.43]{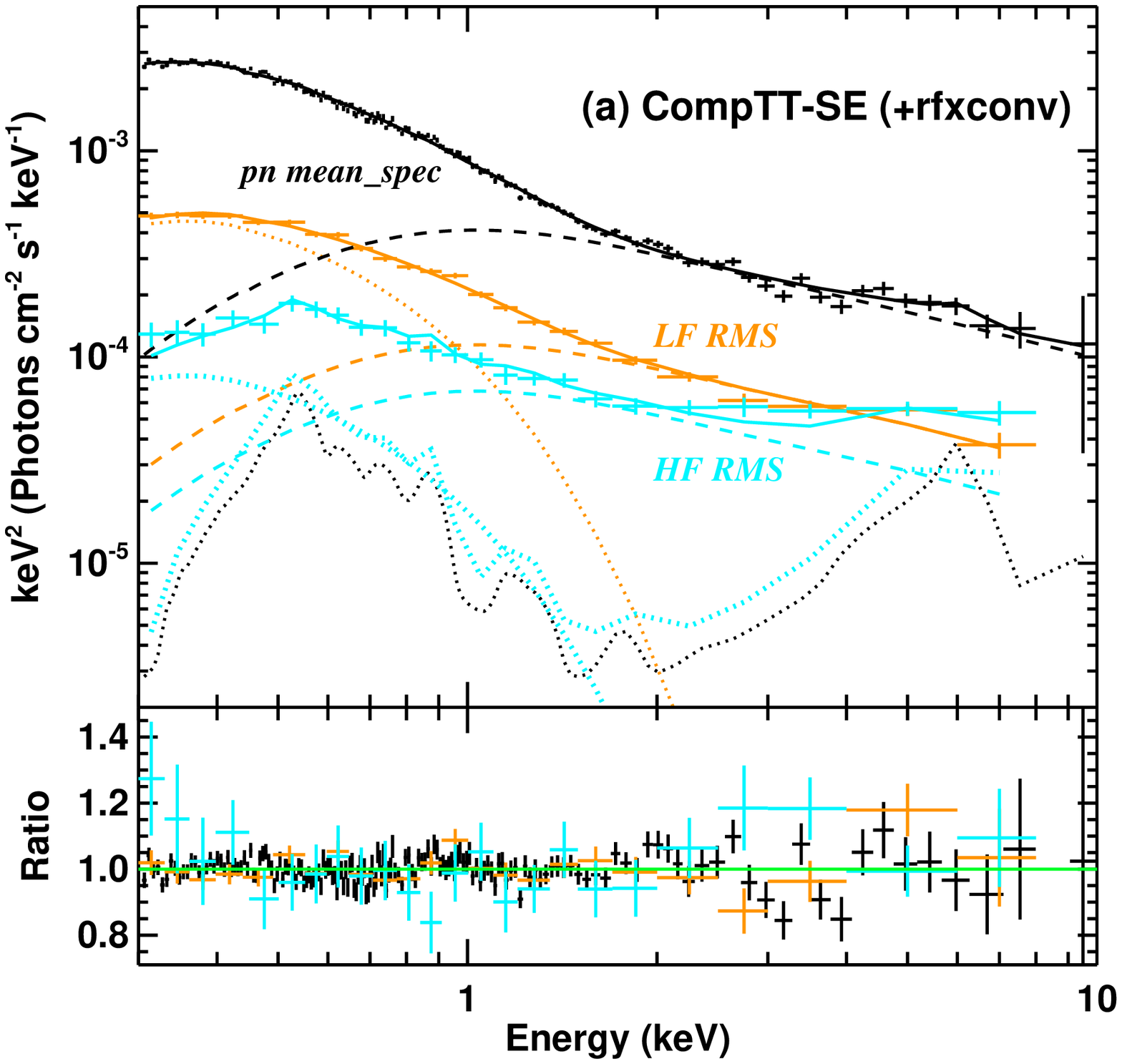} &
\includegraphics[bb=30 124 540 612,scale=0.43]{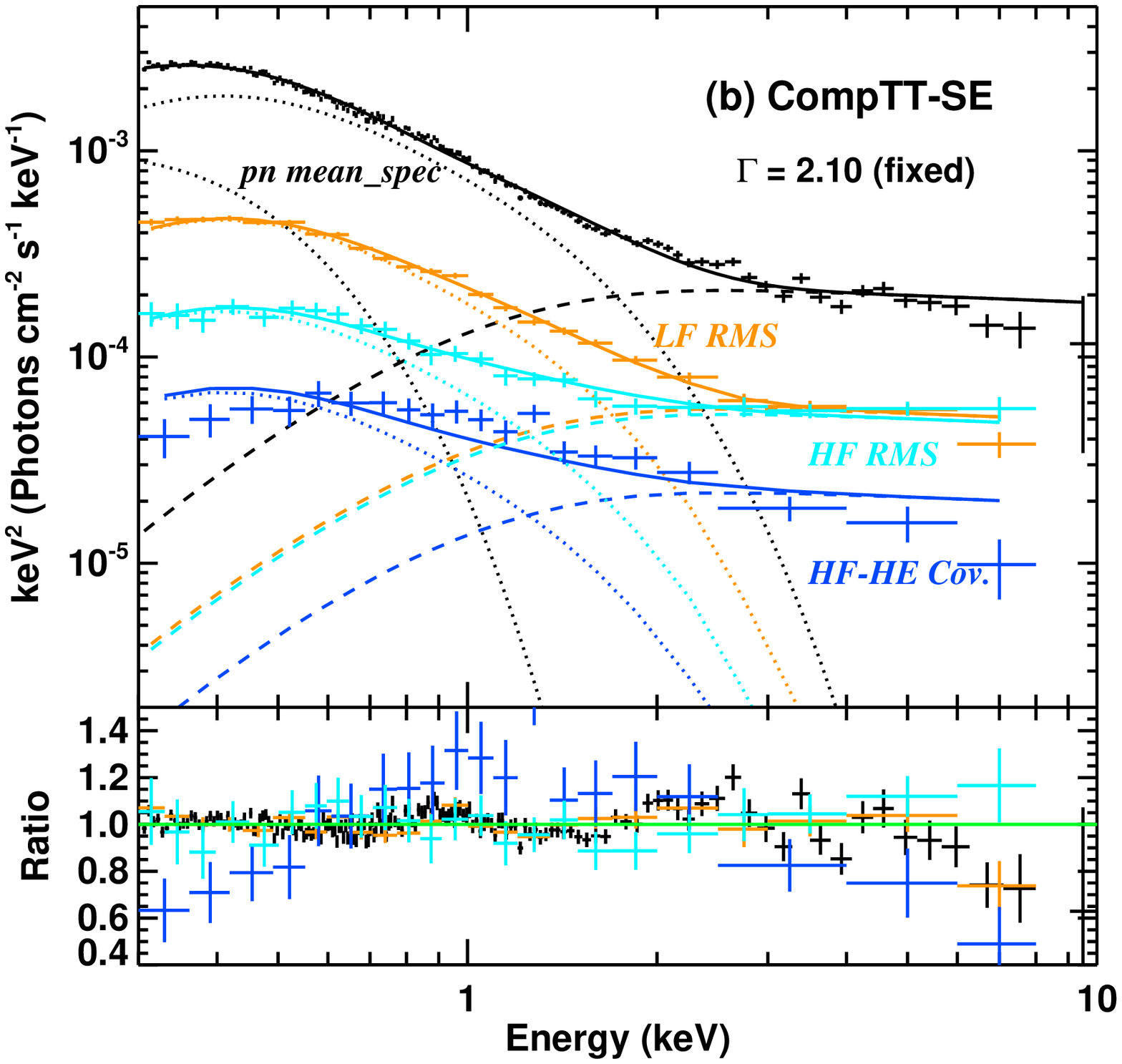}\\
\end{tabular}
\caption{Fitting the EPIC time-averaged spectra, LF and HF RMS spectra with the CompTT-SE model. Panel-a shows the fitting with the CompTT-SE plus reflection model in Fig.\ref{fig-comptrefl}b (but with no lower limit for $R_{\rm in}$), assuming the reflection component can contribute extra variability. For clarity we only plot the time-averaged spectrum from EPIC-pn and the two RMS spectra. Panel-b shows the CompTT-SE model fitting with the hard X-ray photon index fixed at 2.1. We also plot the HF-HX covariance spectrum which cannot be well fitted in this case.}
\label{fig-rms-compttse}
\end{figure*}

\subsection{Modelling the Covariance Spectra}
In the two Comptonisation-dominated models, each of them contains two Comptonisation components, plus a disc component and a weak reflection component. We assume the disc component does not vary within the 120 ks observing time, so it does not contribute to the covariance spectra. The reflection component is not well constrained as it is overwhelmed by the other components across the entire 0.3-10 keV band, so we ignore its contribution in the covariance spectra unless the other major components are not sufficient. The two Comptonisation components can both vary but with different timing properties (as implied by the PSD and RMS spectra), and so they should have different contributions in different types of covariance spectra. Therefore, we assume that their spectral shapes do not change in the covariance spectra, but their normalisations are free to change.

Firstly, we apply the CompTT-SE model to the EPIC-pn time-averaged spectrum and all four covariance spectra, simultaneously. Fig.\ref{fig-cov-compttse} shows that the CompTT-SE model can produce a good fit to all spectra with a total $\chi^2_v = 842/721$. The key parameters are all consistent with those found previously from fitting the time-averaged spectra alone (see Table~\ref{tab-cov-compttse}). The HF-HX covariance spectrum is totally dominated by the hard X-ray Comptonisation component, but there is also a small contribution from the soft excess component, indicating the presence of some reprocessed hard X-ray emission in the soft X-ray region, which is also consistent with the presence of a weak reflection component in Fig.\ref{fig-comptrefl}. The LF covariance spectra consist of significant contributions from both soft and hard X-ray Comptonisation components, indicating a strong LF coherence between these two components (also see Section~\ref{sec-timelag}). Moreover, all covariance spectra are well fitted below $\sim$0.5 keV, confirming that the non-variable accretion disc component only contributes to the soft excess on longer timescales. 

Then we apply the CompTT-Disc model to all the spectra. In this case the hard X-ray Comptonisation component appears too flat compared to the curvature of the HF-HX covariance spectrum in the soft X-ray band (Fig.\ref{fig-cov-compttdisc}). Below $\sim$1.5 keV we find $\chi^2_v = 20/12$ for the HF-HX covariance spectrum, which is clearly much worse compared to $\chi^2_v = 10/12$ in the case of the CompTT-SE model. This suggests that the temperature of the seed photons from the inner disc is too low. The same result was found in PG 1244+026 (J13), implying that this is probably a common property among similar NLS1s with high mass accretion rates.

In the two reflection-dominated models, the spectrum mainly consists of an underlying Comptonisation component from the hot corona which is the primary varying component, and an ionised reflection component. The variability of the reflection component also comes from the hot corona, with some smearing and time lag. In the Reflection-Disc model, the disc is not varying on these short timescales, so only the corona and reflection components contribute to the covariance spectra. Fig.\ref{fig-cov-refl}a shows that this model can roughly match the HF-FX covariance spectrum in the soft X-ray band, with $\chi^2_v = 16/12$ below 1.5 keV.

In the Reflection-Jet model, the jet component can also vary on various timescales, and may be partly correlated with the corona component because the corona may be considered as being the base of the jet (Kara et al. 2014). So we allow the jet, corona and reflection components all to contribute to the covariance spectra. We also find this model cannot fit the HF-HX covariance spectrum, with $\chi^2_v = 40/13$ for the energy band below 1.5 keV (Fig.\ref{fig-cov-refl}b). Therefore, we conclude that the CompTT-Disc model and reflection-dominated models cannot reproduce all covariance spectra, especially the HF-HX covariance spectrum where a clear soft X-ray roll-over is seen, while the two Comptonisation components in the CompTT-SE model can produce a good fit to all four covariance spectra, simultaneously.

\begin{figure*}
\centering
\begin{tabular}{cccc}
\includegraphics[bb=90 144 540 720, scale=0.26]{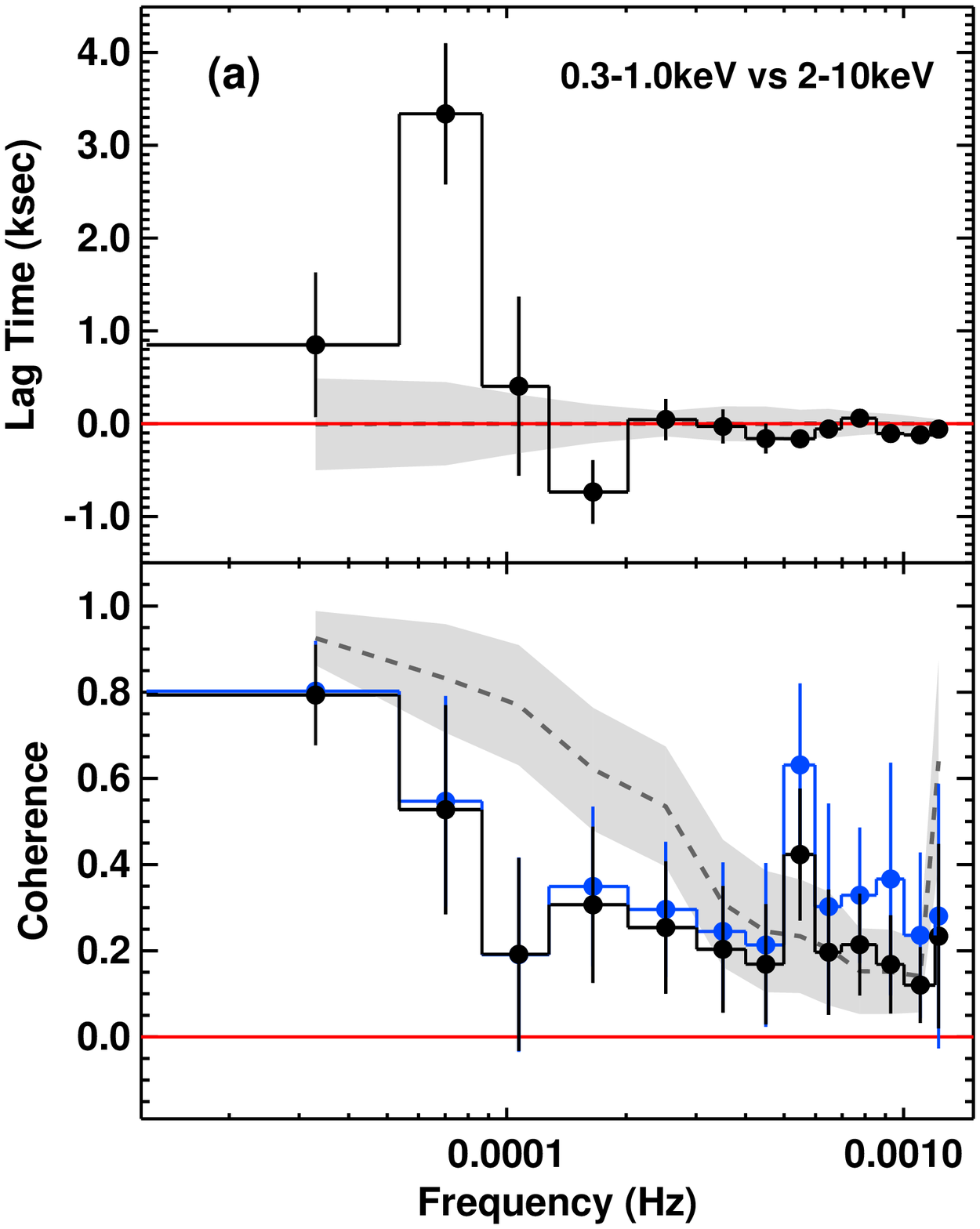} &
\includegraphics[bb=100 144 540 720, scale=0.26]{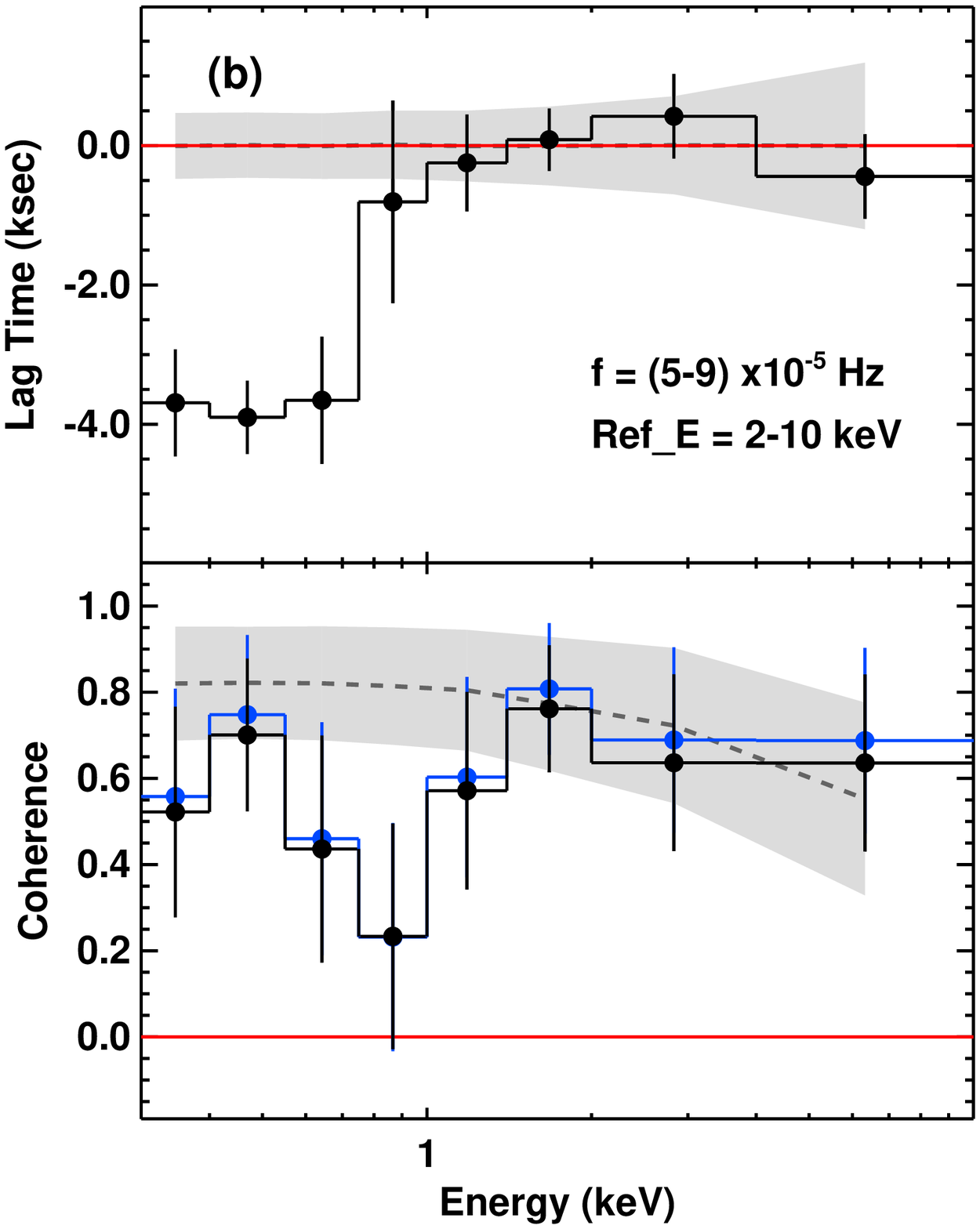} &
\includegraphics[bb=100 144 540 720, scale=0.26]{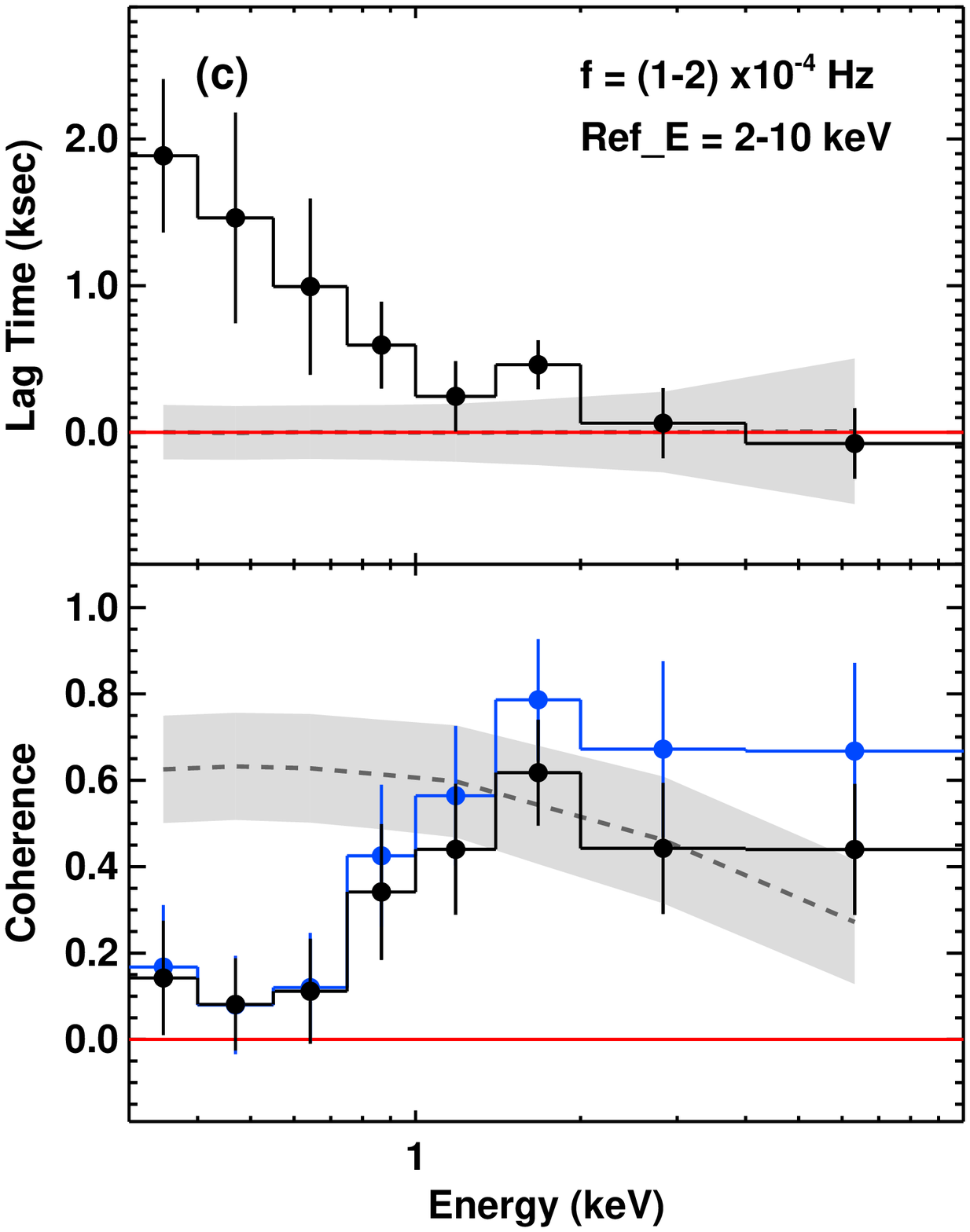} &
\includegraphics[bb=100 144 540 720, scale=0.26]{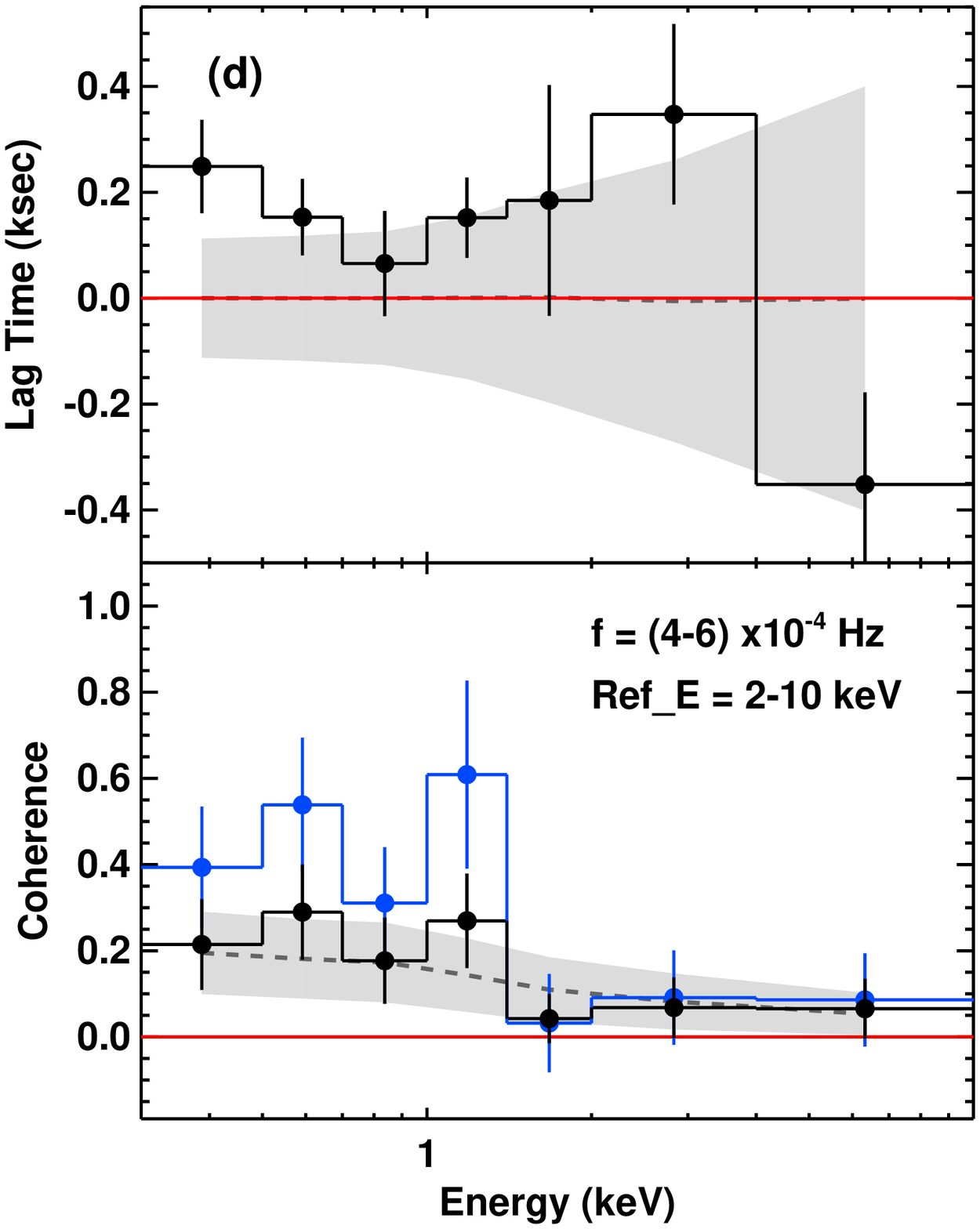} \\
\end{tabular}
\caption{Energy and frequency dependent time-lag and coherence. A negative time-lag indicates the soft X-ray emission  lags the hard X-ray. In the coherence panel, the blue points are the coherences after Poisson noise correction (Vaughan \& Nowak 1997). In every panel, the dash curve and shaded region are the mean and $\pm1\sigma$ dispersion from our simulations assuming an intrinsically full-correlated zero-lag inter-band correlation (see Section~\ref{sec-timelag}). The reference band for the time-lag is 2-10 keV, and the simulation is based on the PSD of 2-10 keV. Results using different reference bands and PSDs can be found in Fig.\ref{app-fig-cohlag1}, \ref{app-fig-cohlag3}, \ref{app-fig-cohlag3}.}
\label{fig-cohlag}
\end{figure*}

\subsection{Modelling the RMS Spectra}
Although among the four models the CompTT-SE model is the best one that can reproduce the time-averaged and all covariance spectra, we identify a problem when applying it to the RMS spectra. The hard X-ray slope measured from the time-averaged spectrum is steeper than the HF RMS spectra, so the CompTT-SE model under-predicts the HF RMS above $\sim$2 keV (see the cyan dash line in Fig.\ref{fig-rms-compttse}a). However, we do not see any similar excess variability in the LF band in either the LF RMS spectrum or the covariance spectra (Fig.\ref{fig-rms-cov}b), and so this is an independent variable component in the HF band only.

However, we note that there could also be an ionised reflection component in the CompTT-SE model as shown in Fig.\ref{fig-comptrefl}, and this component can also vary in the HF. Thus we add this component to fit all three time-averaged EPIC spectra and the two RMS spectra simultaneously. This reflection component does improve the fitting to the HF RMS spectrum above 4 keV. The best-fit $\chi^2_v=1361/1151$, which shows an improvement of $\Delta\chi^2=46$ for 3 extra free parameters ($6.2\sigma$ significance). In this fitting, $R_{\rm in}$ also reaches its lower limit of 10 $R_{\rm g}$. If we do not put a lower limit for $R_{\rm in}$, then the best-fit value would be $5.13^{+1.40}_{-1.63}~R_{\rm g}$ and $\chi^2$ improves by 10 for 1 extra free parameter ($3.2\sigma$ significance, Fig.\ref{fig-rms-compttse}a). The lack of HF covariance between this reflection component and the primary hard X-ray corona component can be explained as due to the smearing of signal during the reverberation process. However, in this scenario we would expect that compared to the hard X-ray corona component, the relative contribution of the reflection component to the time-averaged spectrum should be more than to the HF RMS spectrum, which is opposite to the fitting seen in Fig.\ref{fig-rms-compttse}a. Moreover, there is still a flux discrepancy within the 2-4 keV band, and the HF RMS spectrum appears smoother than the reflection component below 1 keV. Therefore, we cannot conclude that the reflection component is responsible for the excess HF variability above 4 keV.

Another possibility is that the hard X-ray Comptonisation component is indeed flatter. To test this we fix the photon index $\Gamma$ at 2.1 and rerun the fitting. The new fitting has $\chi^2_v=1484/1153$ (Fig.\ref{fig-rms-compttse}b), which is significantly worse than it was when $\Gamma$ was a free parameter ($\Delta\chi^2=83$ for one less degree of freedom). With such a flatter hard X-ray component, the fitting requires the soft X-ray Comptonisation component to extend further into the hard X-rays with $kT_{\rm e}=0.39^{+0.04}_{-0.03}$ keV and $\tau=10.6^{+0.7}_{-0.8}$. Since the soft X-ray Comptonisation region probably extends over several tens of $R_{\rm g}$ (see Section~\ref{sec-timelag}), the electrons in it may have a wider temperature distribution than a single value, so it is indeed possible for the soft X-ray component to be more extended than a single {\sc comptt} model. However, If we add all four types of covariance spectra to the fitting, the best-fit has $\chi^2_v=1642/1236$, which is much worse than the free $\Gamma$ case. We notice that this model also has difficulties in fitting the HF-HE covariance spectrum (Fig.\ref{fig-rms-compttse}b, blue spectrum). Therefore, this flatter hard X-ray Comptonisation model is not a plausible solution. More observations especially above 4 keV could provide better constraints on the shape of the hard X-ray component and help to identify the origin of this hard X-ray excess variance.

\section{Time Lag Analysis}
\label{sec-timelag}
\subsection{Lag and Coherence Spectra}
The strongly correlated X-ray variability seen in \rxj\ also allows us
to measure the time lags between different energy bands in different
frequency ranges. These time lags can provide crucial information about the
absolute distances between distinct physical regions. The RMS and
covariance spectra have shown that the HF variability seen in \rxj\ is
mainly associated with the hard X-rays, while more LF variability is
found in the soft X-rays. Simply associating these variability
timescales with the dynamic timescale of the accretion flow at
different radii, they suggest the soft X-ray Comptonisation region
should be  more extended than the compact hot corona region. Then
the time lag between the soft and hard X-rays can indicate the
distance between these two physical regions. Note that due to the
small contribution of the hard X-ray component in the soft X-ray band, the
observed time lag is only a diluted measurement of the intrinsic time
lag between the two spectral components (e.g. Uttley et al 2014).

We follow the prescription in Vaughan \& Nowak (1997) and Nowak et
al. (1999) (also see Vaughan et al. 2003, Ar\'{e}valo et al. 2008 for
detailed descriptions) to calculate the cross-spectrum between light
curves (400 s binned) of two energy bins. Since the strong LF
variability can easily introduce a bias via the red noise leak
(e.g. Vaughan, Fabian \& Nandra 2003), we use the whole light curve to
do the Fourier transform instead of dividing it into some
segments. The cross-spectrum is binned in frequency with a geometrical
step of 1.4. Then the coherence and time lag can be derived from the
cross-spectra in different frequency bins (Bendat \& Piersol
1986). Poisson noise correction is applied to the coherence using
the algorithm in Vaughan \& Nowak (1997). The standard conventions are
followed, so that a a positive time lag indicates the soft X-rays lead
the hard X-rays. A zero coherence indicates no correlated variability,
while a coherence of unity indicates a fully correlated variability.

However, the Vaughan \& Nowak (1997) analytic prescription
for the effect of the error bars is only
valid in the regime where the intrinsic power is higher than the
Poisson noise. This is not true at high energies, so we also perform
Monte Carlo simulations to better estimate the effect of errors on the
coherence and time lag spectra (Alston et al. 2014).
Firstly, we use the 2-10 keV PSD and the method of Timmer \& Koenig (1995)
to simulate new light curve realisations of the same length and binning and
intrinsic power spectra. We then add errors based on the observed
measurement errors for each energy band. In this way we obtain light curves
in different energy bins with fully-correlated intrinsic
variability like that seen in 2-10~keV with zero-lag. Then we use the
same method to calculate coherence and lag spectra between these
simulated light curves. We repeat this simulation for 10000 times to
measure the random fluctuation of the coherence and lag spectra (Fig.\ref{fig-cohlag}).

We notice that the result of simulation much also depend on the input PSD. Fig.\ref{fig-powspec} has shown that the 2-10 keV PSD contains much more HF variability than the 0.3-1 keV PSD (similar to the 0.3-10 keV PSD as most of the X-ray counts come from the 0.3-1 keV band). This means that the same Poisson errors will have bigger impact on the light curves simulated from the 0.3-1 keV PSD than the 2-10 keV. Thus we also run the same simulations for the 0.3-1 keV PSD and use them as comparison (Fig.\ref{app-fig-cohlag1}-\ref{app-fig-cohlag3}).

Fig.\ref{fig-cohlag}a upper panel shows the lag vs. frequency between 0.3-1 keV
(SX) and 2-10 keV (HX). The shape of this lag-frequency spectrum is
commonly observed in many other AGN (De Marco et al. 2013; Kara et
al. 2016), with a hard X-ray lag in the LF band and a weak soft X-ray
lag in the HF band. The shaded regions indicate the $\pm1\sigma$
fluctuation of these spectra caused by Poisson errors as determined
from the Monte-Carlo simulations. Fig.\ref{fig-cohlag}a lower panel shows that the observed coherence-frequency spectrum is slightly lower than the simulation based on the 2-10 keV PSD, but is more consistent with the simulation based on the 0.3-1 keV PSD (Fig.\ref{app-fig-cohlag1}a). This means that the additional power at high frequencies in the 2-10 keV PSD introduces an intrinsic decoherence especially above $5\times 10^{-4}$~Hz.

While the error bars clearly have a large impact on the data, we do
observe one significant positive lag of $3.4\pm0.8$ ks within
$(5-9)\times10^{-5}$ Hz with high coherence, and an enhanced coherence
within $(4-6)\times10^{-4}$ Hz with a weak negative lag. There is also a negative lag within $(1-2)\times10^{-4}$ Hz with marginal significance. Therefore, we select the these bands to calculate the lag/coherence spectra relative to the 2-10
keV band. Fig.\ref{fig-cohlag}b,c,d show the results (equivalent results relative to the 0.3-1 keV band can be found in
Fig.\ref{app-fig-cohlag2}, \ref{app-fig-cohlag3}).

For the $(5-9)\times10^{-5}$ Hz band (Fig.\ref{fig-cohlag}b and Fig.\ref{app-fig-cohlag2}b), there is a high coherence in
the soft X-ray band with no time lag (lower panel), indicating that a
single component dominates the soft excess, consistent with the
spectral decomposition of the CompTT-SE model. Compared to the
fully-correlated zero-lag simulation (shaded region), the coherence
seems to drop below 0.5 keV (although with large error bars), which is
likely due to the dilution of the non-variable disc
component. Furthermore, the coherence also seems to be lower than the
simulation within 0.8-2 keV, indicating that there might be a
transition of the dominated spectral component within this energy
range, i.e. consistent with the transition between the soft and hard
X-ray Comptonisation components. The most noticeable feature is the
increasing time lag from the soft to hard X-rays (Fig.\ref{fig-cohlag}b upper panel). The
time-lag below 1 keV is $\sim$-4 ks with a moderate coherence of $\sim$0.6,
and it is more than 4$\sigma$ away from the random fluctuation of the
zero-lag simulation. In the $(1-2)\times10^{-4}$ Hz and $(4-6)\times10^{-4}$ Hz bands (Fig.\ref{fig-cohlag}c, d), there
appears to be a 0.3-1 keV lag of $\sim$2 ks and 200 s, separately. But the corresponding coherences in these energy bands are $\lesssim$ 0.2, and so we do not consider them as robust time lag detections. However, it is worth noting that a small soft X-ray lag was found in the high frequency band in PG 1244+026 with higher S/N (Alston, Done \& Vaughan 2014).

\begin{table}
 \centering
   \caption{The assumed time lags in every model simulation in Fig.\ref{fig-lag-mo}. In each model, the time lags are relative to the disc or jet component.}
     \begin{tabular}{@{}lccc@{}}
\hline
  CompTT-SE & disc & hard X-ray corona & soft X-ray Compton \\
  &  0 s & 300 s & 5000 s \\
\hline
  CompTT-Disc & disc & hard X-ray corona & soft X-ray Compton\\
  &  0 s & 300 s & 5000 s \\
\hline
  Reflection-Disc & disc & hard X-ray corona & reflection\\
  &  0 s & 300 s & 5000 s \\
\hline
  Reflection-Jet & jet & hard X-ray corona & reflection\\
  &  0 s & 300 s & 5000 s \\
\hline
   \end{tabular}
 \label{tab-lagsim}
\end{table}

\begin{figure}
\includegraphics[bb=60 216 540 612, scale=0.48]{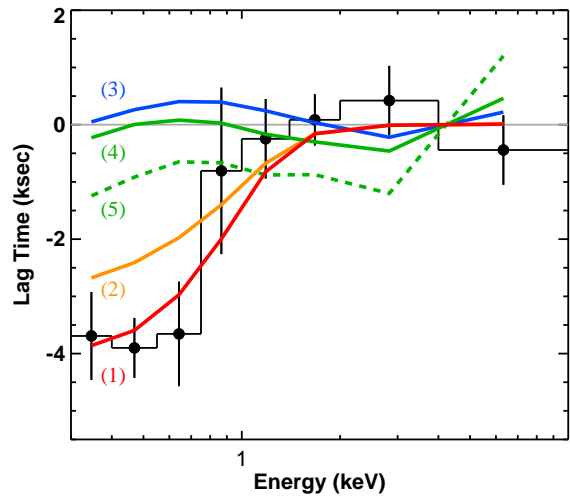}
\caption{Comparison between the LF lag spectrum (the same as Fig.\ref{fig-cohlag}b) and the time-lag predictions of various models. Curves are simulated from the spectral decompositions in Fig.\ref{fig-specfit1} with (1) CompTT-SE model, (2) CompTT-Disc model, (3) Reflection-Disc model and (4) Reflection-Jet model, respectively. The CompTT-SE model shows the best match to the lag spectrum. The green dash curve (5) is also based on the Reflection-Jet model but with the new fitting in Fig.\ref{fig-refljet2}. }
\label{fig-lag-mo}
\end{figure}

\subsection{Modelling the LF Lag Spectrum}
Comptonisation and reflection dominated models have different
predictions on the time lag. In the Comptonisation-dominated models, a
fluctuation propagates from the soft X-ray region to the hard, thereby
introducing a hard X-ray time lag. In the reflection-dominated model,
the reflection component lags behind the primary corona component,
thereby producing a time lag around the broad K$\alpha$ band and a
soft X-ray lag. However, the mixed contributions from different
components in every energy bin can easily dilute the time lag signal
between different energy bands. Therefore, we also perform a light
curve simulation to show the predicted time lags in every model and
compare them to the LF lag spectrum.

We simulate light curves from the 2-10 keV PSD in the same way as in
the previous section. Similar to previous studies (e.g. Kara et
al. 2014; Alston, Done \& Vaughan 2014), we assume identical light
curves for every spectral component, but shift them by an assumed time
lag, and then combine them in every energy bin according to the
fractional contribution of each component as derived from the fitting
of the time-averaged spectra in Fig.\ref{fig-specfit1}. This method
does not take into account of any intrinsic difference in the PSD of the
various spectral components, but it can still reveal the effect of
multi-component contributions within the lag spectrum. We examined
various time lag assumptions between different spectral components in
all models, in order to obtain the best match to the observed LF lag
spectrum. The time lag assumptions are listed in
Table~\ref{tab-lagsim}. Note that 5000 s is approaching the upper
limit for the detectable time lag in the (5-9) $\times10^{-5}$ Hz
frequency band. For every model, the simulation was repeated for
10000 times to derive the mean lag spectrum.

\begin{figure}
\includegraphics[bb=30 140 570 620, scale=0.45]{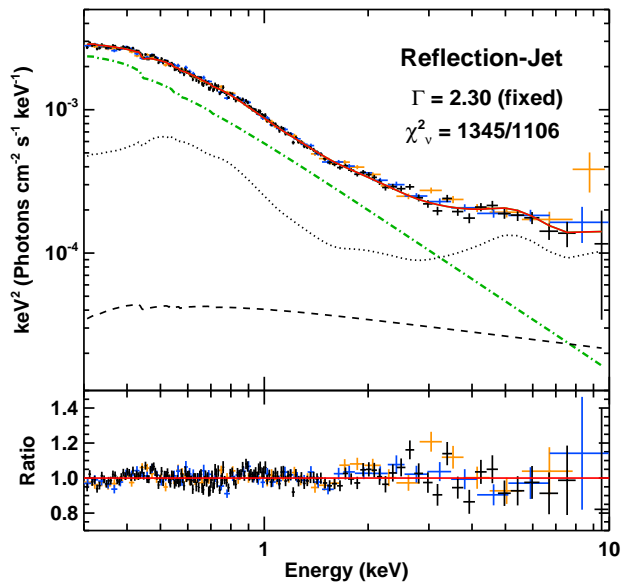}
\caption{Refitting the time-averaged spectrum with the Reflection-Jet model, but with the hard X-ray $\Gamma$ fixed at 2.3 in order to reduce the contribution of the reflection component in the soft excess. Spectral components are the same as those in Fig.\ref{fig-specfit1}d. The best-fit parameters are listed in Table~\ref{app-tab-refljet2}.}
\label{fig-refljet2}
\end{figure}

Fig.\ref{fig-lag-mo} shows that it is difficult for the
reflection-dominated models to reproduce the LF lag spectrum with the
reverberation lag (also see Fig.\ref{app-fig-cohlag1}d for the equivalent result relative to the 0.3-1 keV band).
This is mainly because the reflection component has
strong contributions across the entire 0.3-10 keV band, especially
dominating the 0.3-1 keV band, and so the time lag relative to the 2-10 keV
band is heavily diluted, and can only produce small lag signals around zero.
We also tried fixing the hard X-ray $\Gamma$ at 2.3 in
order to produce a more similar spectral decomposition to that
obtained for PG 1244+026 in Kara et al. (2014)
(Fig.\ref{fig-refljet2}). For this new fit, we also tried to add a weak disc component to the soft excess similar to the CompTT-SE model, but its normalisation was found to be consistent with zero within less than 2$\sigma$, so we no longer consider any disc emission in it. This new fit is considerably worse than the
free $\Gamma$ case by $\Delta\chi^2=46$ for one less free
parameter. But this spectral decomposition can produce a negative lag
at soft X-rays (green dash curve in Fig.\ref{fig-lag-mo}), because now the
0.3-1 keV is dominated by the jet component instead of the reflection
component. However, even the assumption of 5 ks time lag can only
produce $\sim$-1 ks lag in the soft X-ray band, and so this model still
fails to reproduce the observed LF lag spectrum.

On the other hand, it is much easier for the Comptonisation-dominated
models to reproduce the LF lag spectrum with the propagation
lag. Fig.\ref{fig-lag-mo} shows that both CompTT-Disc model and
CompTT-SE model can reproduce the broad lag profile below 1 keV, but
CompTT-Disc model slightly under-predicts the time-lag below 1 keV
because the hard X-ray component extends into the 0.3-1 keV band more
significantly than in the CompTT-SE model. The LF lag spectrum favours
a separate variable component dominating the soft excess such as in
the CompTT-SE model.

A full spectral timing modelling should consider all properties
including the PSD, RMS, covariance, coherence and lag spectra, and
treat the variability propagation between different spectral
components self-consistently. Gardner \& Done (2014) performed this
study for PG 1244+026. Their results also support the spectral
decomposition of the CompTT-SE model, and indicate a weak reflection
component in the soft X-ray region (similar as in
Fig.\ref{fig-comptrefl}b) which is mainly required by the HF soft
X-ray lag detected in PG 1244+026. Considering the spectral similarity
between PG 1244+026 and \rxj, we speculate that similar results could
be obtained for \rxj\ using the same analysis, but this is beyond the
scope of this paper.

\begin{table}
 \centering
   \caption{Best-fit parameters in 
     Fig.\ref{fig-refljet2}. `low' indicates the parameter reaches its lower limit separately during the fitting. {\sc powerlaw} normalisation is given in units of $\rm photons~keV^{-1}cm^{-2}s^{-1}$ at 1 keV.}
\begin{tabular}{@{}clll@{}}
\hline
  Model & Component & Parameter & Value\\
  \hline
{\sc Reflection}   & (Fig.\ref{fig-refljet2}) & \multicolumn{2}{l}{$\chi^{2}_{\rm \nu}~=~1345/1106~=~1.22$} \\
{\sc  -Jet  }    & {\sc tbnew}     & $N_{\rm H}$ (10$^{22}$ cm$^{-2}$) & 0.023~$^{+0.008}_{-0.008}$\\
	     & {\sc powerlaw1}     & $\Gamma$                   & 3.61~$^{+0.08}_{-0.07}$\\
             & {\sc powerlaw1}     & norm                       & 6.13~$^{+0.27}_{-0.24}\times 10^{-4}$\\
             & {\sc powerlaw2}   & $\Gamma$                   & 2.30 fixed\\
             & {\sc powerlaw2}   & norm                       & 4.27~$^{+0.58}_{-0.38}\times 10^{-5}$\\
             & {\sc kdblur}    & Index                      & 5.76~$^{+0.56}_{-0.53}$\\
             & {\sc kdblur}    & $R_{\rm in}$ ($R_{\rm g}$)           & 2.69~$^{+0.16}_{-0.13}$\\
             & {\sc rfxconv}   & rel\_refl ($\Omega/{2\pi}$)         & -10.0~$^{+1.2}_{-0~low}$\\
             & {\sc rfxconv}   & $Fe_{\rm abund}$ (Solar)    & 1.29~$^{+0.16}_{-0.16}$\\
             & {\sc rfxconv}   & log $\xi$                   & 2.57~$^{+0.12}_{-0.09}$\\
             & {\sc const}     & (MOS1) & 0.996~$^{+0.008}_{-0.008}$ \\
             & {\sc const}     & (MOS2) & 1.000~$^{+0.008}_{-0.008}$ \\
\hline
\end{tabular}
 \label{app-tab-refljet2}
\end{table}

\section{Discussion}
\label{sec-discussion}
\subsection{Origin of the Soft Excess in \rxj}
\label{sec-discussion-geo}
The time-averaged X-ray spectra are degenerate to both Comptonisation
and reflection dominated models. But the frequency-dependent RMS and
covariance spectra clearly suggest that the soft excess in
\rxj\ is dominated by a separate component with strong LF
variability, while the hard X-ray component contains stronger HF
variability. The covariance spectra also indicates a weak non-variable
component in the soft excess which is consistent with an origin in the
inner disc. Moreover, the HF covariance spectra suggest that the seed
photons for the hard X-ray Comptonisation in the hot corona should
have a higher temperature than the inner accretion disc, and so are
more likely to come from the soft X-ray Comptonisation region. These
results clearly favour the CompTT-SE model. Then our time lag analysis
shows that in the LF band the soft X-rays
lead the hard by $\sim$ 4 ks and the lag spectrum has a broad profile in
the hard X-ray band, which is most plausibly modelled as the
propagation lag in the CompTT-SE model.

We note that our results are also consistent with the study of another
example of a high mass accretion rate `simple' NLS1 PG 1244+026
(J13; Alston, Done \& Vaughan 2014; Gardner \& Done 2014). Although
PG 1244+026 is radio-quiet, it does have some weak radio emission, so
a contribution from the jet component was proposed as the origin of its
soft X-ray excess (Kara et al. 2014). However, \rxj\ has not been 
detected in any radio surveys, including the Parkes-MIT-NRAO (PMN)
survey (Wright et al. 1994). This indicates that its radio emission is
below the PMN's detection limit of $\lesssim$ 50 mJy at 4.85 GHz,
implying $L_{\rm 4.85GHz}<4\times10^{41}$ erg~s$^{-1}$. Compared to its
optical flux of $L_{5100\AA}=2\times10^{44}$ erg~s$^{-1}$ (Grupe et
al. 2004), we can conclude that \rxj\ has no significant radio
emission. This makes a jet origin much less likely for the soft X-ray excess
in \rxj, yet the spectrum clearly has many similarities to PG 1244+026.
More compellingly, the jet origin for the soft excess completely fails to fit
the low frequency lag spectra of Fig.\ref{fig-lag-mo}. 

Based on all the above results, we can conclude that the soft excess
in \rxj\ is most likely produced in an extended intermediate region between
the inner thermal disc and the compact hard X-ray corona (see Fig.\ref{fig-innerdisc1}).
The low temperature electrons in this intermediate region are optically thick to the
Compton up-scattering of photons from the inner thermal disc, thereby producing the
soft excess. Shielding the inner thermal disc, this intermediate region then
provides soft X-ray seed photons for the high temperature, optically
thin electrons in the hot corona to produce the hard X-ray
Comptonisation emission. Part of the emission from the hard X-ray
region also reaches the soft X-ray region and may even penetrate into
the thermal disc region behind, which is then partially reprocessed
(thermalised) at the temperature close to the soft excess and
partially reflected as a weak reflection component, thereby causing
the HF variability in the soft excess.

We emphasise that in this paper we mainly compared extreme and simplest models where either Comptonisation or reflection dominates the soft excess. But it is possible to have several components contributing the X-ray emission simultaneously, such as thermal disc, Comptonisation, reflection, weak jet (but \rxj\ has no radio emission) with more complex geometries (e.g. patchy corona, Wilkins \& Gallo 2015). Our Fig.\ref{fig-comptrefl} and Fig.\ref{fig-innerdisc1} already show an example of having both reflection and Comptonisation in the spectra. What we report is that the CompTT-SE model can provide the simplest plausible explanation to all the spectral-timing properties of \rxj, while the other models cannot. Certainly we cannot rule out the possibility of having more complex geometry designs in order to make other scenarios work in \rxj, but such study is beyond the scope of this work, and we prefer the simplest solution.

\subsection{Properties of the Soft X-ray Emitting Region}
\begin{figure}
\includegraphics[bb=0 10 612 620, scale=0.335]{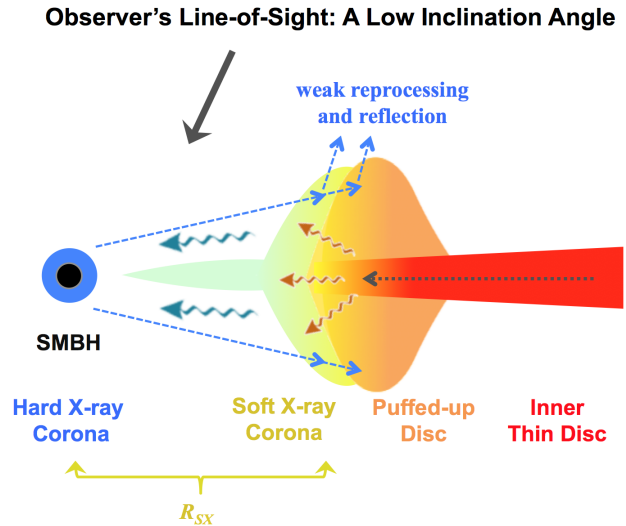}
\caption{A schematic diagram of the inferred structure of the inner accretion flow in \rxj\ and other super-Eddington `simple' NLS1s. The soft X-ray corona is likely to be associated with the puffed-up inner disc region in these super-Eddington sources (see Section~\ref{sec-discussion-geo}). $R_{\rm sx}$ is the distance between the extended soft X-ray corona and the compact hard X-ray corona estiamted from the $\sim$ 4 ks time-lag in \rxj, which is $\sim 80$ $R_{\rm g}$ for $M=10^{7}$ M$_{\odot}$ and $\sim 200$ $R_{\rm g}$ for $M=4\times10^{6}$ M$_{\odot}$. A low inclination angle of $30\degr$ is adopted for \rxj\ and is favoured by the clean line-of-sights in all the `simple' NLS1s.}
\label{fig-innerdisc1}
\end{figure}

The radial distance of this soft X-ray region to the black hole (i.e. $R_{\rm sx}$ in Fig.\ref{fig-innerdisc1}) is still unknown. But if the hard X-ray corona is fairly compact and close to the black hole (Galeev, Rosner \& Vaiana 1979; Haardt \& Maraschi 1991), and the soft X-ray region provides seed photons to the hard X-ray corona, then the 4 ks soft/hard X-ray time-lag can be used to infer $R_{\rm sx}$. For the single-epoch mass of $3.9\times10^{6}$ M$_{\odot}$ (Grupe et al. 2004), the light travel distance would imply $R_{\rm sx}\simeq200$ $R_{\rm g}$. This mass contains a factor of few uncertainty (e.g. Woo \& Urry 2002), thus a black hole mass range of $10^{6}-10^{7}$ M$_{\odot}$ would correspond to a $R_{\rm sx}$ range of $80-800~R_{\rm g}$. However, this $R_{\rm sx}$ is estimated using the speed of light, but it is also possible for some of the soft X-ray variability to propagate into the hard X-ray corona in the accretion flow (but we don't have a clear picture about it), so that this $R_{\rm sx}$ may be an upper limit for a fixed mass.

We also note that Grupe et al. (2004) reported $L/L_{\rm Edd}=12.9$ for their mass estimate, then $M=10^{6}$ M$_{\odot}$ would imply $L/L_{\rm Edd}=50.3$, which is much higher than all other AGN known, so it is probably not likely for \rxj\ to have such a small black hole mass (see our Paper-II for more detailed study), and so $R_{\rm sx}$ may not be as large as  $800~R_{\rm g}$. Anyway, what the observation requires is an extended and geometrically thick soft X-ray region, sitting between the compact hard X-ray corona and the inner thermal disc. Since $R_{\rm sx} \propto M^{-1}$, $\dot{m} \propto M^{-2}$ for an observed optical luminosity (Davis \& Laor 2011), and the puffed-up disc radius ($R_{\rm pf}$) is roughly proportional to $\dot{m}$ (Poutanen et al. 2007), reducing $M$ will increase $R_{\rm sx}$ linearly, but increase $R_{\rm pf}$ quadratically, thus the condition of $R_{\rm sx} \lesssim R_{\rm pf}$ would be easier to satisfy.

We point out that the soft X-ray region is likely to be geometrically thick, which is because it needs to shield the hot corona from the thermal disc photons from several tens or hundreds of $R_{\rm g}$ away. Meanwhile, a so-called `puffed-up' disc region has been proposed to explain the weak optical/UV emission lines in weak-line quasars where high Eddington ratios are also observed (e.g. Madau 1988; Leighly 2004; Luo et al. 2015). Indeed, it is known that the accretion disc begins to deviate from a standard thin disc (Shakura \& Sunyaev 1973) when the mass accretion rate approaches Eddington limit, in which case the disc becomes slim (e.g. Abramowicz et al. 1988; Wang \& Netzer 2003), and is accompanied by significant advection and radiation driven disc wind (e.g. Ohsuga \& Mineshige 2011; Jiang, Stone \& Davis 2014; S\c{a}dowski \& Narayan 2015; Hashizume et al. 2015; Hagino et al. 2016; DJ16). The shielding mechanism provided by the soft X-ray region is very similar to the function of the puffed-up inner disc region. In quasars of $M\gtrsim10^{8}$ M$_{\odot}$, the puffed-up region needs to be high enough and very close to the X-ray corona ($\sim10~R_{\rm g}$, Luo et al. 2015 and references therein), but the disc in \rxj\ must be much hotter because of its smaller black hole mass and higher mass accretion rate, and so the puffed-up region can be farther away, which enhances the possibility for the link between the soft X-ray region and the puffed-up inner disc region. We will report more detailed study on this in a subsequent paper on the multi-wavelength spectrum of \rxj\ (Paper-II).

\section{Conclusions}
\label{sec-conclusions}
In this paper we report results of our X-ray analysis of an
unobscured, highly super-Eddington QSO \rxj\ based on a recent 133 ks
\xmm\ observation proposed by us. We show that the soft excess and
hard X-rays of \rxj\ exhibit different spectral timing properties,
which allows us to distinguish between Comptonisation and
reflection-dominated models for the spectral decomposition. We find
that the reflection-dominated models can be ruled out for \rxj, and
the X-ray emission from this source is most plausibly explained as the
combination of the inner disc emission in the very soft X-ray band,
the hard X-ray corona emission, and the soft X-ray Comptonisation
emission dominating the soft excess, which is produced by a low
temperature, optically thick electron population at several tens or hundreds of
$R_{\rm g}$ radii (implied by the $\sim$4 ks soft X-ray leading the hard),
which receives seed photons from the inner
disc and also provides soft X-ray seed photons for the hard X-ray corona.
This soft X-ray region is likely to be geometrically thick and associated with
a puffed-up inner disc region.

While the general picture is becoming clearer, we note a significant difference
between the high frequency RMS and covariance spectra which implies
that there is an additional uncorrelated fast variability component
above 4~keV. The origin and properties of this component are not yet
known, and it will require future observations with better
signal-to-noise to constrain its properties.

\section*{Acknowledgements}
We thank the anonymous referee for providing useful comments to improve the quality of the paper.
CJ acknowledges the support by the Bundesministerium f\"{u}r Wirtschaft und Technologie/Deutsches Zentrum f\"{u}r Luft- und Raumfahrt (BMWI/DLR, FKZ 50 OR 1408 and FKZ 50 OR 1604) and the Max Planck Society.
CD and MJW acknowledge STFC funding under grant ST/L00075X/1.
This work is based on a recent observation conducted by \xmm, an ESA
science mission with instruments and contributions directly funded by
ESA Member States and the USA (NASA). This research has made use of the NASA/IPAC Extragalactic Database (NED) which is operated by the Jet Propulsion Laboratory, California Institute of Technology, under contract with the National Aeronautics and Space Administration.








\appendix

\onecolumn
\centering

\section{Fitting Results for the Reflection Models at $60\degr$ Inclination Angle}
\label{app-60d}
\begin{figure*}
\centering
\begin{tabular}{cc}
\includegraphics[bb=30 124 540 612,scale=0.41]{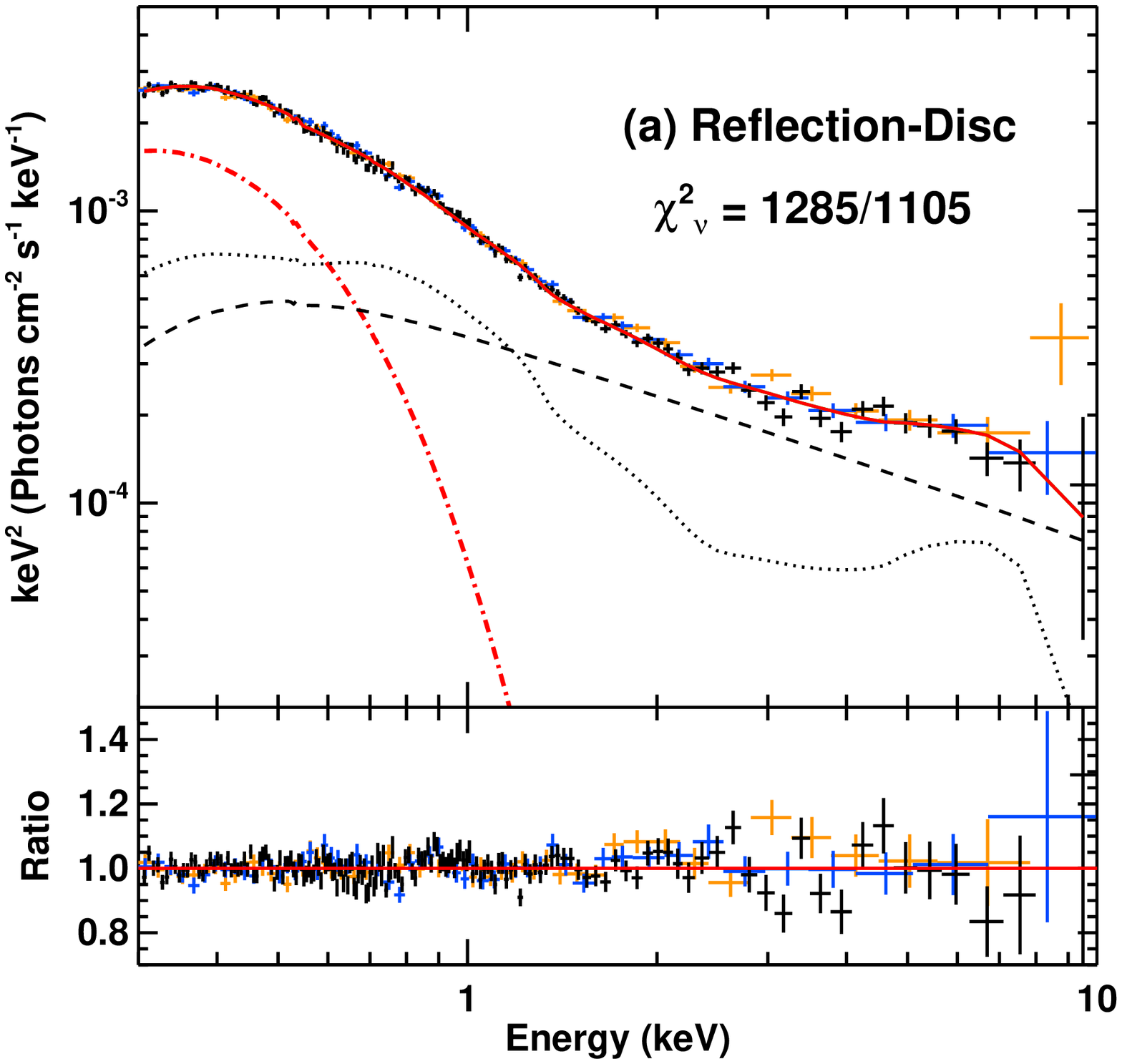} &
\includegraphics[bb=30 124 540 612,scale=0.41]{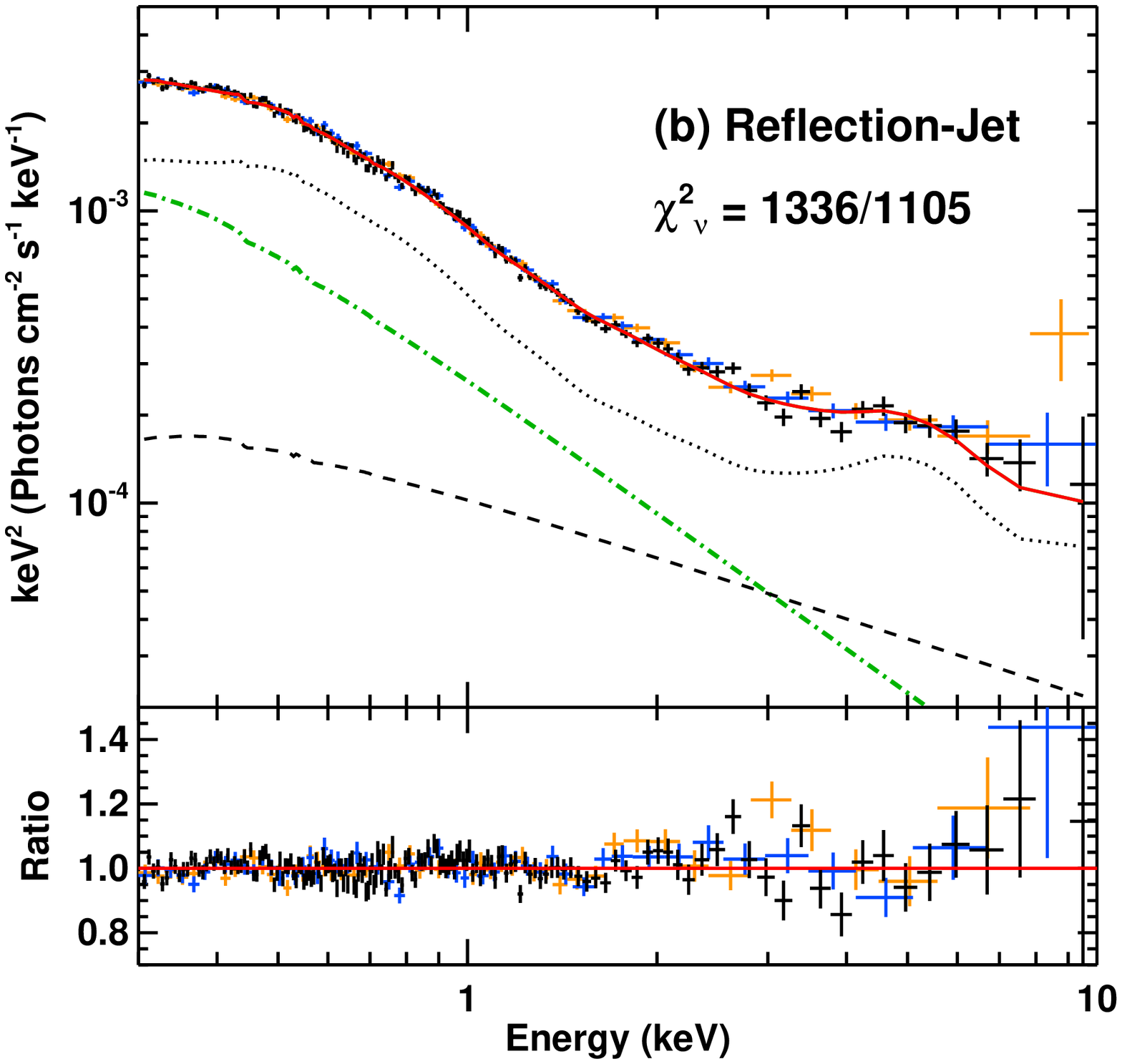} \\
\end{tabular}
\caption{Best-fit results for the two reflection models with the disc inclination angle fixed at $60\degr$, which can be compared to the two fittings in Fig.\ref{fig-specfit1}c,d for $30\degr$. The best-fit parameters can be found in Table~\ref{app-tab-specfit}.}
\label{app-fig-specfit1}
\end{figure*}

\begin{table*}
 \centering
   \caption{Best-fit parameters of the two fittings in Fig.\ref{app-fig-specfit1} with the disc inclination angle fixed at 60$\degr$. These results can be compared to the best-fit results in Table~\ref{tab-specfit}c,d. The upper and lower limits are for the 90\% confidence range. {\it low} ({\it up}) indicates the parameter's lower (upper) limit. {\sc diskbb} normalisation is defined as $(R_{\rm in}/D_{10})^2~cos~\theta$, where $R_{\rm in}$ is the apparent inner disc radius in km, $D_{10}$ is the source distance in units of 10 kpc, $\theta$ is the disc inclination angle. {\sc nthcomp} and {\sc powerlaw} normalisations are given in units of $\rm photons~keV^{-1}cm^{-2}s^{-1}$ at 1 keV.}
     \begin{tabular*}{\textwidth}{@{}cc@{}}
     
\begin{tabular}{@{}clll@{}}
\hline
  Model & Component & Parameter & Value\\
  \hline
{\sc Reflection}   & (Fig.\ref{app-fig-specfit1}a) & \multicolumn{2}{l}{$\chi^{2}_{\rm \nu}~=~1285/1105~=~1.16$} \\
{\sc -Disc  }   & {\sc tbnew}     & $N_{\rm H}$ (10$^{22}$ cm$^{-2}$) & 0~$^{+0.008}_{-0~low}$\\
             & {\sc diskbb}     & $T_{\rm in}$ (keV)                   & 0.105~$^{+0.003}_{-0.004}$\\
             & {\sc diskbb}     & norm                       & 2.53~$^{+0.25}_{-0.14}\times 10^{3}$\\
             & {\sc nthcomp}   & $\Gamma$                   & 2.69~$^{+0.04}_{-0.04}$\\
             & {\sc nthcomp}   & $kT_{\rm seed}$ (keV)          & tied to $kT_{\rm in}$\\
             & {\sc nthcomp}   & norm                       & 3.77~$^{+0.57}_{-0.59}\times 10^{-4}$\\
             & {\sc kdblur}    & Index                      & 5.67~$^{+0.80}_{-0.53}$\\
             & {\sc kdblur}    & $R_{\rm in}$ ($R_{\rm g}$)           & 1.58~$^{+0.12}_{-0.04}$\\
             & {\sc rfxconv}   & rel\_refl ($\Omega/{2\pi}$)                & -0.17~$^{+0.03}_{-0.17}$\\
             & {\sc rfxconv}   & $Fe_{\rm abund}$ (Solar)            & 10.0~$^{+0~up}_{-0.9}$\\
             & {\sc rfxconv}   & log $\xi$                   & 3.43~$^{+0.05}_{-0.04}$\\
             & {\sc const}     & (MOS1) & 0.993~$^{+0.008}_{-0.008}$ \\
             & {\sc const}     & (MOS2) & 0.996~$^{+0.009}_{-0.008}$ \\
\hline
\end{tabular} 

&
\begin{tabular}{@{}clll@{}}
\hline
  Model & Component & Parameter & Value\\
  \hline
{\sc Reflection}   & (Fig.\ref{app-fig-specfit1}b) & \multicolumn{2}{l}{$\chi^{2}_{\rm \nu}~=~1336/1105~=~1.21$} \\
{\sc  -Jet  }    & {\sc tbnew}     & $N_{\rm H}$ (10$^{22}$ cm$^{-2}$) & 0~$^{+0.003}_{-0~low}$\\
             & {\sc powerlaw1}     & $\Gamma$                   & 3.44~$^{+0.34}_{-0.07}$\\
             & {\sc powerlaw1}     & norm                       & 3.21~$^{+0.31}_{-1.46}\times 10^{-4}$\\
             & {\sc powerlaw2}   & $\Gamma$                   & 2.61~$^{+0.04}_{-0.02}$\\
             & {\sc powerlaw2}   & norm                       & 1.11~$^{+0.18}_{-0.07}\times 10^{-4}$\\
             & {\sc kdblur}    & Index                      & 10.0~$^{+0~up}_{-0.14}$\\
             & {\sc kdblur}    & $R_{\rm in}$ ($R_{\rm g}$)           & 1.474~$^{+0.025}_{-0.004}$\\
             & {\sc rfxconv}   & rel\_refl ($\Omega/{2\pi}$)                    & -10.0~$^{+1.0}_{-0~low}$\\
             & {\sc rfxconv}   & $Fe_{\rm abund}$ (Solar)                & 0.99~$^{+0.07}_{-0.09}$\\
             & {\sc rfxconv}   & log $\xi$                   & 2.99~$^{+0.05}_{-0.07}$\\
             & {\sc const}     & (MOS1) & 0.968~$^{+0.008}_{-0.007}$ \\
             & {\sc const}     & (MOS2) & 0.972~$^{+0.007}_{-0.007}$\\
\hline
\end{tabular}  \\

   \end{tabular*}
 \label{app-tab-specfit}
\end{table*}

\section{Time-lag and Coherence Spectra with the 2-10 keV Reference Band}

\begin{figure*}
\centering
\begin{tabular}{cccc}
\includegraphics[bb=90 144 540 720, scale=0.26]{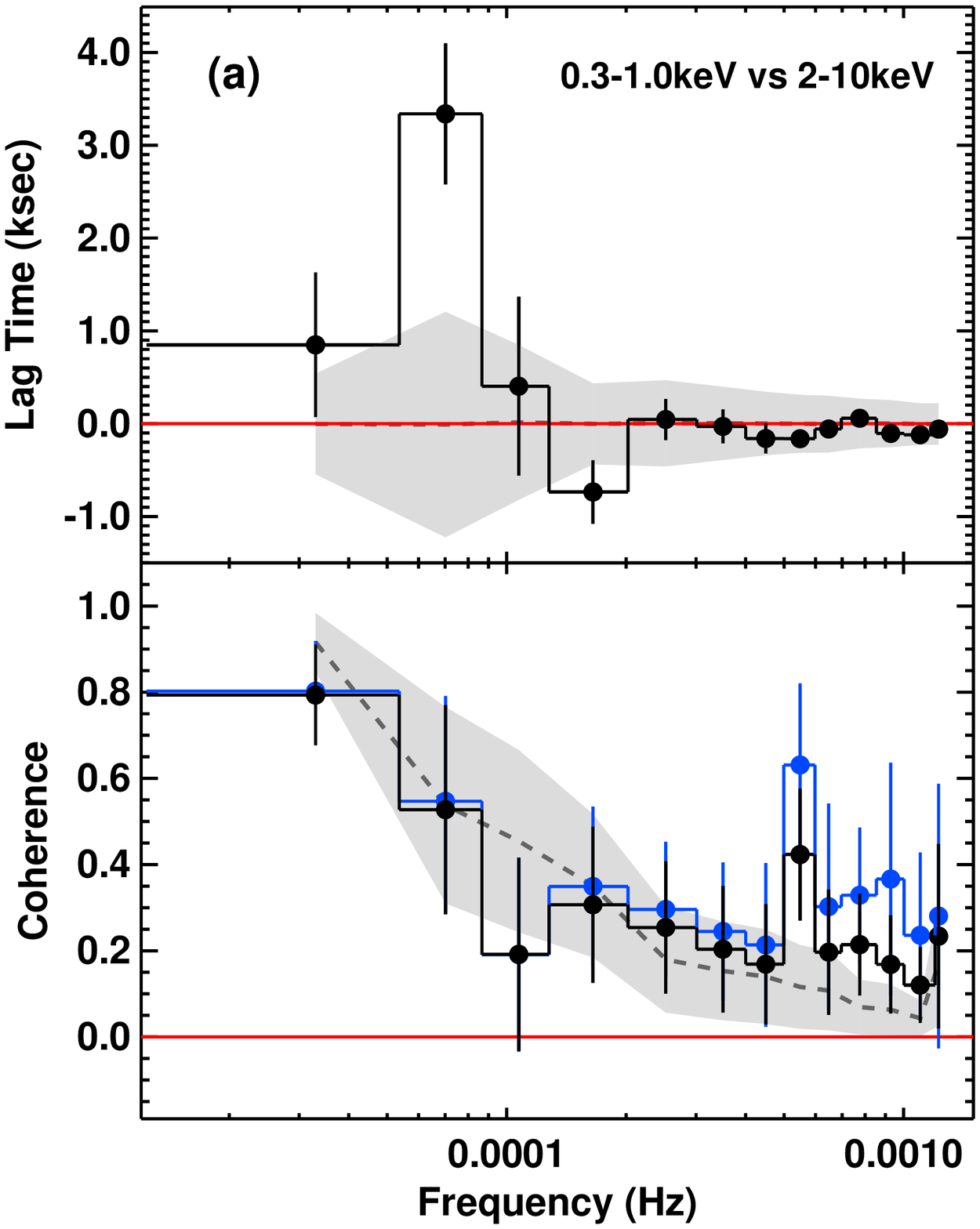} &
\includegraphics[bb=100 144 540 720, scale=0.26]{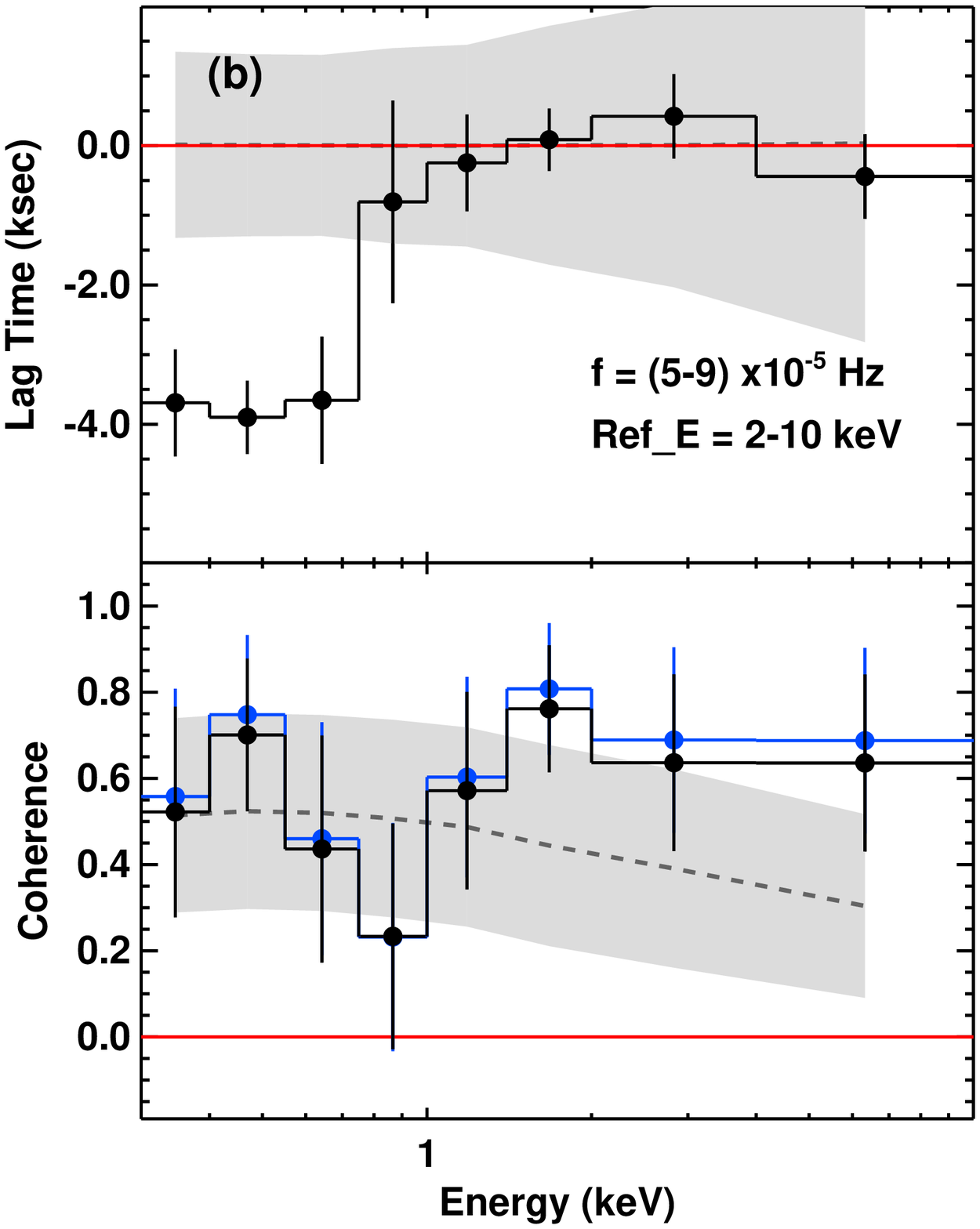} &
\includegraphics[bb=100 144 540 720, scale=0.26]{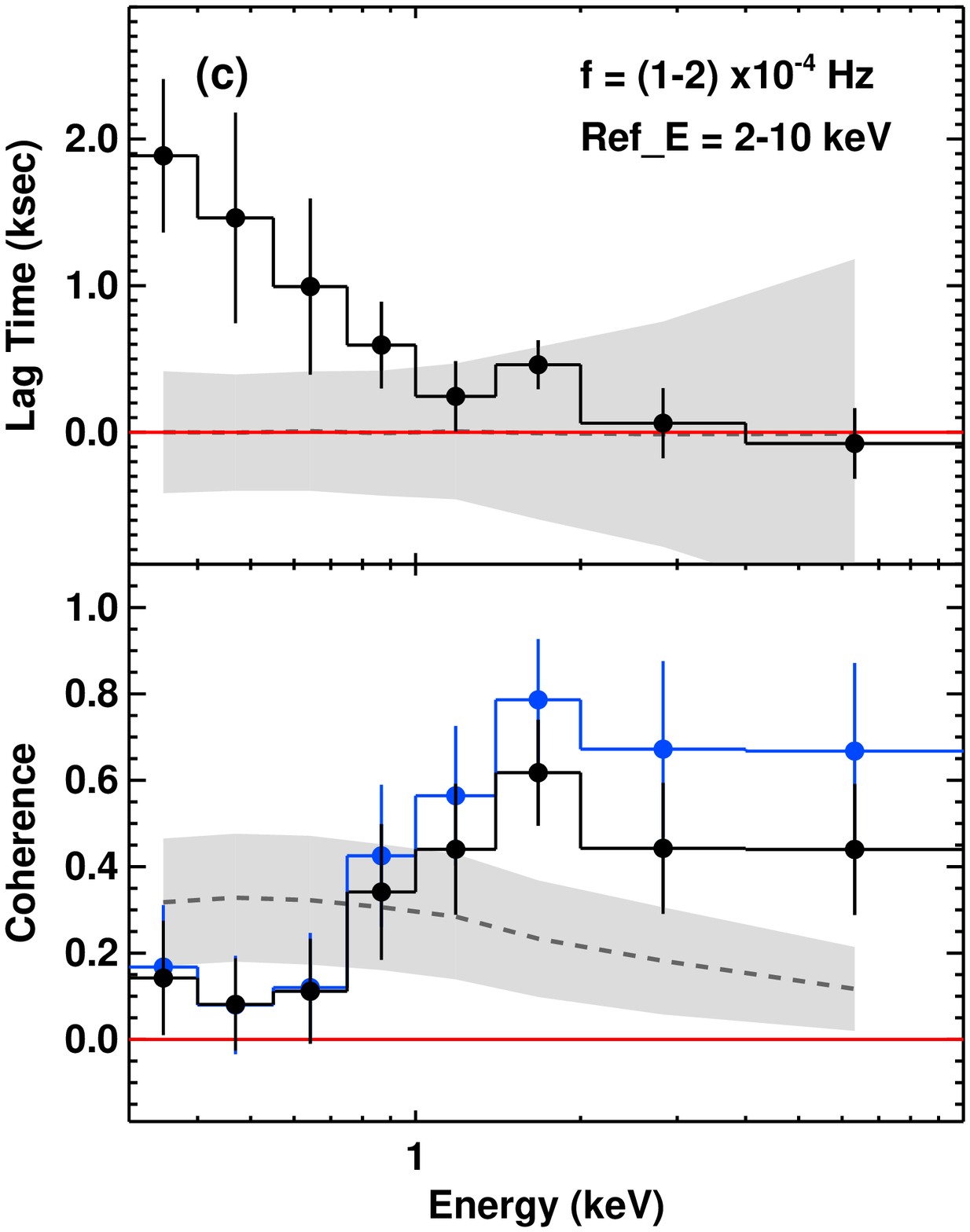} &
\includegraphics[bb=100 144 540 720, scale=0.26]{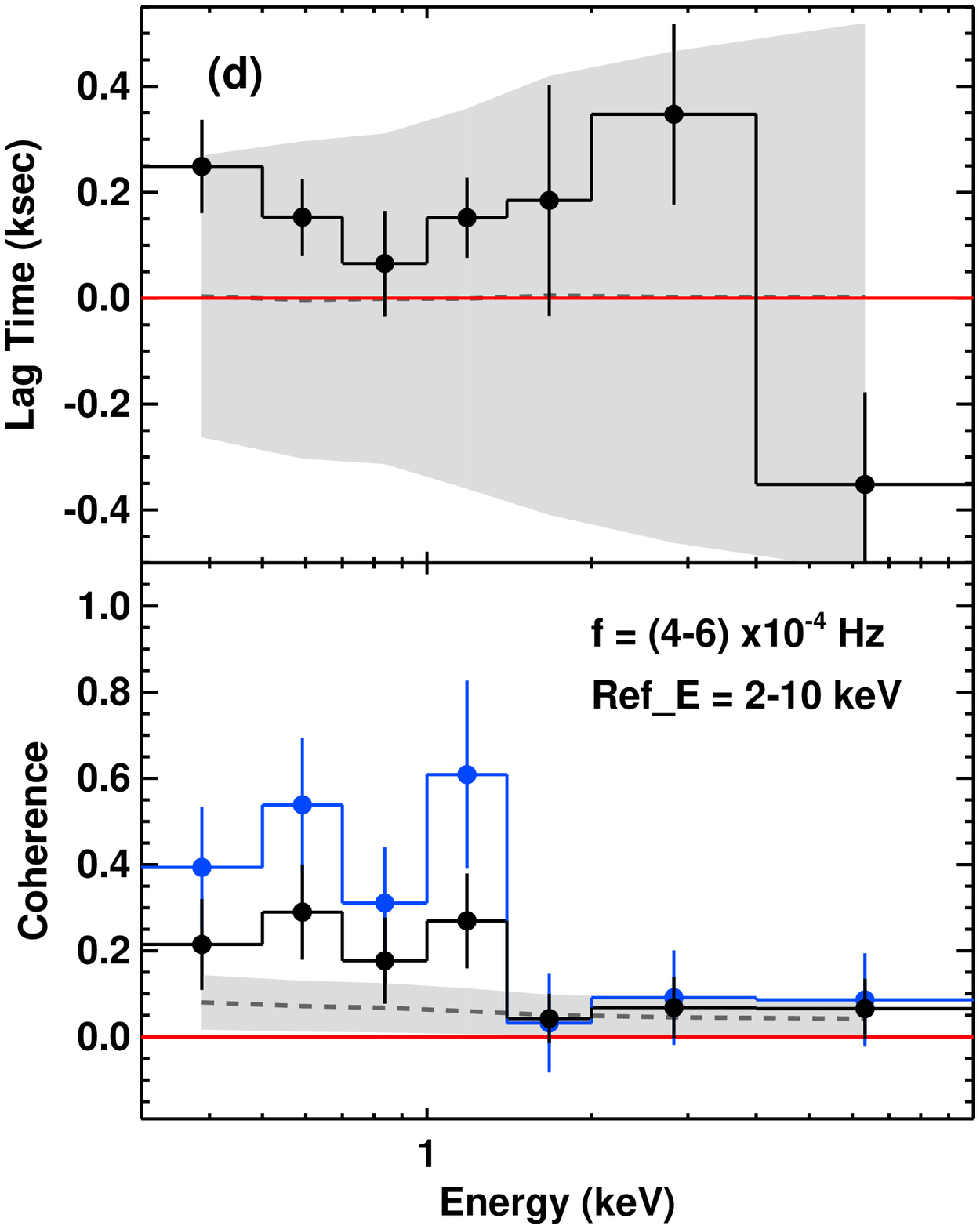} \\
\end{tabular}
\caption{Similar to Fig.\ref{fig-cohlag}, but the simulation is based on the PSD of 0.3-1 keV (see Section~\ref{sec-timelag}).}
\label{app-fig-cohlag1}
\end{figure*}

\begin{figure*}
\centering
\begin{tabular}{cccc}
\includegraphics[bb=90 144 540 720, scale=0.26]{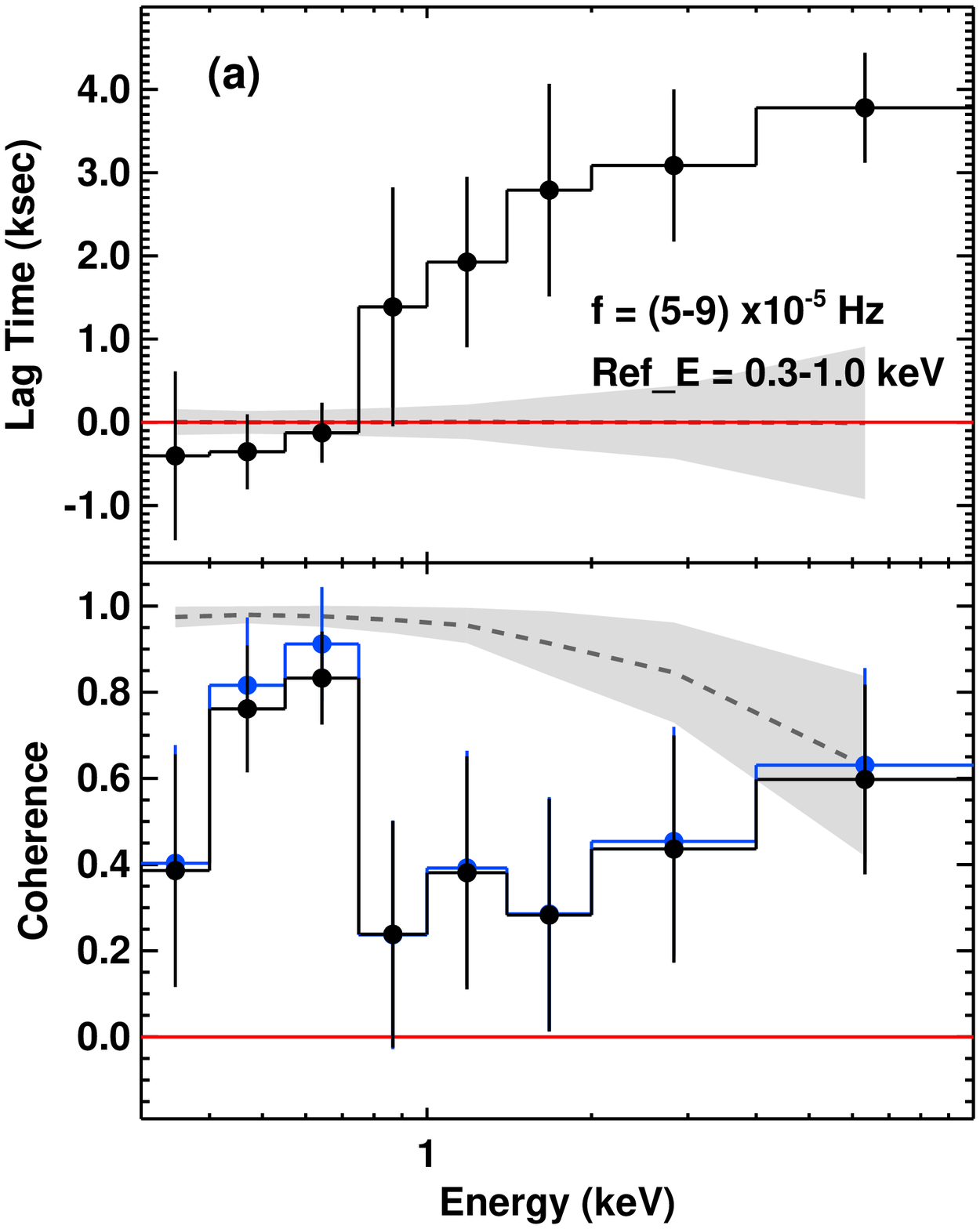} &
\includegraphics[bb=110 144 540 720, scale=0.26]{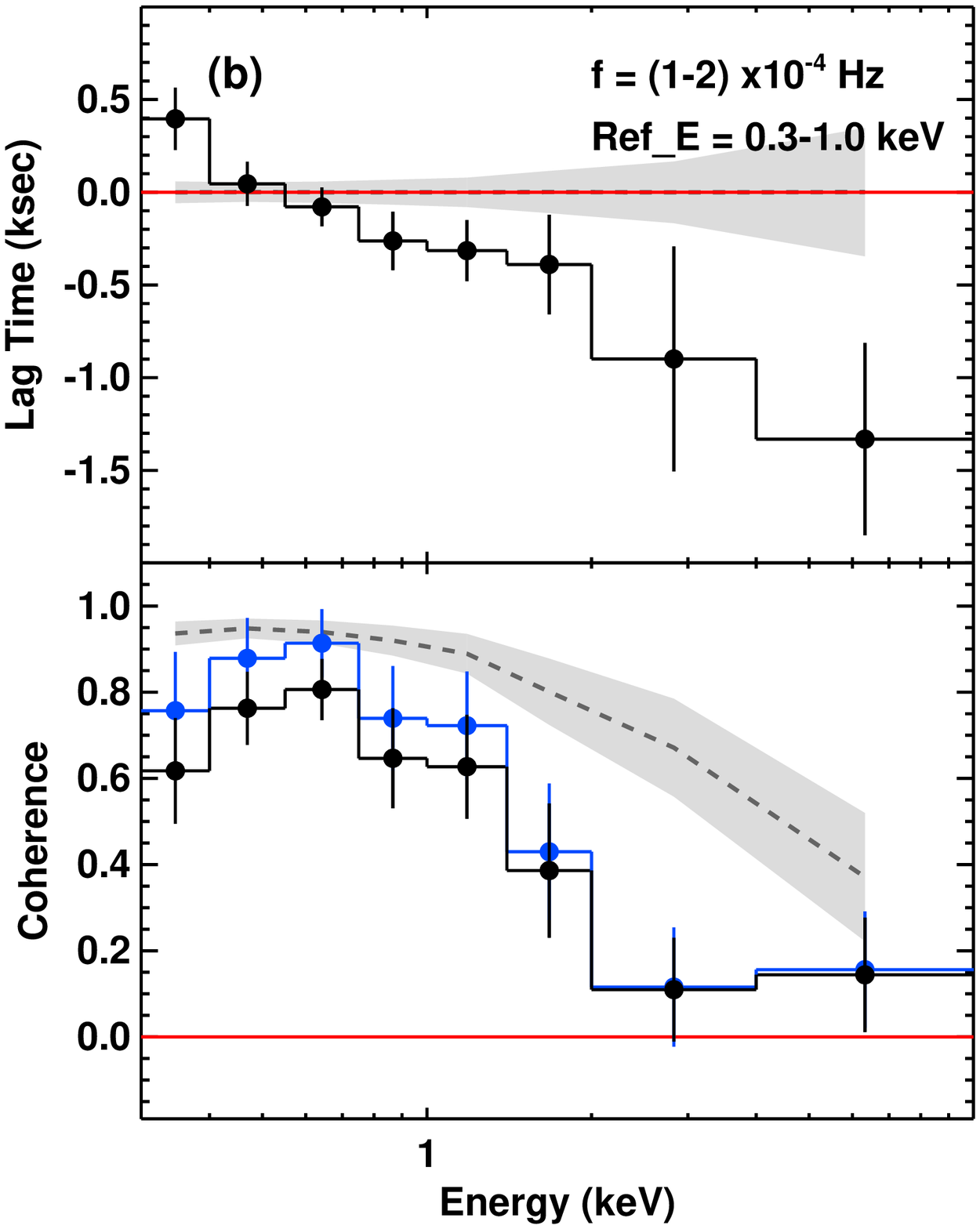} &
\includegraphics[bb=110 144 540 720, scale=0.26]{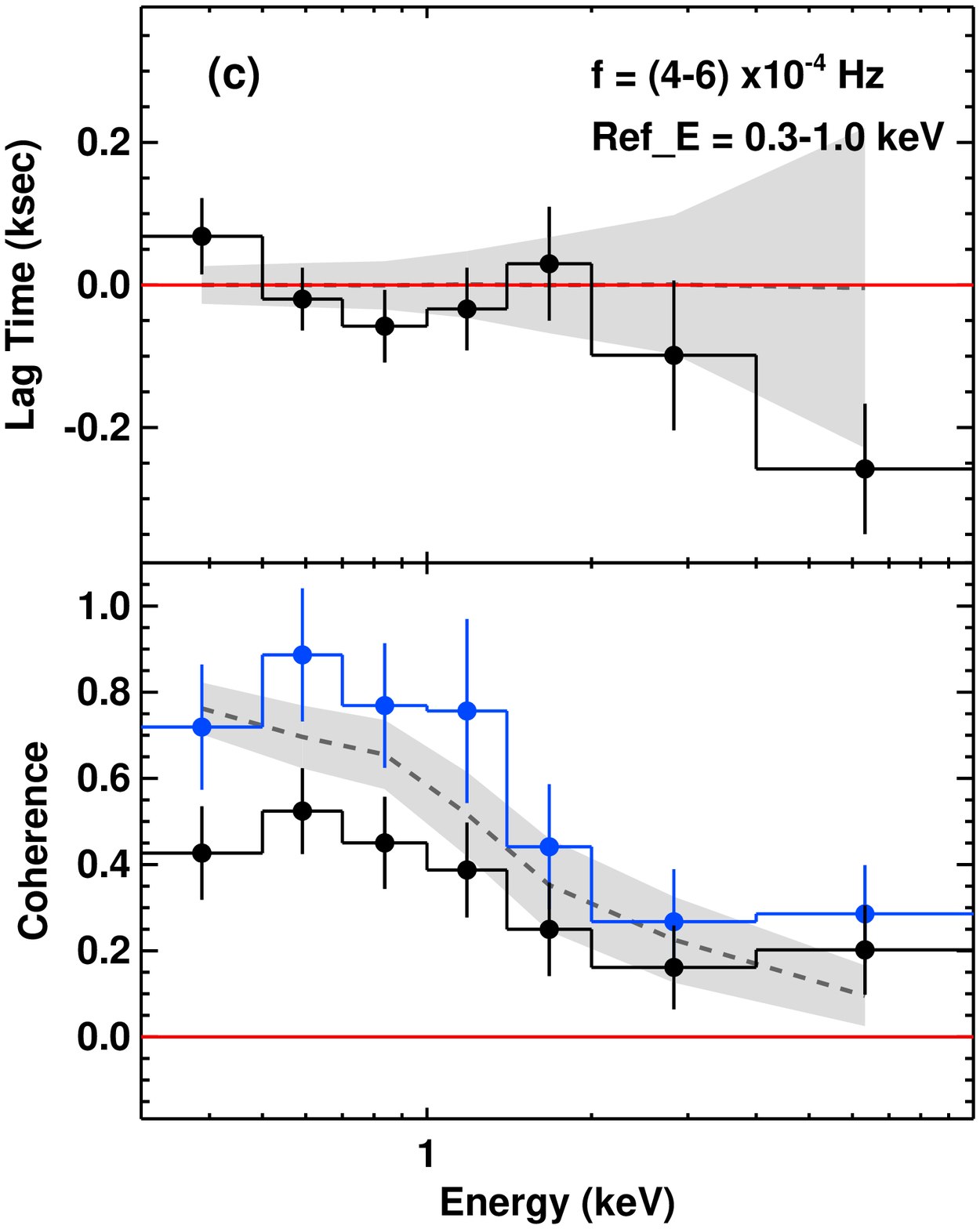} &
\includegraphics[bb=120 216 540 612, scale=0.29]{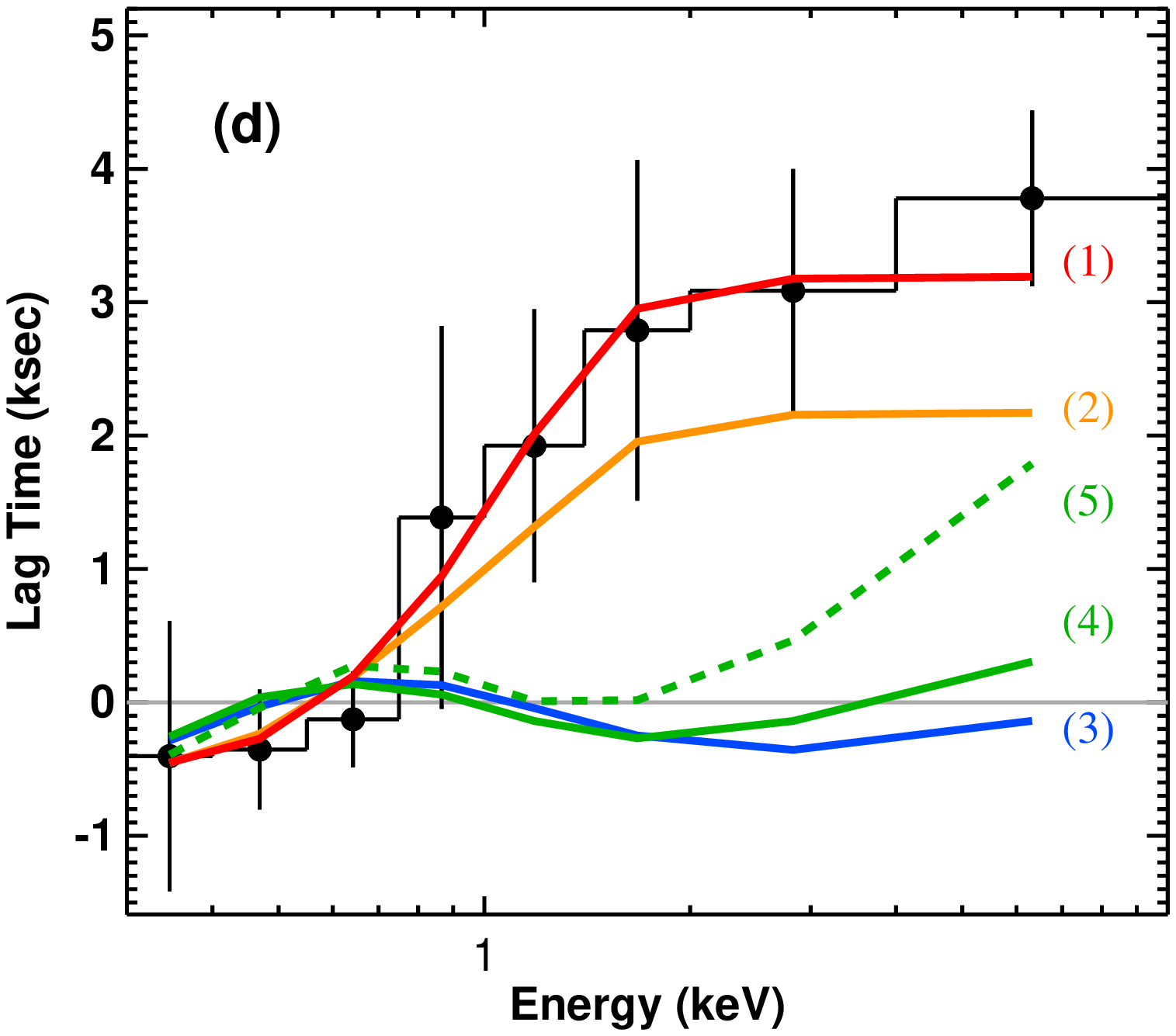}\\
\end{tabular}
\caption{Panel-a, b, c are similar to Fig.\ref{fig-cohlag}b, c, d, and Panel-d is similar to Fig.\ref{fig-lag-mo}. But these results are based on the reference band of 0.3-1 keV.}
\label{app-fig-cohlag2}
\end{figure*}

\begin{figure*}
\centering
\begin{tabular}{cccc}
\includegraphics[bb=100 144 540 720, scale=0.26]{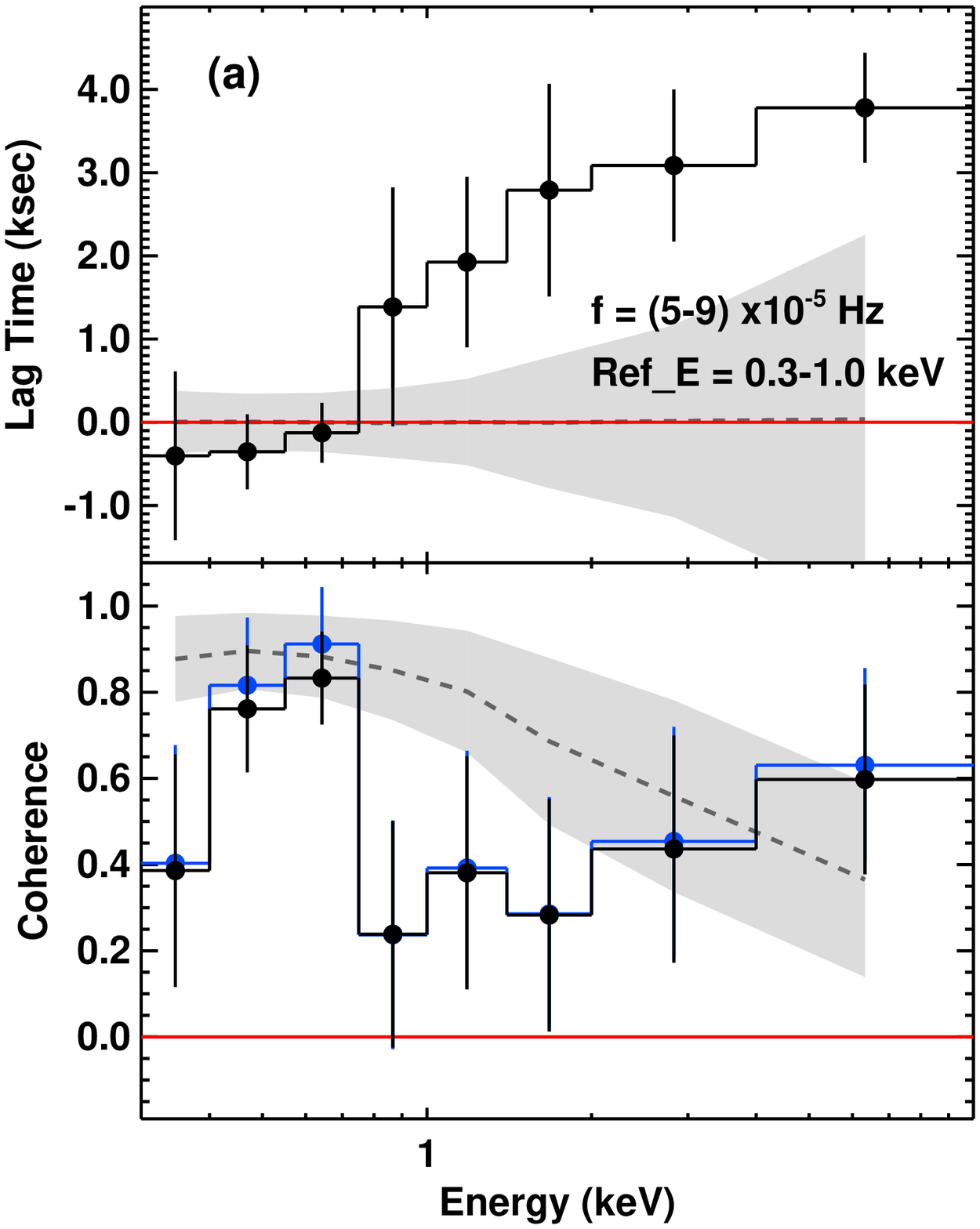} &
\includegraphics[bb=100 144 540 720, scale=0.26]{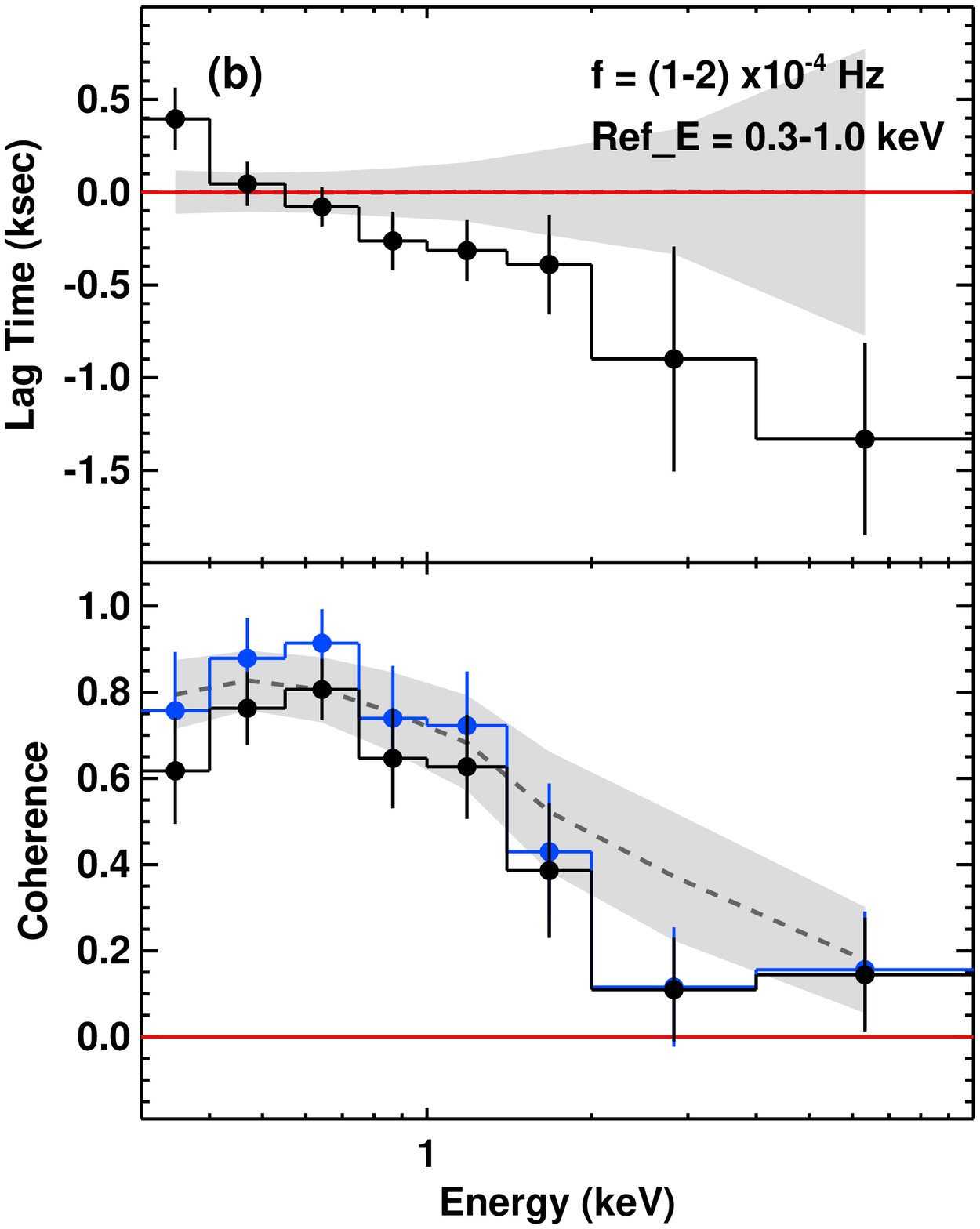} &
\includegraphics[bb=100 144 540 720, scale=0.26]{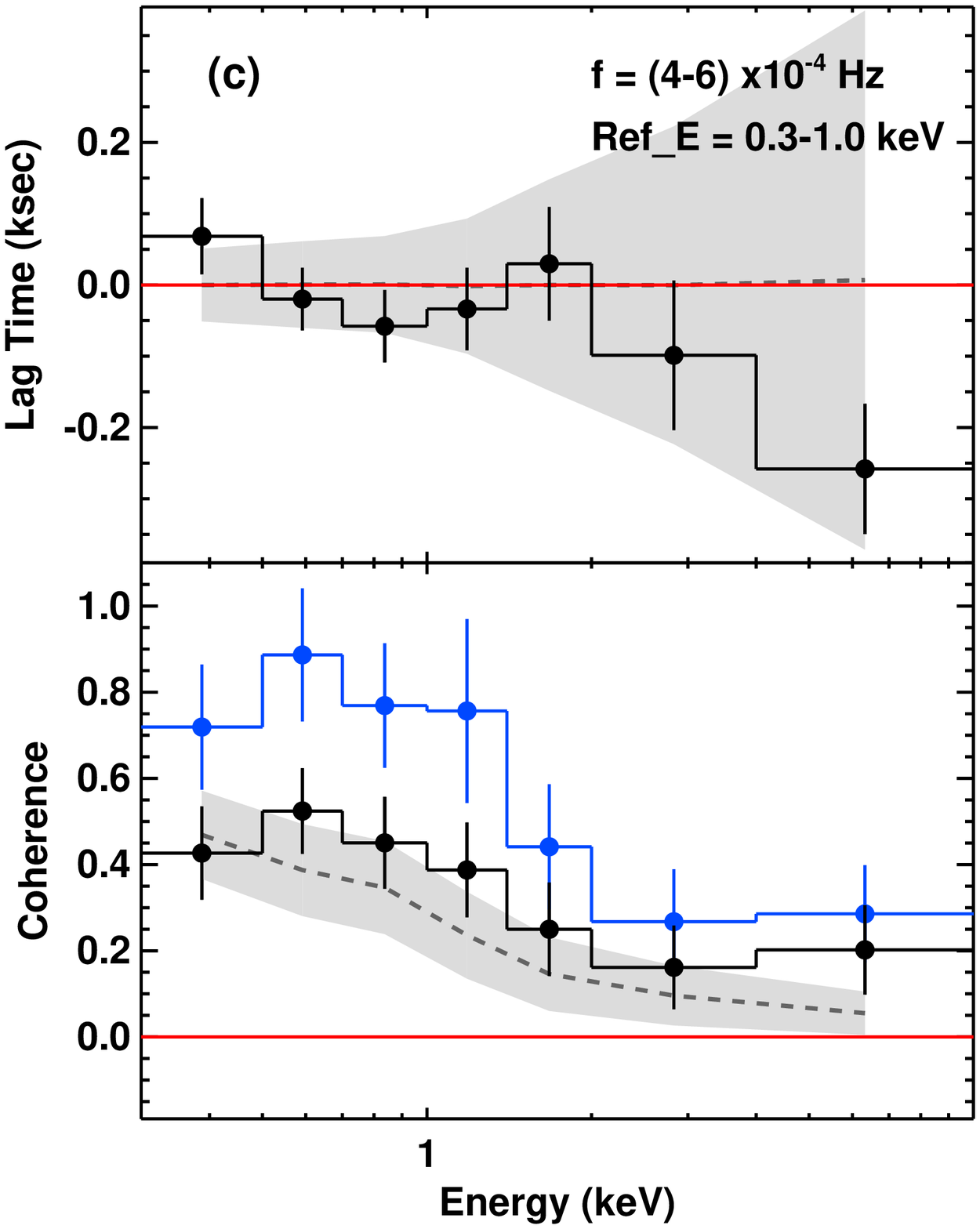} \\
\end{tabular}
\caption{Similar to Fig.\ref{app-fig-cohlag2}a, b, c, but the simulation is based on the PSD of 0.3-1 keV.}
\label{app-fig-cohlag3}
\end{figure*}



\bsp	
\label{lastpage}
\end{document}